\documentclass[a4paper,11pt]{article}
\usepackage{jheppub} 
\usepackage{float}
\usepackage{amssymb} 
\usepackage{amsmath}
\usepackage{booktabs}  
\usepackage{array}
\usepackage{amsfonts}
\usepackage{slashed}
\title{\boldmath NLO Angular Impulse and Leading Singularities to all orders in spin for Kerr Black Holes}

\makeatother

\author[a]{Lucas Haiashi}
\author[a]{and Gabriel Menezes}
\affiliation[a]{Departamento de Física Matemática - Instituto de Física, Universidade de São Paulo,\\R. do Matão, 1371, 05508-090 São Paulo -- SP, Brasil}

\emailAdd{lucashaiashi@usp.br, gsm@if.usp.br}

\abstract{
We compute the next-to-leading order (NLO) angular impulse (spin kick) in the scattering of Kerr black holes using the Kosower--Maybee--O'Connell (KMOC) formalism. Our approach is based on on-shell scattering amplitudes and leading singularities, allowing for a direct extraction of classical observables from quantum amplitudes. We derive compact expressions for the spin-dependent angular impulse valid to all orders in the spin variables at the integrand level, and show that these results reduce to known expressions in the appropriate limits. We perform detailed consistency checks: the conservative result preserves both the covariant spin supplementary condition and the spin magnitude through 2PM order, and the quadratic-in-spin conservative result agrees with existing radial-action results after translating between the direct KMOC spin kick and the radial-action observable. In addition, we extract the corresponding non-relativistic gravitational potential from the triangle leading singularities, obtaining a representation that resums spin effects and reproduces the known spin-orbit interaction at linear order. 
Our results provide further evidence for the efficiency of amplitude-based methods in classical gravitational dynamics, and highlight the KMOC formalism as a powerful framework for computing spin-dependent observables in binary black hole scattering.
}
\tikzfeynmanset{warn luatex=false}

\begin{document}

\maketitle
\flushbottom

\section{Introduction}
\label{sec:intro}

The detection of gravitational waves by the LIGO/VIRGO collaboration, arising from the merger of compact objects~\cite{LIGOScientific:2016aoc,LIGOScientific:2017vwq, LIGOScientific:2016sjg, LIGOScientific:2017ycc, LIGOScientific:2025slb}, marked a milestone in physics. This achievement and the perspective of future detectors~\cite{Punturo:2010zz,LISA:2017pwj,Ballmer:2022uxx,ET:2025xjr} have propelled the scientific community to the goal of increasing the precision for the two-body problem in general relativity~\cite{Buonanno:1998gg,Buonanno:2000ef,Damour:2000we,Damour:2001tu,Pretorius:2005gq,Goldberger:2004jt, Blanchet:2013haa,Porto:2016pyg}. 

This current trend redirected efforts into obtaining higher orders of precision to the two-body problem dynamics. On the one hand, the post-Newtonian (PN) expansion, which applies to the non-relativistic and weak-field regime, has been instrumental in this endeavor, with calculations now reaching 6PN accuracy. Concurrently, the post-Minkowskian (PM) expansion, which captures the relativistic weak-field regime, has also seen significant progress, with the 4PM correction marking the current state of the art. Achieving high precision in both frameworks is critical for improving our ability to model and detect gravitational-wave events with accuracy and reliability~\cite{Neill:2013wsa,Bjerrum-Bohr:2013bxa,Bjerrum-Bohr:2014lea,Bjerrum-Bohr:2014zsa,Levi:2016ofk,Bjerrum-Bohr:2016hpa,Bini:2017xzy,Vines:2017hyw,Bini:2018ywr,Bjerrum-Bohr:2018xdl,Moynihan:2020gxj,Chung:2020rrz,Cristofoli:2020uzm,Parra-Martinez:2020dzs,Haddad:2020tvs,AccettulliHuber:2020oou,Moynihan:2020ejh,Sahoo:2020ryf,delaCruz:2020bbn,Manu:2020zxl,Bonocore:2020xuj,Mogull:2020sak,Cheung:2020gbf,Mougiakakos:2020laz,Carrasco:2020ywq,Kim:2020cvf,Bjerrum-Bohr:2020syg,Gonzo:2020xza,delaCruz:2020cpc,Emond:2020lwi,Bern:2021dqo,Herrmann:2021lqe,DiVecchia:2021bdo,Bjerrum-Bohr:2021vuf,Brandhuber:2021kpo,Bjerrum-Bohr:2021din,Aoude:2021oqj,Brandhuber:2021eyq,Chen:2021kxt,Bautista:2021llr,Brandhuber:2021bsf,Bautista:2021wfy,Cho:2022syn,Alessio:2022kwv,Bern:2022kto,FebresCordero:2022jts,Bini:2024tft,Bini:2024ijq, Brunello:2025gpf, Bern:2025zno, Bern:2025wyd}.

In this context the modern scattering-amplitudes program has proven invaluable in advancing these expansions to higher orders. Specifically, these methods have been instrumental in refining higher-order corrections for a variety of two-body observables within the PM framework, including the momentum impulse, spin kick, scattering waveform, and mass absorption, among others~\cite{Porto:2005ac,Porto:2006bt,Goldberger:2007hy,Kol:2007bc,Porto:2008tb,Porto:2008jj,Levi:2008nh,Goldberger:2009qd,Porto:2010tr,Porto:2010zg,Levi:2010zu,Levi:2011eq,Porto:2012as,Foffa:2013qca,Levi:2014gsa,Levi:2014sba,Levi:2015msa,Levi:2015uxa,Levi:2015ixa,Foffa:2016rgu,Maia:2017yok,Maia:2017gxn,Kalin:2020mvi,Levi:2020kvb,Levi:2020uwu,Kosower_2019,Maybee:2019jus,Cachazo:2017jef,Liu:2021zxr,Cristofoli:2021vyo,Cristofoli:2021jas,Cristofoli:2022phh,Adamo:2022rmp, Driesse:2024xad, Bohnenblust:2025gir, Kim:2024svw}. In this paper, we focus on the Kosower--Maybee--O’Connell (KMOC) formalism~\cite{Kosower_2019,Maybee:2019jus}, a highly effective approach that establishes a direct correspondence between scattering amplitudes and physical observables, providing a powerful tool for exploring two-body dynamics in gravitational systems.

Gravitational amplitudes have a deep relationship with Yang-Mills theory amplitudes through the so-called double copy. This map asserts that gravitational amplitudes can be derived by combining two Yang-Mills amplitudes, or, in other words, tree-level gravitational amplitudes can be expressed as a sum of products involving two color-ordered tree-level Yang-Mills amplitudes. Its early form dates back to the work of Kawai, Lewellen, and Tye who originally derived a similar relation between open-string amplitudes and closed-string amplitudes~\cite{Kawai:1985xq}. In a groundbreaking result, Bern, Carrasco, and Johansson showed that gravitational numerators can be seen as the square of the kinematic numerators in Yang-Mills theory~\cite{Bern:2008qj,Bern:2010ue,Bern:2019prr}. This relationship has significant implications for the study of gravitational waves, offering a new perspective on their structure and interactions. The double-copy approach establishes a direct link between classical solutions in Yang-Mills theory and gravity, where point charges correspond to point sources in gravitational fields~\cite{Monteiro:2014cda,Luna:2016due,Goldberger:2016iau,Luna:2016hge}. Furthermore, the double copy has the potential to enhance our understanding of bound states and spinning particles~\cite{Goldberger:2017vcg,Goldberger:2017ogt,Plefka:2018dpa,Li:2018qap,Kalin:2019rwq,Kalin:2019inp,Cho:2021arx}. These results underscore the importance of scattering-amplitudes techniques in advancing the field of classical gravitational physics, providing valuable insights.

Classical observables derived from 2--to--2 scattering amplitudes within the PM expansion include the scattering angle, impulse, and spin kick~\cite{Guevara:2018wpp,Bini:2018zxp,Chung:2019duq,Bjerrum-Bohr:2019kec,Cristofoli:2019neg,Goldberger:2020wbx,Kol:2021jjc,Bjerrum-Bohr:2021wwt,Jakobsen:2022fcj,Jakobsen:2022zsx,Aoude:2022thd,Menezes:2022tcs,Damgaard:2022jem,Alessio:2023kgf,Bautista:2023szu,Gatica:2023iws,Gatica:2024mur,Luna_2024,Akpinar:2024meg,Bohnenblust:2024hkw, Alessio:2025flu}, and also facilitated the study of radiation-reaction effects, higher-order waveforms, beyond-GR phenomena, tidal interactions, understanding of high-energy limits, quantum effects, and more~\cite{Caron-Huot:2018ape,KoemansCollado:2018hss,Burger:2019wkq,Emond:2019crr,Cristofoli:2019ewu,Brandhuber:2019qpg,KoemansCollado:2019lnh,Moynihan:2019bor,Chung:2019yfs,Aoude:2020ygw,Kalin:2020lmz,Kim:2020dif,Haddad:2020que,Cheung:2020sdj,Bini:2020flp,Heissenberg:2022tsn,Mougiakakos:2022sic,Adamo:2020qru,Cristofoli:2020hnk,AccettulliHuber:2020dal,Abreu:2020lyk,Riva:2021vnj,Mougiakakos:2021ckm,Crawley:2021auj,Carrillo-Gonzalez:2021mqj,Monteiro:2021ztt,Chen:2021huj,Saketh:2021sri,Manohar:2022dea,Kalin:2022hph,Bini:2022enm,Dlapa:2022lmu,Kim:2022iub,Adamo:2022qci,Adamo:2022rob,Gonzo:2022tjm, Bautista:2022wjf,Jones:2022aji,Cangemi:2022abk,Brandhuber:2023hhy,Georgoudis:2023lgf,Georgoudis:2023ozp,Caron-Huot:2023vxl,Caron-Huot:2023ikn,Bini:2023fiz,Davis:2023zqv,Kosmopoulos:2023bwc,Jones:2023tgz,Georgoudis:2023eke,Crawley:2023brz,Guevara:2023wlr,Adamo:2023fbj,Adamo:2023cfp,Bianchi:2023lrg,Hoogeveen:2023bqa,Cangemi:2023bpe, Cangemi:2023ysz, Georgoudis:2024pdz,Bini:2024rsy,Buonanno:2024byg,Bhattacharyya:2024aeq,Fransen:2024fzs,Bautista:2024agp,Brandhuber:2024bnz,Brandhuber:2024qdn,Adamo:2024oxy,Gambino:2024uge,Akhtar:2024mbg,Aoki:2024boe}.

It is crucial that spin effects are embodied in the theory aiming to describe astrophysical black holes, as they significantly influence their dynamics and interactions. By employing scattering amplitudes for massive spinning particles, one can extract spin multipoles, which are expressed in terms of the spin vector $S^\mu=ma^\mu$ for a Kerr black hole~\cite{Vaidya:2014kza}. A class of generic spin-$s$ amplitudes describing three-point Kerr black hole amplitudes was later derived~\cite{Arkani-Hamed:2017jhn,Johansson:2019dnu,Guevara:2018wpp,Chung:2018kqs}. Along with generalized unitarity and other techniques, this approach helped to achieve several results in the 2-to-2 scattering of spinning objects~\cite{Guevara:2017csg,Guevara:2019fsj,Chung:2019duq,Arkani-Hamed:2019ymq,Guevara:2020xjx,Aoude:2020mlg,Bautista:2021wfy,Bautista:2022wjf,Cangemi:2022bew,Aoude:2022thd,Aoude:2022trd,Aoude:2023vdk,Aoude:2023dui,Cangemi:2023ysz,Cangemi:2023bpe,Chen:2024mmm,Bohnenblust:2024hkw,Bjerrum-Bohr:2025lpw}.


We structure our work as follows. In section \ref{review} we briefly review the KMOC formalism that will be used to compute the classical observables from the amplitudes. In section \ref{NLO} we compute the KMOC formula for the LO and NLO angular impulse, extending the work in~\cite{Maybee:2019jus}. In section \ref{amplitudes} we calculate all tree and one-loop amplitudes needed in the spin-kick formulas. In particular, we calculate in detail the one-loop leading singularity for our set-up. In section \ref{classical} we calculate the 2PM spin kick for the 2-to-2 scattering of a binary Kerr black-hole system. We conclude at last in section \ref{conclusion}. Appendix~\ref{app:NR_LS} addresses the calculation of the classical non-relativistic gravitational potential from our triangle leading singularity, Appendix~\ref{app:born-cutbox} explains the Born-iteration interpretation of the conservative cut box, and Appendix~\ref{app:C} discusses the spin expansion of the NLO spin kick and its comparison with ref.~\cite{Alessio:2025flu}.

\section{Review}
\label{review}

Before we proceed, let us briefly review the formalism and establish some conventions. We’ll start with the basics of the KMOC formalism, in which the main idea revolves around extracting computable classical observables from scattering amplitudes, while we incorporate spin into the framework.

\subsection{General discussion}

The change of an observable $\mathcal{O}$ that undergoes a scattering process, will be defined as the difference of the expectation values of the corresponding quantum operator $\mathbb{O}$, associated with the observable, in the incoming states and outgoing states
\begin{equation}
    \Delta\mathcal{O} = {}_\textrm{out}\bra{\Psi}\mathbb{O}\ket{\Psi}_\textrm{out} - {}_\textrm{in}\bra{\Psi}\mathbb{O}\ket{\Psi}_\textrm{in}
\end{equation}
with the outgoing states being related to the incoming states by the $S$-matrix. By employing unitarity of the $S$-matrix, \emph{i.e.} $S=\mathbf{1}+iT$, the change in the observables can be written as\footnote{Some other works~\cite{Aoude:2021oqj} use the definition $\Delta\mathcal{O} = i\bra{\Psi}\mathbb{O}T - T^\dagger\mathbb{O}\ket{\Psi} + \bra{\Psi}T^\dagger\mathbb{O}T\ket{\Psi}$. At leading order, only the first term contributes and we have $T^\dagger=T$}
\begin{equation}
    \Delta\mathcal{O} = i\bra{\Psi}\left[\mathbb{O}, T\right]\ket{\Psi} + \bra{\Psi}T^\dagger\left[\mathbb{O}, T\right]\ket{\Psi} = \Delta^{(1)}\mathcal{O} + \Delta^{(2)}\mathcal{O},
\end{equation}
where the terms are known as ``virtual'' and ``real'' parts of the observable, and we dropped out the $in$ subscript. The appropriate incoming states for the scattering of two massive particles is given by~\cite{Kosower_2019}
\begin{equation}
    \ket{\Psi} = \sum_{a_1,a_2}\int d\Phi(p_1) d\Phi(p_2)\phi_1(p_1)\phi_2(p_2)\xi_{a_1}\xi_{a_2}e^{\frac{ib\cdot p_1}{\hbar}}\ket{p_1,a_1;p_2,a_2}
    \label{2.3}
\end{equation}
where we have considered wavefunctions of the form $\phi(p) \xi_i$, $i$ being a little-group index, with normalization
\begin{equation}
    \int  d\Phi(p) |\phi(p)|^2 = \sum_i |\xi_i|^2 = 1 .
\end{equation}
Here the measure reads
\begin{equation}
    \int d\Phi(p_i) \equiv \int \hat{d}^4p_i\hat{\delta}^{(+)}(p^2_i-m^2),
\end{equation}
where $b$ is the impact parameter between the particles' wave packets, whose details are discussed in~\cite{Kosower_2019}, and
\begin{equation}
    \hat{d}^np_i=\frac{d^np}{(2\pi)^n}, \qquad \hat{\delta}^{(+)}(p^2_i-m^2) = 2\pi\Theta(p^0)\delta^{(n)}(p^2-m^2). 
\end{equation}
With that in hand, we are able to express the classical observable as 
\begin{equation}
    \Delta \mathcal{O} = \sum_{a_1,a_1'}\sum_{a_2,a_2'}\int d\Phi(p_1)d\Phi(p_2)d\Phi(p_1')d\Phi(p_2')\xi_{a_1}\xi^*_{a_1'}\xi_{a_2}\xi^*_{a_2'}e^{\frac{ib\cdot(p_1-p_1')}{\hbar}}\left(K_v + K_r\right),\label{2.6}
\end{equation}
where
\begin{equation}
    K_v = i\bra{p_1',a_1';p_2',a_2'}\left[\mathbb{O}, T\right]\ket{p_1,a_1;p_2,a_2}, \quad K_r=\bra{p_1',a_1';p_2',a_2'}T^\dagger\left[\mathbb{O}, T\right]\ket{p_1,a_1;p_2,a_2}
\end{equation}

are the virtual and real kernels respectively. When taking the classical limit, the wave packets, $\phi(p_i)$ and $\phi(p_i')$, should peak sharply around their classical values. As for the $\xi_{a_i}$ $\xi_{a_i}^*$, they should be coherent in spin states, minimizing the uncertainty, as the spin tensor peaks sharply about the classical value $S_i^\munu$~\cite{Kosower_2019,Maybee:2019jus,Cristofoli:2021jas,Aoude:2021oqj}. Using those simplifications and defining $q^\mu=p_1^{\prime\mu}-p_1^\mu$, we can express \eqref{2.6} as
\begin{equation}
    \Delta \mathcal{O}_{cl} = \int \hat{d}^4q\hat{\delta}(2p_1\cdot q)\hat{\delta}(2p_2\cdot q)e^{-\frac{ib\cdot q}{\hbar}}\llangle[\bigg]\left(K_\mathrm{v} + K_\mathrm{r}\right)\rrangle[\bigg]
\end{equation}
where the double angle brackets represent the classical limit being taken and
\begin{equation}
    \begin{aligned}
    \llangle[\bigg] f(p_1,p_2,\ldots) \rrangle[\bigg] &\equiv \sum_{a_1,a'_1}\sum_{a_2,a'_2} \int d\Phi(p_1)d\Phi(p_2) \abs{\phi_1(p_1)}^2 \abs{\phi_2(p_2)}^2 \\
    &\hspace{4cm} \times \xi_{a'_1}^* \xi_{a'_2}^*f^{a'_1a'_2a_1a_2}(p_1,p_2,\ldots)\xi_{a_1}\xi_{a_2} .
    \label{2.9}
    \end{aligned}
\end{equation}
In the above we also used the notation $\hat{\delta}(p) \equiv 2 \pi \delta(p)$. For more details concerning notation and terminology, see the original paper~\cite{Kosower_2019}.

\subsection{Spin}

When dealing with spin change, \emph{i.e.} angular impulse, in the formalism we need to define the quantum analog of the classical spin pseudo vector $s^\mu$. As in~\cite{Maybee:2019jus}, we will consider the Pauli-Lubanski operator as this quantum counterpart, thus by manipulating this operator, we can express the angular impulse in a form analogous to that of the linear impulse.

In General Relativity, a black hole is uniquely characterized by its mass $m$ and spin $s$~\cite{Israel:1967wq, Carter:1971zc, Robinson:1975bv}. This uniqueness is mathematically encapsulated in the structure of its multipole moments. While a generic object possesses mass multipoles $\mathcal{I}_\ell$ and current (spin) multipoles $\mathcal{J}_\ell$ that vary according to its specific internal composition, the case for a Kerr black hole is quite different. For such an object, all these moments are locked together by the Hansen relation~\cite{Hansen:1974zz, Geroch:1970cd}
\begin{equation}
    \mathcal{I}_{\ell} + i\mathcal{J}_{\ell} = m (ia)^{\ell}
\end{equation}
where $\ell$ is the order of the multipole, $m$ is the mass of the black hole, and $a$ is the ring radius, or spin parameter, defined as $a=s/m$, where $s$ is the magnitude of the spin. This equation tells us that if you know the mass and the spin, you know the entire infinite tower of gravitational moments. The goal is to reproduce this specific structure using scattering amplitudes.

We now establish the classical kinematics for an isolated body, e.g. a black hole, in a scattering event. A body has a constant linear momentum $p^\mu$ and a total angular momentum tensor $J^{\mu\nu}(x)$. The angular momentum $J^{\mu\nu}(x)$ is always defined with respect to a reference point. If one calculates angular momentum $J^{\mu\nu}$ about a point $x$, and then calculates it about a different point $x'$, the two are related by the equation
\begin{equation}
    J^{\mu\nu} (x') = J^{\mu\nu}(x) + x^{\prime\mu} p^\nu - x^{\prime\nu} p^\mu - \left(x^{\mu} p^\nu - x^{\nu} p^\mu\right) = J^{\mu\nu}(x) + 2p^{[\mu}\left(x^{\prime} - x\right)^{\nu]}
\end{equation}
the brackets $\left[\ldots\right]$ denote antisymmetrization. We define the proper center of mass worldline as the set of points $z$ where the mass-dipole moment vanishes in the rest frame. Mathematically, this condition is $J^{\mu\nu}(z)p_\nu=0$.

Let $S^{\mu\nu}$ be the angular momentum tensor calculated at a point $z$ on this worldline. We call this the intrinsic spin tensor. Substituting $x'\to x$ and the reference point $x\to z$ 
into the equation, we arrive at
\begin{equation}
    J^{\mu\nu}(x) = 2p^{[\mu}(x-z)^{\nu]} + S^{\mu\nu}
\end{equation}
the $2p^{[\mu}(x-z)^{\nu]}$ part is the orbital angular momentum $L^{\mu\nu}$ relative to the point $z$
\begin{equation}
    2p^{[\mu}(x-z)^{\nu]} = p^\mu(x-z)^\nu - p^\nu(x-z)^\mu = (x-z)^\nu p^\mu - (x-z)^\mu p^\nu.
\end{equation}
$S^{\mu\nu}$ satisfies the Tulczyjew--Dixon spin supplementary condition
\begin{equation}
    S^{\mu\nu}p_{\nu} = 0 .
\end{equation}
This implies $S^{\mu0}=0$. $S^{\mu\nu}$ only has spatial components $S^{ij}$. This ensures $S^{\mu\nu}$ describes purely intrinsic rotation, not a boosted mass dipole.

While the spin tensor is useful, a 4-vector representation of spin is often easier to handle in scattering amplitudes. We define the spin pseudo-vector $s_\mu$
\begin{equation}
    s_{\mu} = \frac{1}{2m} \epsilon_{\mu\nu\rho\sigma} p^{\nu} S^{\rho\sigma}
\end{equation}
In the rest frame $p=(m,0)$, $s_0=0$ and $s_i$ corresponds to the 3D spin angular momentum vector. We can express the tensor $S^{\mu\nu}$ entirely in terms of the vector $s^\mu$ and momentum $p^\mu$.
\begin{equation}
    S_{\mu\nu} = \frac{1}{m} \epsilon_{\mu\nu\sigma\rho} p^{\sigma} s^{\rho}
\end{equation}
To calculate changes in this vector using scattering amplitudes, we must identify the corresponding quantum operator. It seems natural to identify this operator as the Pauli-Lubanski pseudovector 
$\mathbb{W}^\mu$. We define the classical spin vector $s^\mu$ as the expectation value of the Pauli-Lubanski operator
\begin{equation}
    \langle s^{\mu} \rangle \equiv \frac{1}{m} \langle \mathbb{W}^{\mu} \rangle \qquad\text{where}\quad\mathbb{W}_{\mu} = \frac{1}{2} \epsilon_{\mu\nu\rho\sigma} \mathbb{P}^{\nu} \mathbb{J}^{\rho\sigma}
\end{equation}

To compute amplitudes, we need to know how $\mathbb{W}^\mu$ acts on momentum eigenstates $\ket{p,i}$. Since $\mathbb{W}$ commutes with $\mathbb{P}$, it does not change the momentum $p$, it only mixes the spin indices $i,j$
\begin{equation}
    \langle p', j | \mathbb{W}^{\mu} | p, i \rangle \equiv m s_{ij}^{\mu}(p) \hat{\delta}_{\Phi}(p-p')\label{2.15}
\end{equation}
where the matrix $s_{ij}^{\mu}(p)$ is the quantum analog of the classical spin vector, and it carries the spin indices. The expectation value of the spin vector reads
\begin{equation}
    \left\langle s^\mu\right\rangle=\int d \Phi(p)|\phi(p)|^2 \xi_i^* s_{i j}^\mu \xi_j
\end{equation}
The spin polarization vector, denoted as $s_{i j}^\mu(p)$, inherits the commutation relations of the Pauli-Lubanski vector. Consequently, these relations hold for the spin polarization vector.
\begin{equation}
    \left[s^\mu(p), s^\nu(p)\right]=-i \frac{\hbar}{m} \epsilon^{\mu \nu \rho \sigma} s_\rho(p) p_\sigma .
\end{equation}

\section{Angular Impulse}
\label{NLO}

Now we are ready to proceed and define the observable that we want, namely the angular impulse (or spin kick) of a particle. With the goal of calculating it up to next-to-leading order (NLO), we begin with the leading-order (LO) contribution. This first-order effect is purely conservative, meaning it describes an interaction that can be derived from a potential, without any dissipative effects like gravitational radiation. 

Following the set-up in~\cite{Maybee:2019jus}, we consider the scattering of two particles with spin $S_1$ and $S_2$ which are separated by an impact parameter $b^{\mu}$. The relevant incoming two-particle state was given above. The initial spin vector of particle $1$ reads
\begin{equation}
    \langle s^{\mu}_{1} \rangle = \frac{1}{m_1} \langle \Psi| \mathbb{W}^{\mu}_1 |\Psi \rangle
\end{equation}
where $\mathbb{W}^{\mu}_1$ is the Pauli-Lubanski operator of the field corresponding to particle $1$. Since the S-matrix is the time evolution operator from the far past to the far future, the final spin vector of particle $1$ is
\begin{equation}
    \langle s^{\prime \mu}_{1} \rangle = \frac{1}{m_1} \langle \Psi| S^{\dagger} \mathbb{W}^{\mu}_1 S |\Psi \rangle .
\end{equation}
We define the angular impulse on particle $1$ as the difference between these quantities:
\begin{equation}
    \langle \Delta s^{\mu}_{1} \rangle = \frac{1}{m_1} \langle \Psi| S^{\dagger} \mathbb{W}^{\mu}_1 S |\Psi \rangle -  \frac{1}{m_1} \langle \Psi| \mathbb{W}^{\mu}_1 |\Psi \rangle.
\end{equation}
Writing $S = 1 + i T$ and using unitarity of the $S$-matrix yields
\begin{equation}
    \langle \Delta s^{\mu}_{1} \rangle = \frac{i}{m_1} \langle \Psi| [\mathbb{W}^{\mu}_1, T] |\Psi \rangle + \frac{1}{m_1} \langle \Psi| T^{\dagger} [\mathbb{W}^{\mu}_1, T] |\Psi \rangle = \langle \Delta s^{(1), \mu}_{1} \rangle + \langle \Delta s^{(2), \mu}_{1} \rangle
\end{equation}
We are interested in both contributions. Only $\langle \Delta s^{(1), \mu}_{1} \rangle$ produces the LO contribution, so let us first focus on this particular term.

\subsection{Leading-Order}

For convenience, we here reproduce the steps as given in ref.~\cite{Maybee:2019jus}. The derivation begins with the formal quantum mechanical expression for the change in spin, which involves the Pauli-Lubanski operator and the T-matrix. Replacing the incoming state of equation \eqref{2.3} in the virtual kernel equation leads us to
\begin{equation}
    \begin{aligned}
        \langle \Delta s_1^{(1),\mu} \rangle = \frac{i}{m_1}\sum_{a_1',a_1}\sum_{a_2',a_2}\prod_{i=1,2} \int& d\Phi(p'_i)d\Phi(p_i) \phi_i^*(p_i')\phi_i(p_i)\xi^*_{i,a_i'}\xi_{i,a_i} e^{\frac{ib\cdot(p_1-p_1')}{\hbar}} \\
        &\times\bra{p_1',a_1';p_2',a_2'}\left[\mathbb{W}^\mu, T\right]\ket{p_1,a_1;p_2,a_2}.
    \end{aligned}
\end{equation}
To resolve the commutator $\left[\mathbb{W}^\mu, T\right]$, we insert a complete set of states between the operators 
\begin{equation}
    {\mathbf 1}  = \sum_{b_1,b_2} \int d\Phi(r_1)d\Phi(r_2)\ket{r_1,b_1;r_2,b_2}\bra{r_1,b_1;r_2,b_2}
\end{equation}
These states can be viewed as the initial state of a scattering process. The resulting matrix elements are written in terms of scattering amplitudes $\mathcal{A}$ and spin matrices $s^\mu$, turning the operator expression into a phase-space integral. After some manipulation we obtain
\begin{equation}
    \begin{aligned}
        \langle \Delta s_1^{(1),\mu} \rangle =& \frac{i}{m_1}\sum_{a_1,a_1'}\sum_{a_2,a_2'}\sum_{b_1,b_2} \prod_{i=1,2} \int d\Phi(r_i) d\Phi(p'_i) d\Phi(p_i) \phi_i(p_i)\phi_i^*(p'_i)\xi_{a_i}\xi^*_{a'_i}e^{ib\cdot\frac{(p_1-p'_1)}{\hbar}}
        \\
        &\times \big(\bra{p'_1,a'_1;p'_2,a'_2}\mathbb{W}^\mu_1\ket{r_1,b_1;r_2,b_2}\bra{r_1,b_1;r_2,b_2}T\ket{p_1,a_1;p_2,a_2}
        \\
        &-\bra{p'_1,a'_1;p'_2,a'_2}T\ket{r_1,b_1;r_2,b_2}\bra{r_1,b_1;r_2,b_2}
        \mathbb{W}^\mu_1\ket{p_1,a_1;p_2,a_2}\big)
        \\
        =& i\sum_{a_1,a_1'}\sum_{a_2,a_2'}\sum_{b_1,b_2}\prod_{i=1,2} \int d\Phi(p_i+q_i) d\Phi(p_i)\phi_i(p_i)\phi_i^*(q_i+p_i) \xi^*_{a'_i} \xi_{a_i} e^{ib\cdot\frac{(p_1-p'_1)}{\hbar}}
        \\
        &\times \big(s_{1b_1a'_1}^\mu(p_1+q_1) \mathcal{A}_{b_1a'_2a_1a_2}(p_1,p_2\rightarrow p_1+q_1,p_2+q_2)+ 
        \\
        & - \mathcal{A}_{a'_1a'_2b_1a_2}(p_1,p_2\rightarrow p_1+q_1,p_2+q_2)s_{1b_1a_1}^\mu(p_1)\big)\hat{\delta}^{(4)}(q_1+q_2)
        \label{2.8}
    \end{aligned}
\end{equation}
where we introduced the momentum mismatch $q_i= p'_i-p_i$ and used the definition of the matrix elements of $\mathbb{W}$ on the states of a given momentum, given by \eqref{2.15}, as well as the usual definition of the scattering matrix element
\begin{equation}
    \bra{p'_1,a'_1;p'_2,a'_2}T\ket{p_1,a_1;p_2,a_2} = \mathcal{A}_{a'_1a'_2a_1a_2}(p_1,p_2 \rightarrow p'_1,p'_2)\hat{\delta}^{(4)}(p_1+p_2 - p'_1-p'_2).
\end{equation}
The following steps simplify the phase-space integrals by using momentum-conserving delta functions, $\hat{\delta}^{(4)}(q_1+q_2)$. This delta allows us to perform the integral over one of the momentum transfers (e.g., $q_2$), as it constrains it to be $q_2=-q_1$. We can then relabel the remaining momentum transfer $q_1\to q$. 
By systematically applying these constraints, the multi-particle integral is reduced to an integral over a single, independent momentum transfer variable, $q$. The resulting expression for the contribution 
$\langle \Delta s_1^{(1),\mu} \rangle$ is
\begin{equation}
    \begin{aligned}
        \langle \Delta s_1^{(1),\mu} \rangle = i \sum_{a_1, a_1'} \sum_{a_2, a_2'} \sum_{b_1}
        \prod_{i=1,2} &\int d\Phi(p_i) \hat{d}^4q \Theta(p_1^0+q^0)\hat{\delta}\big(2q\cdot p_1+q^2\big)\Theta(p_2^0-q^0)\\
        &\times \hat{\delta}\big(2q\cdot p_2-q^2\big)\phi_i(p_i) \phi_1^*(p_1+q) \phi_2^*(p_2-q) \xi_{a_i}\xi^*_{a'_i}e^{ib\cdot\frac{(p_1-p'_1)}{\hbar}}\\
        &\times \bigg(s_{1b_1a'_1}^\mu(p_1+q) \mathcal{A}_{b_1a'_2a_1a_2}(p_1,p_2\rightarrow p_1+q,p_2-q)+\\
        &-\mathcal{A}_{a'_1a'_2b_1a_2}(p_1,p_2\rightarrow p_1+q,p_2-q)s_{1b_1a_1}^\mu(p_1)\bigg),
    \end{aligned}
\end{equation}

We now extract the classical limit from this quantum result through a series of approximations. The first is to re-scale the momentum transfer $q$ by Planck's constant, $\hbar$, introducing a classical wave number $\bar{q} = q/\hbar$. This allows us to organize the expression in powers of $\hbar$ and systematically isolate the leading term in the classical limit (\emph{i.e.} $\hbar\to0$). While on this limit we can make some key approximations: in the on-shell delta functions, the $q^2$ (now $\hbar^2\bar{q}^2$) is negligible with respect to $2p\cdot q$ (now $2\hbar p\cdot q$)\footnote{A similar argument allows us to neglect the $\hbar\bar{q}^0$ term inside the positive-energy theta function}; and the small momentum shift $q$ can be ignored in parts of the expression, such as the wave functions. With that, we arrive at the following classical limit~\cite{Maybee:2019jus}
\begin{equation}
    \begin{aligned}
        \langle \Delta s_1^{(1),\mu} \rangle = i\llangle[\bigg] &\int d^4\bar{q} \hat{\delta}(2\bar{q}\cdot p_1) \hat{\delta}(2\bar{q}\cdot p_2)e^{-ib\cdot\bar{q}}\\
        & \times\bigg(s^\mu_{1}(p_1+\hbar\bar{q}) \mathcal{A}(p_1,p_2\rightarrow p_1+q_1,p_2+q_2)\\
        &-\mathcal{A}(p_1,p_2\rightarrow p_1+q_1,p_2+q_2)s_{1}^\mu(p_1)\bigg)\rrangle[\bigg].
    \end{aligned}
\end{equation}
It remains to determine how the spin vector $s^\mu(p+\hbar q)$ transforms. For a small momentum transfer the change is an infinitesimal Lorentz transformation, whose generator $\omega^{\mu\nu}$ is fixed by $\hbar q$
\begin{equation}
\omega^\munu = -\frac{\hbar}{m_1^2}(p^\mu_1\bar{q}^\nu - p^\nu_1\bar{q}^\mu).
\end{equation}

The spin polarization vector $s^\mu_{1b_1a'_1}(p_1+\hbar\bar{q})$ is just the Lorentz boost of $s^\mu_{1b_1a'_1}(p_1)$. In the classical limit, $q$ is small and hence we are dealing with an infinitesimal Lorentz transformation, $\Lambda^\mu_{  \nu} = \delta^\mu_\nu + \omega^\mu_{  \nu}$. In the present case $\omega^\mu_{  \nu}p^\nu_1 = \hbar\bar{q}^\mu$, with $\omega^{\mu\nu}$ the generator written above. Applying this transformation to the initial spin vector tells us exactly how it changes due to the interaction. Using the spin-supplementary condition ($s(p)\cdot p=0$), the transformed spin vector is
\begin{equation}
    s^\mu(p_1+\hbar\bar{q}) = \Lambda^\mu_{  \nu}s^\nu(p_1) = s^\mu_{1b_1a'_1}(p_1) - \frac{\hbar}{m_1^2}p^\mu_1\bar{q}_\nu s^\nu(p_1)
\end{equation}
This describes how the spin direction tilts as the particle recoils.

Substituting the boosted spin vector back into the integral gives the classical expression for the contribution 
$\langle \Delta s_1^{(1),\mu} \rangle$
\begin{equation}
    \begin{aligned}
        \langle \Delta s_1^{(1),\mu} \rangle = i\llangle[\bigg]\hbar^2  \int \hat{d}^4\bar{q}\hat{\delta}(2\bar{q}\cdot p_1 + \hbar\bar{q}^2)&\hat{\delta}(2\bar{q}\cdot p_2 - \hbar\bar{q}^2)e^{-ib\cdot\bar{q}}\\
        &\times\left(\left[{s^\mu_1(p_1)}{\mathcal{A}(\bar{q})}\right] - \frac{\hbar}{m_1^2}p_1^\mu\bar{q}_\nu s^\nu_1(p_1)\mathcal{A}(\bar{q})\right)\rrangle[\bigg] 
    \end{aligned}
\end{equation}
where we have suppressed the spin indices, so one must remember that the spin vector and the amplitude are matrices with spinor indices. Furthermore, there is an implicit summation over the contracted index $b_1$. The terms inside the parentheses have distinct and important physical interpretations: The Commutator $\left[{s^\mu_1}{A(\bar{q})}\right]$ describes the spin precession. It captures the rotation of the spin vector around an axis determined by the interaction, without changing the spin's magnitude; and the term $-\frac{\hbar}{m_1^2}p_1^\mu\bar{q}_\nu s^\nu_1A(\bar{q})$ arises directly from the Lorentz boost associated with the particle's recoil. It represents the change in the spin vector's components due to the change in the particle's own momentum.

In order to obtain the LO spin kick, we replace in the above formula the expression $A^{(0)}$ for a tree-level amplitude. In addition, one can also neglect the term $\hbar \bar{q}^2$ term in comparison with 
$\bar{q} \cdot p_{1,2}$ inside the delta functions. Moreover, a factor of $g/\sqrt{\hbar}$ for every interaction vertex should be removed, where $g$ is a generic coupling constant. We denote the amplitude stripped of coupling-constant factors as $\mathcal{A}$. We find
\begin{equation}
    \begin{aligned}
        \langle \Delta s_1^{\mu,\text{LO}} \rangle = i g^2
        \llangle[\bigg]\hbar  \int& \hat{d}^4\bar{q}\hat{\delta}(2\bar{q}\cdot p_1 )
        \hat{\delta}(2\bar{q}\cdot p_2 )e^{-ib\cdot\bar{q}}\\
        &\times\left(\left[{s^\mu_1}{\mathcal{A}^{(0)}(\bar{q})}\right] - \frac{\hbar}{m_1^2}p_1^\mu\bar{q}_\nu s^\nu_1(p_1)\mathcal{A}^{(0)}(\bar{q})\right)\rrangle[\bigg]
    \end{aligned}
\end{equation}

\subsection{Next-to-Leading-Order}

With the first contribution understood, we are now ready to derive an explicit expression for the spin-kick at the next-to-leading-order (NLO). For this we will have contributions from $\langle \Delta s^{(1), \mu}_{1} \rangle$ as well as $\langle \Delta s_1^{(2),\mu} \rangle$. So we need to derive an expression for the latter as well. In this case, one must bear in mind that the scattering is conservative at LO and at NLO. At NNLO, however, we must take radiative effects into account. This back-reaction is entirely described by $\langle \Delta s^{(2), \mu}_{1} \rangle$. In order to incorporate this new piece of information into our framework, the mathematical formalism must be expanded to allow for the creation of new particles (e.g. emitted gravitons). This is achieved by inserting a more general complete set of states into our calculations, which takes the form
\begingroup
\allowdisplaybreaks
\begin{equation}
    \mathbf{1} = \sum_{b_1,b_2,b_X} \sum_X \int d\Phi(r_1)d\Phi(r_2)\ket{r_1,b_1;r_2,b_2;X,b_x}\bra{r_1,b_1;r_2,b_2;X,b_x}.
\end{equation}
By inserting this identity between the $T^\dagger$ and the $[\mathbb{W},T]$ operators in the NLO impulse definition, we are effectively cutting the diagram. The term where $X$ is empty corresponds to the elastic cut (virtual corrections), and the terms where $X$ contains gravitons correspond to the inelastic cut (radiative back-reaction). This ensures that the loss of angular momentum to the radiation field is accounted for, which is necessary to cancel the IR divergences from the virtual loop diagrams

We then begin our derivation by introducing this set of states into the expression for angular impulse. This first step translates the abstract operator equation into a more concrete, albeit lengthy, integral involving scattering amplitudes and the Pauli-Lubanski spin operator. After expanding the operators into amplitudes and spin matrices we get
\begin{equation}
    \begin{aligned}
        \langle \Delta s^{(2), \mu}_{1} \rangle =& \frac{1}{m_1}
        \sum_{b_1, b_2, b_X} \sum_{X} \int d\Phi(r_1) d\Phi(r_2) 
        \langle \Psi| T^{\dagger} | r_1 r_2 X ; b_1 b_2 b_X \rangle 
        \langle r_1 r_2 X ; b_1 b_2 b_X | [\mathbb{W}^{\mu}_1, T] |\Psi \rangle
        \\
        =& \frac{1}{m_1}
        \sum_{b_1, b_2, b_X} \sum_{b_1', b_2'} \sum_{a_1, a_1'} \sum_{a_2, a_2'} 
        \sum_{X} 
        \\
        &\times 
        \int d\Phi(r_1') d\Phi(r_2')  d\Phi(r_1) d\Phi(r_2) d\Phi(p_1') d\Phi(p_2') d\Phi(p_1) d\Phi(p_2)
        \\
        &\times \phi^{*}_{1}(p_1') \phi^{*}_{2}(p_2') \phi_{1}(p_1) \phi_{2}(p_2) 
        \xi^{*}_{a_1'} \xi^{*}_{a_2'} \xi_{a_1} \xi_{a_2}
        e^{i b \cdot (p_1 - p_1')/\hbar}
        \\
        &\times 
        A^{*}_{a_1' a_2' b_1 b_2 b_X}(p_1' p_2' \to r_1 r_2 r_X) \hat{\delta}^{(4)}( p_1' + p_2' -  r_1 - r_2 - r_X )
        \\
        &\times 
        \Bigg( \sum_{b_X'} \sum_{X'}
        \langle r_1 r_2 X ; b_1 b_2 b_X | \mathbb{W}^{\mu}_1 | r_1' r_2' r_X'; b_1' b_2' b_X' \rangle
        \\
        &\times
        \mathcal{A}_{b_1'  b_2' b_X' a_1 a_2}(p_1 p_2 \to r_1' r_2' r_X') \hat{\delta}^{(4)}( p_1 + p_2 -  r_1' - r_2' - r_X')
        \\
        & - A_{b_1 b_2 b_X b_1' b_2'}(r_1' r_2' \to r_1 r_2 r_X)  
        \langle r_1' r_2' ; b_1' b_2' | \mathbb{W}^{\mu}_1  | p_1 p_2 ; a_1 a_2 \rangle
        \\
        &\hspace{6cm}\times
        \hat{\delta}^{(4)}( r_1' + r_2' - r_1 - r_2 - r_X )
        \Bigg)
    \end{aligned}
\end{equation}
which simplifies to
\begin{equation}
    \begin{aligned}
        \langle \Delta s^{(2), \mu}_{1} \rangle=& \sum_{b_1, b_2, b_X} \sum_{b_1'} \sum_{a_1, a_1'} \sum_{a_2, a_2'} 
        \sum_{X} \\
        &\times\int d\Phi(r_1) d\Phi(r_2)  d\Phi(p_1) d\Phi(p_2)  \hat{d}^4 q
        \\
        &\times  (2\pi) \theta(q^{0}+p^{0}_{1}) \delta( 2 q \cdot p_1 + q^2)
        (2\pi) \theta(-q^{0}+p^{0}_{2}) \delta(2 q \cdot p_2 - q^2)
        \\
        &\times \phi^{*}_{1}(p_1 + q) \phi^{*}_{2}(p_2 - q) \phi_{1}(p_1) \phi_{2}(p_2) 
        \xi^{*}_{a_1'} \xi^{*}_{a_2'} \xi_{a_1} \xi_{a_2}
        e^{- i b \cdot q/\hbar}
        \\
        &\times \mathcal{A}^{*}_{a_1' a_2' b_1 b_2 b_X}(p_1 + q, p_2 - q \to r_1, r_2, r_X) 
        \hat{\delta}^{(4)}( p_1 + p_2 -  r_1 - r_2 - r_X)
        \\
        &\times 
        \Big( s^{\mu}_{1 b_1' b_1}(r_1) \mathcal{A}_{b_1'  b_2 b_X a_1 a_2}(p_1 p_2 \to r_1 r_2 r_X) 
        \\
        &\hspace{5cm}- \mathcal{A}_{b_1 b_2 b_X b_1' a_2}(p_1 p_2 \to r_1 r_2 r_X)  s^{\mu}_{1 a_1 b_1' }(p_1) \Big)
    \end{aligned}
\end{equation}
where we used previous arguments and definitions. In this expression, $r_X$ denotes the total momentum carried by particles in $X$. Now  instead of integrating over the momenta $r_i$, we change to the momentum transfers, $w_i=r_i-p_i$. This is done because the momentum transfers are the small quantities that directly characterize the strength of the interaction and are the natural variables for analyzing the classical limit. This process results in the following expression
\begin{equation}
    \begin{aligned}
        \langle \Delta s^{(2), \mu}_{1} \rangle =& 
        \sum_{b_1, b_2, b_X} \sum_{b_1'} \sum_{a_1, a_1'} \sum_{a_2, a_2'} 
        \sum_{X} \int d\Phi(p_1) d\Phi(p_2)
        \hat{d}^4 w_1 \hat{d}^4 w_2 \hat{d}^4 q 
        \\
        &\times (2\pi) \theta(w^{0}_{1}+p^{0}_{1}) \delta(2 w_1 \cdot p_1 + w^2_{1})
        (2\pi) \theta(w^{0}_{2}+p^{0}_{2}) \delta(2 w_2 \cdot p_2 + w^2_{2})
        \\
        &\times 
        (2\pi) \theta(q^{0}+p^{0}_{1}) \delta( 2 q \cdot p_1 + q^2)
        (2\pi) \theta(-q^{0}+p^{0}_{2}) \delta(2 q \cdot p_2 - q^2)
        \\
        &\times \phi^{*}_{1}(p_1 + q) \phi^{*}_{2}(p_2 - q) \phi_{1}(p_1) \phi_{2}(p_2) 
        \xi^{*}_{a_1'} \xi^{*}_{a_2'} \xi_{a_1} \xi_{a_2}
        e^{- i b \cdot q/\hbar}
        \\
        &\times \mathcal{A}^{*}_{a_1' a_2' b_1 b_2 b_X}(p_1 + q, p_2 - q \to p_1 + w_1, p_2 + w_2, r_X) 
        \hat{\delta}^{(4)}( w_1 + w_2 + r_X)
        \\
        &\times 
        \Big( s^{\mu}_{1 b_1' b_1}(p_1+w_1) \mathcal{A}_{b_1'  b_2 b_X a_1 a_2}(p_1,p_2 \to p_1+ w_1, p_2 + w_2, r_X) 
        \\
        &- \mathcal{A}_{b_1 b_2 b_X b_1' a_2}(p_1,p_2 \to p_1+w_1, p_2+w_2, r_X)  s^{\mu}_{1 a_1 b_1' }(p_1) \Big) .
    \end{aligned}
\end{equation} 
Once again, the bridge between quantum and classical physics is constructed by systematically taking the classical limit, which corresponds to the limit where Planck’s constant $\hbar$ approaches zero. In this process, we re-scale all momentum transfer variables, $q$, $w_1$, and $w_2$, by $\hbar$. This re-scaling organizes the entire expression in powers of $\hbar$, enabling us to isolate the dominant classical behavior. After this re-scaling, our expression takes the form
\begin{equation}
    \begin{aligned}
        \langle \Delta s^{(2), \mu}_{1} \rangle = &\llangle[\bigg] \hbar^{8}
        \sum_{b_1, b_2, b_X} \sum_{b_1'} 
        \sum_{X} \int \hat{d}^4 \bar{w}_1 \hat{d}^4 \bar{w}_2 \hat{d}^4 \bar{q} 
        \\
        &\times (2\pi) \theta(w^{0}_{1}+p^{0}_{1})  \delta(2 \bar{w}_1 \cdot p_1 + \hbar \bar{w}^2_{1})
        (2\pi) \theta(w^{0}_{2}+p^{0}_{2}) \delta(2 \bar{w}_2 \cdot p_2 + \hbar \bar{w}^2_{2})
        \\
        &\times 
        (2\pi) \theta(q^{0}+p^{0}_{1}) \delta( 2 \bar{q} \cdot p_1 + \hbar \bar{q}^2)
        (2\pi) \theta(-q^{0}+p^{0}_{2}) \delta(2 \bar{q} \cdot p_2 - \hbar \bar{q}^2)
        e^{- i b \cdot \bar{q}}
        \\
        &\times 
        \hat{\delta}^{(4)}( \hbar \bar{w}_1 + \hbar \bar{w}_2 + r_X)
        \mathcal{A}^{*}_{a_1' a_2' b_1 b_2 b_X}(p_1 + \hbar \bar{q}, p_2 - \hbar \bar{q} \to p_1 + \hbar \bar{w}_1, p_2 + \hbar \bar{w}_2, r_X) 
        \\
        &\times 
        \Big( s^{\mu}_{1 b_1' b_1}(p_1+\hbar \bar{w}_1) 
        \mathcal{A}_{b_1'  b_2 b_X a_1 a_2}(p_1,p_2 \to p_1+ \hbar \bar{w}_1, p_2 + \hbar \bar{w}_2, r_X) 
        \\
        &-  
        \mathcal{A}_{b_1 b_2 b_X b_1' a_2}(p_1,p_2 \to p_1+\hbar\bar{w}_1, p_2+\hbar\bar{w}_2, r_X)  
        s^{\mu}_{1 a_1 b_1' }(p_1) \Big) 
        \rrangle[\bigg] .
    \end{aligned}
\end{equation} 
The most crucial physical insight in this step, just as in the leading-order case, revolves around the behavior of the spin vector $s_1^\mu(p_1+\hbar\bar{w}_1)$ when subjected to a slight change on momentum. This change is described by an infinitesimal Lorentz transformation. This enables us to approximate the final spin vector as the initial spin vector plus a small correction term that depends linearly on the momentum transfer $\bar{w}_1$. Utilizing this approximation, we arrive at
\begin{equation}
    \begin{aligned}
        \langle \Delta s^{(2), \mu}_{1} \rangle = \llangle[\bigg]& \hbar^{8}
        \sum_{X} \int \hat{d}^4 \bar{w}_1 \hat{d}^4 \bar{w}_2 \hat{d}^4 \bar{q} 
        \\
        &\times (2\pi) \theta(w^{0}_{1}+p^{0}_{1})  \delta(2 \bar{w}_1 \cdot p_1 + \hbar \bar{w}^2_{1})
        (2\pi) \theta(w^{0}_{2}+p^{0}_{2}) \delta(2 \bar{w}_2 \cdot p_2 + \hbar \bar{w}^2_{2})
        \\
        &\times 
        (2\pi) \theta(q^{0}+p^{0}_{1}) \delta( 2 \bar{q} \cdot p_1 + \hbar \bar{q}^2)
        (2\pi) \theta(-q^{0}+p^{0}_{2}) \delta(2 \bar{q} \cdot p_2 - \hbar \bar{q}^2)
        e^{- i b \cdot \bar{q}}
        \\
        &\times 
        \hat{\delta}^{(4)}( \hbar \bar{w}_1 + \hbar \bar{w}_2 + r_X) \mathcal{A}^{*}(\bar{q}, \bar{w}_1, \bar{w}_2, r_X)
        \\
        &\times \left( [s^{\mu}_{1}(p_1),  \mathcal{A}(\bar{w}_1, \bar{w}_2, r_X)] 
        - \frac{\hbar}{m_1^2} p_{1}^{\mu} \bar{w}_{1\nu} s^{\nu}_{1}(p_1) 
        \mathcal{A}(\bar{w}_1, \bar{w}_2, r_X) \right) 
        \rrangle[\bigg] .
    \end{aligned}
\end{equation} 

Let us rewrite this expression in a more convenient way. For brevity, let us denote
\begin{equation}
    \begin{aligned}
        \mathcal{A}^{*} &\equiv \mathcal{A}^{*}(\bar{q}, \bar{w}_1, \bar{w}_2, r_X),
        \\
        \mathcal{A} &\equiv \mathcal{A}(\bar{w}_1, \bar{w}_2, r_X),
        \\
        \alpha^{\mu}{}_{\nu} &\equiv
        \frac{\hbar}{m_1^2} p_1^\mu \bar{w}_{1\nu} .
    \end{aligned}
\end{equation}
Then the integrand of the real kernel can be written exactly as
\begin{equation}
    \begin{aligned}
        \mathcal{A}^{*}\big( \left[s^{\mu}_{1}(p_1),\mathcal{A}\right] - \alpha^{\mu}{}_{\nu} s^\nu_1(p_1) A \big).
    \end{aligned}
\end{equation}
Using
\begin{equation}
    \begin{aligned}
        \mathcal{A}^{*}\left[s^{\mu}_{1}(p_1),\mathcal{A}\right] =&
        \left[s^{\mu}_{1}(p_1),\mathcal{A}^{*}\mathcal{A}\right] - \left[s^{\mu}_{1}(p_1),\mathcal{A}^{*}\right]A,
        \\
        \mathcal{A}^{*} s^\nu_1(p_1) \mathcal{A}
        =&
        s^\nu_1(p_1) \mathcal{A}^{*}\mathcal{A} - \left[s^\nu_1(p_1),\mathcal{A}^{*}\right]\mathcal{A},
    \end{aligned}
\end{equation}
we obtain the exact identity
\begin{equation}
    \begin{aligned}
        \mathcal{A}^{*}\left( [s^{\mu}_{1}(p_1),\mathcal{A}] - \alpha^{\mu}{}_{\nu} s^\nu_1(p_1) \mathcal{A} \right) =& 
        \left[s^{\mu}_{1}(p_1),\mathcal{A}^{*}\mathcal{A}\right]
        - \alpha^{\mu}{}_{\nu} s^\nu_1(p_1) \mathcal{A}^{*}\mathcal{A}
        \\
        &- [s^{\mu}_{1}(p_1),\mathcal{A}^{*}]\mathcal{A}
        + \alpha^{\mu}{}_{\nu}[s^\nu_1(p_1),\mathcal{A}^{*}]\mathcal{A}
        \\
        =&
        \left[s^{\nu}_{1}(p_1),\mathcal{A}^{*}\mathcal{A}\right]\delta^\mu_{\nu}
        - \alpha^{\mu}{}_{\nu} s^\nu_1(p_1) \mathcal{A}^{*}\mathcal{A}
        \\
        & - \left[s^{\nu}_{1}(p_1),\mathcal{A}^{*}\right]\mathcal{A}
        \left( \delta^\mu_{\nu} - \alpha^{\mu}{}_{\nu} \right)
    \end{aligned}
\end{equation}
Therefore
\begin{equation}
    \begin{aligned}
        \langle \Delta s^{(2), \mu}_{1} \rangle =
        \llangle[\bigg]& \hbar^{8}
        \sum_{X} \int \hat{d}^4 \bar{w}_1 \hat{d}^4 \bar{w}_2 \hat{d}^4 \bar{q}
        \\
        &\times (2\pi) \theta(w^{0}_{1}+p^{0}_{1})  \delta(2 \bar{w}_1 \cdot p_1 + \hbar \bar{w}^2_{1})
        (2\pi) \theta(w^{0}_{2}+p^{0}_{2}) \delta(2 \bar{w}_2 \cdot p_2 + \hbar \bar{w}^2_{2})
        \\
        &\times
        (2\pi) \theta(q^{0}+p^{0}_{1}) \delta( 2 \bar{q} \cdot p_1 + \hbar \bar{q}^2)
        (2\pi) \theta(-q^{0}+p^{0}_{2}) \delta(2 \bar{q} \cdot p_2 - \hbar \bar{q}^2)
        e^{- i b \cdot \bar{q}}
        \\
        &\times
        \hat{\delta}^{(4)}( \hbar \bar{w}_1 + \hbar \bar{w}_2 + r_X)
        \Biggl[
        [s^{\nu}_{1}(p_1),\mathcal{A}^{*}\mathcal{A}]\delta^\mu_{\nu}
        -
        \frac{\hbar}{m_1^2} p_1^\mu \bar{w}_{1\nu} s^\nu_1(p_1) A^{*}A
        \\
        &-
        [s^{\nu}_{1}(p_1),\mathcal{A}^{*}]\mathcal{A}
        \left( \delta^\mu_{\nu} - \frac{\hbar}{m_1^2} p_1^\mu \bar{w}_{1\nu} \right)
        \Biggr]
        \rrangle[\bigg] .
    \end{aligned}
\end{equation}
This step is exact and does not require rewriting the coherent-state matrix element as a trace. At this stage the structures relevant for the cancellation of superclassical terms are already visible, namely $\left[s^{\mu}_{1},\mathcal{A}^{*}\mathcal{A}\right]$ and $\bar{w}_{1\nu}\mathcal{A}^{*}\mathcal{A}$. We recall here the classical spin commutator identity~\cite{Guevara:2019fsj}
\begin{equation}
    \left[{s^\mu_1},\mathcal{A}\right] = \frac{i\hbar}{m_1}\epsilon^{\mu\nu\rho\sigma}p_{1\nu}s_{1\rho}\frac{\partial \mathcal{A}}{\partial s^\sigma_1},
    \qquad
    \left[{s^\mu_1},\mathcal{A}^*\right] = \frac{i\hbar}{m_1}\epsilon^{\mu\nu\rho\sigma}p_{1\nu}s_{1\rho}\frac{\partial \mathcal{A}^*}{\partial s^\sigma_1}.
\end{equation}
The total NLO spin kick is then
\begin{equation}
    \langle \Delta s^{\mu,\text{NLO}}_{1} \rangle =
    i g^4 \llangle[\bigg]
    \int  \hat{d}^4 \bar{q}   \hat{\delta}(2 \bar{q} \cdot p_1)
    \hat{\delta}(2 \bar{q} \cdot p_2) e^{- i b \cdot \bar{q}} \mathscr{A}^{\mu(1)}_{\text{gen}}
    \rrangle[\bigg] ,
\end{equation}
where
\begin{equation}
    \begin{aligned}
        \mathscr{A}^{\mu(1)}_{\text{gen}} =&  \left[s^{\mu}_{1}(p_1),\mathcal{A}^{(1)}(\bar{q})\right]
        - \frac{\hbar}{m_1^2} p_{1}^{\mu} \bar{q}_{\nu} s^{\nu}_{1}(p_1) \mathcal{A}^{(1)}(\bar{q})
        \\
        &- i \hbar^{2}
        \sum_{X} \int \hat{d}^4 \bar{w}_1 \hat{d}^4 \bar{w}_2 
        \hat{\delta}(2 \bar{w}_1 \cdot p_1 + \hbar \bar{w}^2_{1})
        \hat{\delta}(2 \bar{w}_2 \cdot p_2 + \hbar \bar{w}^2_{2})
        \\
        &\times
        \hat{\delta}^{(4)}( \hbar \bar{w}_1 + \hbar \bar{w}_2 + r_X)
        \Biggl\{
        [s^{\nu}_{1}(p_1),
        \mathcal{A}^{(0) *}_{X}(\bar{q}, \bar{w}_1,\bar{w}_2,r_X)
        \mathcal{A}^{(0)}_{X}(\bar{w}_1,\bar{w}_2,r_X)]\delta^\mu_{\nu}
        \\
        &-
        \frac{\hbar}{m_1^2} p_{1}^{\mu} \bar{w}_{1\nu} s^{\nu}_{1}(p_1)
        \mathcal{A}^{(0) *}_{X}(\bar{q}, \bar{w}_1,\bar{w}_2,r_X)
        \mathcal{A}^{(0)}_{X}(\bar{w}_1,\bar{w}_2,r_X)
        \\
        &-
        \left[s^{\nu}_{1}(p_1),\mathcal{A}^{(0) *}_{X}(\bar{q}, \bar{w}_1,\bar{w}_2,r_X)\right]
        \mathcal{A}^{(0)}_{X}(\bar{w}_1,\bar{w}_2,r_X)
        \left( \delta^\mu_{\nu} - \frac{\hbar}{m_1^2} p_{1}^{\mu} \bar{w}_{1\nu} \right)
        \Biggr\} .
    \end{aligned}
\end{equation}
So far everything is exact. Now we can use the relation
\begin{equation}
    \left[s^{\nu}_{1}(p_1),\mathcal{A}^{*}\right]\mathcal{A} = \frac{1}{2}\left[s^{\nu}_{1}(p_1),\mathcal{A}^{*}\mathcal{A}\right]
\end{equation}
which is not a general operator identity. Rather, it becomes valid only after restricting to the conservative elastic sector, $X=\emptyset$, and taking the classical limit. In that limit the stripped tree amplitude may be written as
\begin{equation}
    \mathcal{A}^{(0)}(\bar{q},\bar{w}) = e^{i\chi(\bar{q},\bar{w})}F(\bar{q},\bar{w};s_1,s_2),
\end{equation}
where $\chi$ is independent of the spins and $F$ is real. It follows that
\begin{equation}
    \begin{aligned}
        \frac{\partial \mathcal{A}^{(0) *}}{\partial s^\sigma_1}\mathcal{A}^{(0)}
        =
        \mathcal{A}^{(0) *}\frac{\partial \mathcal{A}^{(0)}}{\partial s^\sigma_1},
    \end{aligned}
\end{equation}
and therefore
\begin{equation}
    \left[s^{\nu}_{1}(p_1),\mathcal{A}^{(0) *}\right] \mathcal{A}^{(0)}
    =
    \mathcal{A}^{(0) *}\left[s^{\nu}_{1}(p_1),\mathcal{A}^{(0)}\right]
    =
    \frac{1}{2}\left[s^{\nu}_{1}(p_1),\mathcal{A}^{(0) *}\mathcal{A}^{(0)}\right].
\end{equation}
For the conservative spin kick at 2PM we now set $X=\emptyset$, so that $r_X=0$ and $\bar{w}_2=-\bar{w}_1\equiv-\bar{w}$. The kernel then simplifies to
\begin{equation}
    \begin{aligned}
        \mathscr{A}^{\mu(1)} =&  \left[s^{\mu}_{1}(p_1),\mathcal{A}^{(1)}(\bar{q})\right]
        - \frac{\hbar}{m_1^2} p_{1}^{\mu} \bar{q}_{\nu} s^{\nu}_{1}(p_1) \mathcal{A}^{(1)}(\bar{q})
        \\
        &- i \hbar^{2}
        \int \hat{d}^4 \bar{w}   \hat{\delta}(2 \bar{w} \cdot p_1 + \hbar \bar{w}^2)
        \hat{\delta}(2 \bar{w} \cdot p_2 - \hbar \bar{w}^2)
        \\
        &\times \left( \left[s^{\nu}_{1}(p_1),  \mathcal{A}^{(0) *}(\bar{q}, \bar{w}) \mathcal{A}^{(0)}(\bar{w})\right]
        \left( \frac{1}{2}  \delta^{\mu}_{\nu}  + \frac{\hbar}{2 m_1^2} p_{1}^{\mu} \bar{w}_{\nu} \right)
        \right.
        \\
        &\hspace{4cm}- \left. \frac{\hbar}{m_1^2} p_{1}^{\mu} \bar{w}_{\nu} s^{\nu}_{1}(p_1)
        \mathcal{A}^{(0) *}(\bar{q}, \bar{w}) \mathcal{A}^{(0)}(\bar{w}) \right) .
    \end{aligned}
\end{equation}
However, one should keep in mind that the relation
\begin{equation}
[s^{\nu}_{1}(p_1),\mathcal{A}^{(0) *}] \mathcal{A}^{(0)}
=
\frac{1}{2}[s^{\nu}_{1}(p_1),\mathcal{A}^{(0) *}\mathcal{A}^{(0)}]
\end{equation}
is still an additional simplification. In the present case the two tree amplitudes entering the cut box are evaluated at different momentum transfers, namely $\bar{w}$ and $-\bar{w}+\bar{q}$, so this relation need not hold identically. Therefore, the exact conservative kernel should be written as
\begin{equation}
    \begin{aligned}
\mathscr{A}^{\mu(1)}_{\text{exact}}
=
\mathscr{A}^{\mu(1)}
+
\Delta\mathscr{A}^{\mu(1)}_{\text{extra}},
    \end{aligned}
\end{equation}
where $\mathscr{A}^{\mu(1)}$ is the symmetrized expression above and
\begin{equation}
    \begin{aligned}
        \Delta\mathscr{A}^{\mu(1)}_{\text{extra}}
        =
        - i \hbar^{2}
        &\int \hat{d}^4 \bar{w}   \hat{\delta}(2 \bar{w} \cdot p_1 + \hbar \bar{w}^2)
        \hat{\delta}(2 \bar{w} \cdot p_2 - \hbar \bar{w}^2)
        \\
        &\times
        \left( \delta^{\mu}_{\nu} - \frac{\hbar}{m_1^2} p_{1}^{\mu} \bar{w}_{\nu} \right)
        \Biggl[
        \frac{1}{2}[s^{\nu}_{1}(p_1),\mathcal{A}^{(0) *}(\bar{q}, \bar{w}) \mathcal{A}^{(0)}(\bar{w})]
        \\
        &\hspace{4cm}-
        [s^{\nu}_{1}(p_1),\mathcal{A}^{(0) *}(\bar{q}, \bar{w})] \mathcal{A}^{(0)}(\bar{w})
        \Biggr].
    \end{aligned}
\end{equation}
Notice that again we have pulled out factors of $g/\sqrt{\hbar}$. The first line is associated with the contribution $\langle \Delta s^{(1),\mu}_{1} \rangle$, whereas the remaining terms come from $\langle \Delta s^{(2),\mu}_{1} \rangle$. The expression for $\mathscr{A}^{\mu(1)}$ involves the one-loop four-point amplitude $\mathcal{A}^{(1)}$ and a contribution which looks like a weighted cut of a one-loop amplitude; this part comes entirely from $\langle \Delta s^{(2),\mu}_{1} \rangle$ with an empty $X$. In this conservative sector, the superclassical cancellation is already nearly manifest, since the real kernel is organized directly in terms of $[s^{\mu}_{1},\mathcal{A}^{(0)*}\mathcal{A}^{(0)}]$ and $\bar{w}_{\nu}\mathcal{A}^{(0)*}\mathcal{A}^{(0)}$. On the other hand, for a non-trivial $X$ the second part, involving the integral over $\mathcal{A}^{(0)*}\mathcal{A}^{(0)}$, describes the emission of real messengers (that is, photons or gravitons). It is the source of the dissipative back reaction, capturing the angular momentum lost from the system due to radiation.

The conservative real-kernel contribution displayed above is already organized in a way that makes its relation to the KMOC impulse formula transparent. In particular, the terms involving $[s_1^\mu,\cdots]$ should be viewed as the spin variation of a scalar kernel, whereas the pieces with an explicit factor of $p_1^\mu$ are manifestly longitudinal. This suggests comparing the NLO angular-impulse kernel directly with the corresponding KMOC formula for the NLO linear impulse, which we now recall.

It is useful to relate the NLO angular-impulse kernel to the corresponding KMOC formula for the NLO linear impulse~\cite{Kosower_2019,Maybee:2019jus}. In the same notation used above, the latter may be written as
\begin{equation}
    \langle \Delta p_{1,\text{NLO}}^{\mu} \rangle
    =
    i g^4 \llangle[\bigg]
    \int  \hat{d}^4 \bar{q}   \hat{\delta}(2 \bar{q} \cdot p_1)
    \hat{\delta}(2 \bar{q} \cdot p_2) e^{- i b \cdot \bar{q}}  
    \mathscr{I}^{\mu(1)}_{\text{gen}}
    \rrangle[\bigg] ,
\end{equation}
with
\begin{equation}
    \begin{aligned}
        \mathscr{I}^{\mu(1)}_{\text{gen}}
        =
        \bar q^\mu \mathcal{A}^{(1)}(\bar q) &-
        i \hbar^{2}
        \sum_{X} \int \hat{d}^4 \bar{w}_1 \hat{d}^4 \bar{w}_2 
        \hat{\delta}(2 \bar{w}_1 \cdot p_1 + \hbar \bar{w}^2_{1})
        \hat{\delta}(2 \bar{w}_2 \cdot p_2 + \hbar \bar{w}^2_{2})
        \\
        &\times
        \hat{\delta}^{(4)}( \hbar \bar{w}_1 + \hbar \bar{w}_2 + r_X) 
        \bar w_1^\mu 
        \mathcal{A}^{(0) *}_{X}(\bar{q}, \bar{w}_1,\bar{w}_2,r_X)
        \mathcal{A}^{(0)}_{X}(\bar{w}_1,\bar{w}_2,r_X) .
    \end{aligned}
\end{equation}
If we denote by
\begin{equation}
    \begin{aligned}
        \mathscr{K}^{(1)}_{\text{gen}}
        \equiv
        \mathcal{A}^{(1)}(\bar q) &-
        i \hbar^{2}
        \sum_{X} \int \hat{d}^4 \bar{w}_1 \hat{d}^4 \bar{w}_2 
        \hat{\delta}(2 \bar{w}_1 \cdot p_1 + \hbar \bar{w}^2_{1})
        \hat{\delta}(2 \bar{w}_2 \cdot p_2 + \hbar \bar{w}^2_{2})
        \\
        &\times
        \hat{\delta}^{(4)}( \hbar \bar{w}_1 + \hbar \bar{w}_2 + r_X) 
        \mathcal{A}^{(0) *}_{X}(\bar{q}, \bar{w}_1,\bar{w}_2,r_X)
        \mathcal{A}^{(0)}_{X}(\bar{w}_1,\bar{w}_2,r_X)
    \end{aligned}
\end{equation}
the associated scalar kernel obtained by stripping the explicit momentum insertion. Then, at the level of the conservative NLO kernel, the spin-kick formula can be organized schematically as
\begin{equation}
    \begin{aligned}
        \mathscr{A}^{\mu(1)}_{\text{cons}}
        =&
        \left[s_1^\mu(p_1),\mathscr{K}^{(1)}_{\text{gen}}\right]
        -
        \frac{\hbar}{m_1^2} p_1^\mu s_{1\nu} 
        \mathscr{I}^{\nu(1)}_{\text{gen}}
        +
        \Delta\mathscr{A}^{\mu(1)}_{\text{extra}} .
        \label{eq:spin-kick-from-impulse-kernel}
    \end{aligned}
\end{equation}
Here the first two terms are the direct analogue of the NLO impulse kernel, while $\Delta\mathscr{A}^{\mu(1)}_{\text{extra}}$ denotes the additional exact correction that appears in the conservative elastic sector discussed below. Using the classical spin commutator identity,
\begin{equation}
    \left[s_1^\mu(p_1),\mathscr{K}^{(1)}_{\text{gen}}\right]
    =
    i\hbar \epsilon^{\mu\nu\rho\sigma}u_{1\nu}s_{1\rho}
    \frac{\partial}{\partial s^\sigma_1} 
    \mathscr{K}^{(1)}_{\text{gen}},
\end{equation}
so the second term is manifestly longitudinal, whereas non-longitudinal structures can only arise from the spin variation of the scalar impulse kernel and from the exact correction
$\Delta\mathscr{A}^{\mu(1)}_{\text{extra}}$.

\endgroup

\section{Kerr black hole angular impulse from quantum scattering amplitudes}
\label{amplitudes}

Gravitationally interacting spinning black holes with masses $m_1, m_2$ can be described by a system of two spin-$S_k$ fields $\Phi_k$, $k=1,2$ minimally coupled to gravity:
\begin{equation}
    S = \int d^{4}x \sqrt{-g} \left( \frac{1}{16 \pi G_N} R + {\mathcal L}_{\textrm{int}}[\Phi_{k}, g_{\mu\nu}] \right)
\end{equation}
where the first term is the Einstein-Hilbert contribution. Here we consider the point-particle approximation and we exclude local interactions between matter fields. We are interested in calculating the total change in the classical spin of a Kerr black hole during the collision with another Kerr black hole. We will calculate tree-level and one-loop quantum scattering amplitudes, focusing on the pieces that are relevant to the computation of our classical observable.

\subsection{Classical contribution from scattering amplitudes}

Now we move on to the calculation of the 4-point amplitude involving gravitons and massive spinning particles up to one loop order. We do not calculate the full amplitude; instead we focus on only the pieces that will produce a classical contribution.

\subsubsection{Tree-level amplitudes}

The tree-level 3-point gravity amplitudes can be obtained by considering the KLT relations. One finds\footnote{We restrict our analysis to the S-integer case.}~\cite{Arkani-Hamed:2017jhn,Johansson:2019dnu}
\begin{equation}
    \begin{aligned}
        M_{S}({\mathbf 1}, {\mathbf 2}, 3^{++}) =& \frac{i \kappa}{2} 
        \frac{\langle {\mathbf 1} {\mathbf 2} \rangle^{2S}}{m^{2S-2}} x^2
        = \frac{i \kappa}{2} \frac{1}{m^{2S-2}} x^2
        \bigl[ {\mathbf 2} |^{2S} \exp{ i \frac{p_{3}^{\mu} \epsilon_{+}^{\nu} J_{\mu\nu} }{p \cdot \epsilon_{+}} }  | {\mathbf 1} \bigr]^{2S}
        \\
        M_{S}({\mathbf 1}, {\mathbf 2}, 3^{--}) =&  \frac{i \kappa}{2}
        \frac{\bigl[ {\mathbf 1} {\mathbf 2} \bigr]^{2S}}{m^{2S-2}} \frac{1}{x^2}
        = \frac{i \kappa}{2}
        \frac{1}{m^{2S-2}} \frac{1}{x^2}
        \langle {\mathbf 2} |^{2S} \exp{ i \frac{p_{3}^{\mu} \epsilon_{-}^{\nu} J_{\mu\nu}}{p \cdot \epsilon_{-}} }  | {\mathbf 1} \rangle^{2S}
    \end{aligned}
\end{equation}
where $\kappa^2 = 32 \pi G_N$ and $p = (p_1-p_2)/2$. Moreover, $J_{\mu\nu}$ is the angular-momentum operator; for details of its expression, see ref.~\cite{Guevara:2018wpp}. The four-point amplitude corresponding to the scattering of a spin-$S_1$ particle with mass $m_1$ with a spin-$S_2$ particle with mass $m_2$ with exchange of gravitons is given by
\begin{equation}
    M^{(0)}(1,2 \to 1^{\prime}, 2^{\prime}) = -  m_1^2 m_2^2 \frac{\kappa^2}{4 q^2} 
    \left( \frac{\langle {\mathbf 2} {\mathbf 2}^{\prime} \rangle^{2S_2}}{m_2^{2S_2}} 
    \frac{\bigl[ {\mathbf 1} {\mathbf 1}^{\prime} \bigr]^{2S_1}}{m_1^{2S_1}} e^{- 2 \phi}
    + \frac{\bigl[ {\mathbf 2} {\mathbf 2}^{\prime} \bigr]^{2S_2}}{m_2^{2S_2}} 
    \frac{\langle {\mathbf 1} {\mathbf 1}^{\prime} \rangle^{2S_1}}{m_1^{2S_1}}  e^{2 \phi} \right) 
\end{equation}
where $\phi$ is defined through its relations with the relative Lorentz factor $\gamma = p_1 \cdot p_2/(m_1 m_2)$: $\cosh 2\phi=2\gamma^2-1,    \sinh 2\phi=2\gamma\sqrt{\gamma^2-1},    \sinh \phi=\sqrt{\gamma^2-1} $. The classical contribution can be easily extracted, see for instance ref.~\cite{Menezes:2022tcs}
\begin{equation}
    \begin{aligned}
        {\mathcal M}^{(0)}(1,2 \to 1^{\prime}, 2^{\prime}) =&  
        -  m_1^2 m_2^2 \frac{1}{4 \hbar^2 \bar{q}^2} 
        \left[ e^{ \bar{q} \cdot a_1}   e^{ \bar{q} \cdot a_2}  e^{- 2 \phi} 
        + e^{ -\bar{q} \cdot a_1}   e^{ -\bar{q} \cdot a_2}  e^{ 2 \phi} \right] 
        \\
        =&  -  m_1^2 m_2^2 \frac{1}{2 \hbar^2 \bar{q}^2} 
        \cosh[\bar{q} \cdot (a_1+a_2) - 2 \phi]
        \\
        =&  -  \frac{1}{4 \hbar^2 \bar{q}^2} 
        \biggl[ \Big( ( s - m_1^2 - m_2^2 )^2 - 2 m_1^2 m_2^2 \Big) \cosh[\bar{q} \cdot (a_1+a_2)]  
        \\
        &- ( s - m_1^2 - m_2^2 ) \sqrt{( s - m_1^2 - m_2^2 )^2 - 4 m_1^2 m_2^2} \sinh[\bar{q} \cdot (a_1+a_2)]
        \biggr] 
        \\
    \end{aligned}
\end{equation}
where $a_i^{\mu} = s_i^{\mu}/m_i$ (this is the ring radius in the classical limit) and $s = (p_1 + p_2)^2$. We can recast this amplitude in a more convenient form using the Gram determinant constraint and on-shell conditions; we find that
\begin{equation}
    \begin{aligned}
        & \cosh[ \bar{q} \cdot (a_1+a_2)]  \cosh 2\phi - \sinh[ \bar{q} \cdot (a_1+a_2)]  \sinh 2\phi 
        \\
        & = P_{\mu\nu\alpha\beta} u_{1}^{\mu} u_{1}^{\nu} \left[ 2 u_{2}^{\alpha} u_{2}^{\beta} 
        -  i \epsilon^{(\alpha}_{\ \lambda\rho\sigma} u_{2}^{\beta)} u_{2}^{\lambda} 
        (a_1 + a_2)^{\rho} \bar{q}^{\sigma}
        + \frac{1}{2!} \Big( \bar{q} \cdot (a_1+a_2) \Big)^2 2 u_{2}^{\alpha} u_{2}^{\beta}
        + \cdots \right]
        \\
        & =  P_{\mu\nu\alpha\beta}   u_{1}^{\mu} u_{1}^{\nu} 
        \exp{- i (a_1+a_2)* \bar{q}}^{(\alpha}_{\ \lambda} u_{2}^{\beta)} u_{2}^{\lambda}
    \end{aligned}
\end{equation}
where $(a * \bar{q})_\nu^\mu=\epsilon_{\nu \rho \sigma}^\mu a^\rho \bar{q}^\sigma$ and $P_{\mu\nu\alpha\beta}$ is the usual de Donder trace reverser
\begin{equation}
    P_{\mu\nu\rho\sigma}=\frac{1}{2}\left(
    \eta_{\mu\rho}\eta_{\nu\sigma}+ \eta_{\mu\sigma}\eta_{\nu\rho}-\eta_{\mu\nu}\eta_{\rho\sigma}
    \right).
\end{equation}
Therefore,
\begin{equation}
    {\mathcal M}^{(0)}(1,2 \to 1^{\prime}, 2^{\prime}) = -  \frac{1}{4 \hbar^2 \bar{q}^2}  
    2 P_{\mu\nu\alpha\beta}   p_{1}^{\mu} p_{1}^{\nu} 
    \exp{- i (a_1+a_2)* \bar{q}}^{(\alpha}_{\ \lambda} p_{2}^{\beta)} p_{2}^{\lambda} .
\end{equation}

\subsubsection{One-loop level amplitudes}

For the one-loop computation, the classical contributions will come from box, crossed box and triangle terms. As usual in KMOC expressions, we will also need a so-called cut box. All box-type contributions were calculated in ref.~\cite{Menezes:2022tcs}, which we quote here for convenience
\begin{equation}
    \begin{aligned}
        i M^{(1)}_{\Box} =&
        4 \kappa^4 \int {\hat{d} ^D \ell}  \frac{B_{h}(\ell, p_1,p_2) B_{h}(-\ell-p_1+p_1^{\prime}, p_1,p_2)}
        { \ell^2 [(\ell + p_1)^2 - m_1^2] (\ell+p_1-p_1^{\prime})^2  [ (\ell-p_2)^2 - m_2^2]}
        \\
        i M^{(1)}_{\boxtimes} =& 
        4 \kappa^4 \int {\hat{d} ^D \ell}  
        \frac{ B_{h}(\ell, p_1,p_2) B_{h}(-\ell-p_1+p_1^{\prime}, p_1,p_2) }
        {\ell^2 [(\ell + p_1)^2 - m_1^2](\ell+p_1-p_1^{\prime})^2 [ (\ell+p_2^{\prime})^2 - m_2^2] }
    \end{aligned}
\end{equation}
where
\begin{equation}
    B_h(q, p_1,p_2) = \frac{1}{4} P_{\mu\nu\alpha\beta}   p_{1}^{\mu} p_{1}^{\nu} 
    \exp{- i \frac{(a_1+a_2)}{\hbar}* q }^{(\alpha}_{\ \lambda} p_{2}^{\beta)} p_{2}^{\lambda},
\end{equation}
and we have dropped any explicit term in the numerators that will not produce a classical contribution. The final results are
\begin{equation}
    i {\mathcal M}^{(1)}_{\Box}(1,2 \to 1^{\prime}, 2^{\prime}) \bigg|_{\textrm{Classical}}
    + i {\mathcal M}^{(1)}_{\boxtimes}(1,2 \to 1^{\prime}, 2^{\prime}) \bigg|_{\textrm{Classical}}
    = i {\mathcal M}_{-2} + i {\mathcal M}_{-1} + {\mathcal O}(\hbar^0)
\end{equation}
where
\begin{equation}
    \begin{aligned}
        i {\mathcal M}_{-2} =& 
        - \frac{1}{2\hbar^{2}} 
        \int {\hat{d} ^4 \bar{\ell}}  
        \frac{ B_{h}(\bar{\ell}, p_1,p_2) B_{h}(-\bar{\ell}+\bar{q}, p_1,p_2) }{\bar{\ell}^2 (\bar{\ell} - \bar{q})^2} 
        \hat{\delta}( \bar{\ell} \cdot p_2) \hat{\delta}( \bar{\ell} \cdot p_1) 
        + {\mathcal O}(1/\hbar)
    \end{aligned}
\end{equation}
and
\begin{equation}
    \begin{aligned}
        i {\mathcal M}_{-1}  =&
        \frac{1}{2 \hbar^{1+2\epsilon}} 
        \int {\hat{d} ^D \bar{\ell}}  
        \frac{ B_h(\bar{\ell}, p_1,p_2) B_h(-\bar{\ell}+\bar{q}, p_1,p_2) }{\bar{\ell}^2 (\bar{\ell} - \bar{q})^2 
        ( \bar{\ell} \cdot p_1  + i\varepsilon ) 
        (\bar{\ell} \cdot p_2 - i\varepsilon )} 
        \left( \frac{\bar{\ell}^2}{\bar{\ell} \cdot p_1 + i\varepsilon} \right)
        + \{ 2 \leftrightarrow -2 \}
        \\
        &+ \frac{1}{2\hbar^{1+2\epsilon}} 
        \int {\hat{d} ^D \bar{\ell}}  
        \frac{ B_h(\bar{\ell}, p_1,p_2) B_h(-\bar{\ell}+\bar{q}, p_1,p_2) }{\bar{\ell}^2 (\bar{\ell} - \bar{q})^2 
        ( \bar{\ell} \cdot p_1 + i\varepsilon ) 
        (\bar{\ell} \cdot p_2 - i\varepsilon )}
        \frac{\bar{\ell}^2}{- \bar{\ell} \cdot p_2  + i\varepsilon} 
        \\
        &- \frac{1}{2 \hbar^{1+2\epsilon}} 
        \int {\hat{d} ^D \bar{\ell}}   \frac{ B_h(\bar{\ell}, p_1,p_2) B_h(-\bar{\ell}+\bar{q}, p_1,p_2) }
        {\bar{\ell}^2 (\bar{\ell} - \bar{q})^2 
        ( \bar{\ell} \cdot p_1 + i\varepsilon ) 
        (\bar{\ell} \cdot p_2  + i\varepsilon )}
        \frac{ ( \bar{\ell} - \bar{q})^2 - \bar{q}^2 }{\bar{\ell} \cdot p_2 + i\varepsilon} .
    \end{aligned}
\end{equation}
Hence
\begin{equation}
    \begin{aligned}
        i {\mathcal M}_{-1} + \left[i {\mathcal M}_{-2}\right]_{{\mathcal O}(1/\hbar)} =&
        \frac{i}{2\hbar^{1+2\epsilon}} 
        \int {\hat{d} ^D \bar{\ell}}  
        \frac{ \bar{\ell} \cdot (\bar{\ell} - \bar{q}) \hat{\delta}(\bar{\ell} \cdot p_2) }{\bar{\ell}^2 (\bar{\ell} - \bar{q})^2 
        ( \bar{\ell} \cdot p_1  + i\varepsilon )^2 } 
        B_{h}(\bar{\ell}, p_1,p_2) B_{h}(-\bar{\ell}+\bar{q}, p_1,p_2)
        \\
        &+
        \frac{i}{ 2 \hbar^{1+2\epsilon}} 
        \int {\hat{d} ^D \bar{\ell}}  
        \frac{ \bar{\ell} \cdot (\bar{\ell} - \bar{q}) \hat{\delta}(\bar{\ell} \cdot p_1) }
        {\bar{\ell}^2 (\bar{\ell} - \bar{q})^2 ( \bar{\ell} \cdot p_2  - i\varepsilon )^2 } 
        B_{h}(\bar{\ell}, p_1,p_2) \\
        &\times B_{h}(-\bar{\ell}+\bar{q}, p_1,p_2)- \frac{1}{4 \hbar^{1+2\epsilon}} 
        \int {\hat{d} ^D \bar{\ell}}  
        ( 2 \bar{\ell} \cdot \bar{q} - \bar{\ell}^2 )
        \\
        & \times\frac{ B_{h}(\bar{\ell}, p_1,p_2) B_{h}(-\bar{\ell}+\bar{q}, p_1,p_2) }
        {\bar{\ell}^2 (\bar{\ell} - \bar{q})^2}
        \\
        &\times \Big( \hat{\delta}^{\prime}( \bar{\ell} \cdot p_1) 
        \hat{\delta}( \bar{\ell} \cdot p_2 ) 
        - \hat{\delta}( \bar{\ell} \cdot p_1) 
        \hat{\delta}^{\prime}( \bar{\ell} \cdot p_2 ) \Big) .
    \end{aligned}
\end{equation}
For the cut-box, the analytic expression reads
\begin{equation}
    {\mathcal M}^{(0) *} {\mathcal M}^{(0)}
    = \frac{4}{\hbar^{4}}
    \frac{ B_{h}(\bar{w}, p_1,p_2) B_{h}(-\bar{w}+\bar{q}, p_1^{\prime},p_2^{\prime}) }
    {\bar{w}^2 (\bar{w} - \bar{q})^2} .
\end{equation}
We relabel $w \to \ell$. In the conservative classical reduction, complex conjugation acts by reversing the momentum argument, so we keep writing the second cut-box factor as $B_h(-\bar\ell+\bar q,u_1,u_2)$. However,  we have to distinguish two situations in the formula for the angular impulse, namely the term
$$
\frac{\hbar}{m_1^2} p_{1}^{\mu} s^{\nu}_{1}(p_1)
\int \hat{d}^4 \bar{w}   \hat{\delta}(2 \bar{w} \cdot p_1 + \hbar \bar{w}^2)
\hat{\delta}(2 \bar{w} \cdot p_2 - \hbar \bar{w}^2)    \bar{w}_{\nu}  
 {\mathcal M}^{(0) *}(\bar{q}, \bar{w}) {\mathcal M}^{(0)}(\bar{w}) 
$$
and the commutator
$$
 \int \hat{d}^4 \bar{w}   \hat{\delta}(2 \bar{w} \cdot p_1 + \hbar \bar{w}^2)
\hat{\delta}(2 \bar{w} \cdot p_2 - \hbar \bar{w}^2)  
 [s^{\nu}_{1}(p_1),  {\mathcal M}^{(0) *}(\bar{q}, \bar{w}) {\mathcal M}^{(0)}(\bar{w})] 
\left( \frac{1}{2}  \delta^{\mu}_{\nu}  + \frac{\hbar}{2 m_1^2} p_{1}^{\mu} \bar{w}_{1\nu} \right) .
$$
For the former, which we call $I_1^{\nu}$, we may follow the same general recipe to isolate a contribution that will cancel out the corresponding one from the box contributions, see for instance ref.~\cite{Menezes:2022tcs}. We find that
\begin{equation}
    \begin{aligned}
        I_1^{\nu} =& 
        - i \frac{4}{\hbar^{2}} 
        \int {\hat{d} ^4 \bar{\ell}}  \hat{\delta}( 2 \bar{\ell} \cdot p_1 + \hbar \bar{\ell}^2) 
        \hat{\delta}( 2 \bar{\ell} \cdot p_2 - \hbar \bar{\ell}^2) \bar{\ell}^{\nu}
        \frac{ B_{h}(\bar{\ell}, p_1,p_2) B_{h}(-\bar{\ell}+\bar{q},p_1^{\prime},p_2^{\prime}) }
        {\bar{\ell}^2 (\bar{\ell} - \bar{q})^2} .
    \end{aligned}
\end{equation}
Observe that there is still an additional factor of $\hbar$ in the formula. Expand in $\hbar$, and truncate after order $1/\hbar$, so that
\begin{equation}
    I_1^{\nu} = {\mathcal C}_{-2}^{\nu} + {\mathcal C}_{-1}^{\nu}
\end{equation}
where
\begin{equation}
    {\mathcal C}_{-2}^{\nu}  = - \frac{i}{2 \hbar^{2}} 
    \int {\hat{d} ^4 \bar{\ell}}  \frac{\bar{q}^{\nu}}{\bar{\ell}^2 (\bar{\ell} - \bar{q})^2}
    \hat{\delta}( \bar{\ell} \cdot p_1 ) 
    \hat{\delta}( \bar{\ell} \cdot p_2 ) 
    B_{h}(\bar{\ell}, p_1,p_2) B_{h}(-\bar{\ell}+\bar{q}, p_1,p_2)
    + {\mathcal O}(1/\hbar)
\end{equation}
and
\begin{equation}
    {\mathcal C}_{-1}^{\nu} =  \widetilde{{\mathcal C}}^{(1)\nu} +  \widetilde{{\mathcal C}}^{(2)\nu}
\end{equation}
with
\begin{equation}
    \begin{aligned}
        \widetilde{{\mathcal C}}^{(1)\nu} =& i \frac{1}{4\hbar} 
        \int {\hat{d} ^4 \bar{\ell}}  \hat{\delta}( \bar{\ell} \cdot p_1) 
        \hat{\delta}( \bar{\ell} \cdot p_2) \bar{\ell}^{\nu}
        \frac{ B_{h}(\bar{\ell}, p_1,p_2) }
        {\bar{\ell}^2 (\bar{\ell} - \bar{q})^2} 
        \\
        &\times \Big[ P_{\gamma\delta\alpha\beta}   p_{1}^{\gamma} p_{1}^{\delta} 
        \exp{\bigl(- i (a_1+a_2)* ( -\bar{\ell} + \bar{q} ) \bigr)}^{(\alpha}_{\ \lambda} \bar{q}^{\beta)} p_{2}^{\lambda}
        \\
        &+ P_{\gamma\delta\alpha\beta}   p_{1}^{\gamma} p_{1}^{\delta} 
        \exp{\bigl(- i (a_1+a_2)* ( -\bar{\ell} + \bar{q} ) \bigr)}^{(\alpha}_{\ \lambda} p_{2}^{\beta)} \bar{q}^{\lambda}
        \Big]
        \\
        &- i \frac{1}{4\hbar} 
        \int {\hat{d} ^4 \bar{\ell}}  \hat{\delta}( \bar{\ell} \cdot p_1) 
        \hat{\delta}( \bar{\ell} \cdot p_2) \bar{\ell}^{\nu}
        \frac{ B_{h}(\bar{\ell}, p_1,p_2) }
        {\bar{\ell}^2 (\bar{\ell} - \bar{q})^2}  
        \\
        &\times \Big[ P_{\gamma\delta\alpha\beta}   \bar{q}^{\gamma} p_{1}^{\delta} 
        \exp{\bigl(- i (a_1+a_2)* ( -\bar{\ell} + \bar{q} ) \bigr)}^{(\alpha}_{\ \lambda} p_{2}^{\beta)} p_{2}^{\lambda}
        \\
        &+ P_{\gamma\delta\alpha\beta}   p_{1}^{\gamma} \bar{q}^{\delta} 
        \exp{\bigl(- i (a_1+a_2)* ( -\bar{\ell} + \bar{q} ) \bigr)}^{(\alpha}_{\ \lambda} p_{2}^{\beta)} p_{2}^{\lambda}
        \Big]
    \end{aligned}
\end{equation}
and
\begin{equation}
    \begin{aligned}
        \widetilde{{\mathcal C}}^{(2)\nu} =&
        - \frac{i}{2\hbar}
        \int {\hat{d} ^4 \bar{\ell}}  
        \bar{\ell}^2 \frac{\bar{\ell}^{\nu}}{\bar{\ell}^2 (\bar{\ell} - \bar{q})^2}
        B_{h}(\bar{\ell}, p_1,p_2) B_{h}(-\bar{\ell}+\bar{q}, p_1,p_2)
        \\
        &\times\Big( \hat{\delta}^{\prime}( \bar{\ell} \cdot p_1) 
        \hat{\delta}( \bar{\ell} \cdot p_2 ) 
        - \hat{\delta}( \bar{\ell} \cdot p_1) 
        \hat{\delta}^{\prime}( \bar{\ell} \cdot p_2 ) \Big) .
    \end{aligned}
\end{equation}
We keep the above expression for $\widetilde{{\mathcal C}}^{(1)\nu}$ as is and sum the ${{\mathcal O}(1/\hbar)}$ contributions from $\widetilde{{\mathcal C}}^{(2)\nu} $ and ${\mathcal C}^{\nu}_{-2}$; we get
\begin{equation}
    \begin{aligned}
        \widetilde{{\mathcal C}}^{(2)\nu} + [{\mathcal C}^{\nu}_{-2}]_{{\mathcal O}(1/\hbar)} =& 
        - \frac{i}{2 \hbar} 
        \int {\hat{d} ^4 \bar{\ell}}  
        \bar{\ell} \cdot (\bar{\ell} - \bar{q}) \frac{\bar{\ell}^{\nu}}{\bar{\ell}^2 (\bar{\ell} - \bar{q})^2}
        B_{h}(\bar{\ell}, p_1,p_2) B_{h}(-\bar{\ell}+\bar{q}, p_1,p_2)
        \\
        &\times \Big( \hat{\delta}^{\prime}( \bar{\ell} \cdot p_1) 
        \hat{\delta}( \bar{\ell} \cdot p_2 ) 
        - \hat{\delta}( \bar{\ell} \cdot p_1) 
        \hat{\delta}^{\prime}( \bar{\ell} \cdot p_2 ) \Big)
        \\
        &- \frac{i}{4 \hbar} 
        \int {\hat{d} ^4 \bar{\ell}}  
        ( 2 \bar{\ell} \cdot \bar{q} - \bar{\ell}^2 ) \frac{\bar{q}^{\nu}}{\bar{\ell}^2 (\bar{\ell} - \bar{q})^2}
        B_{h}(\bar{\ell}, p_1,p_2) B_{h}(-\bar{\ell}+\bar{q}, p_2,p_2)
        \\
        &\times \Big( \hat{\delta}^{\prime}( \bar{\ell} \cdot p_1) 
        \hat{\delta}( \bar{\ell} \cdot p_2 ) 
        - \hat{\delta}( \bar{\ell} \cdot p_1) 
        \hat{\delta}^{\prime}( \bar{\ell} \cdot p_2 ) \Big).
    \end{aligned}
\end{equation}
Now let us turn our attentions to the commutator term. Since the commutator will produce an extra factor of $\hbar$, the second term in such a contribution can be easily carried out using similar steps as above; calling this term $I_3^{\nu}$, we find
\begin{equation}
    I_3^{\nu} = - \frac{i}{2 \hbar^{2}} 
    \int {\hat{d} ^4 \bar{\ell}}  \frac{\bar{q}^{\nu}}{\bar{\ell}^2 (\bar{\ell} - \bar{q})^2}
    \hat{\delta}( \bar{\ell} \cdot p_1 ) 
    \hat{\delta}( \bar{\ell} \cdot p_2 ) 
    B_{h}(\bar{\ell}, p_1,p_2) B_{h}(-\bar{\ell}+\bar{q}, p_1,p_2) .
\end{equation}
Now the remaining term is 
\begin{equation}
    \begin{aligned}
        I_2 =& 
        - i \frac{2}{\hbar^{2}} 
        \int {\hat{d} ^4 \bar{\ell}}  \hat{\delta}( 2 \bar{\ell} \cdot p_1 + \hbar \bar{\ell}^2) 
        \hat{\delta}( 2 \bar{\ell} \cdot p_2 - \hbar \bar{\ell}^2) 
        \frac{ B_{h}(\bar{\ell}, p_1,p_2) B_{h}(-\bar{\ell}+\bar{q},p_1^{\prime},p_2^{\prime}) }
        {\bar{\ell}^2 (\bar{\ell} - \bar{q})^2} .
    \end{aligned}
\end{equation}
Again expand in $\hbar$ and keep terms up to $1/\hbar$; we find that
\begin{equation}
    I_2 = {\mathcal D}_{-2} + {\mathcal D}_{-1} 
\end{equation}
where
\begin{equation}
    {\mathcal D}_{-2}  = - \frac{i}{2 \hbar^{2}} 
    \int {\hat{d} ^4 \bar{\ell}}  \frac{1}{\bar{\ell}^2 (\bar{\ell} - \bar{q})^2}
    \hat{\delta}( \bar{\ell} \cdot p_1 ) 
    \hat{\delta}( \bar{\ell} \cdot p_2 ) 
    B_{h}(\bar{\ell}, p_1,p_2) B_{h}(-\bar{\ell}+\bar{q}, p_1,p_2)
\end{equation}
and
\begin{equation}
    {\mathcal D}_{-1} =  \widetilde{{\mathcal D}}^{(1)} +  \widetilde{{\mathcal D}}^{(2)}
\end{equation}
with
\begin{equation}
    \begin{aligned}
        \widetilde{{\mathcal D}}^{(1)} =& i \frac{1}{8\hbar} 
        \int {\hat{d} ^4 \bar{\ell}}  \hat{\delta}( \bar{\ell} \cdot p_1) 
        \hat{\delta}( \bar{\ell} \cdot p_2) 
        \frac{ B_{h}(\bar{\ell}, p_1,p_2) }
        {\bar{\ell}^2 (\bar{\ell} - \bar{q})^2} 
        \\
        &\times \Big[ P_{\gamma\delta\alpha\beta}   p_{1}^{\gamma} p_{1}^{\delta} 
        \exp{\bigl(- i (a_1+a_2)* ( -\bar{\ell} + \bar{q} ) \bigr)}^{(\alpha}_{\ \lambda} \bar{q}^{\beta)} p_{2}^{\lambda}
        \\
        &+ P_{\gamma\delta\alpha\beta}   p_{1}^{\gamma} p_{1}^{\delta} 
        \exp{\bigl(- i (a_1+a_2)* ( -\bar{\ell} + \bar{q} ) \bigr)}^{(\alpha}_{\ \lambda} p_{2}^{\beta)} \bar{q}^{\lambda}
        \Big]
        \\
        &- i \frac{1}{8\hbar} 
        \int {\hat{d} ^4 \bar{\ell}}  \hat{\delta}( \bar{\ell} \cdot p_1) 
        \hat{\delta}( \bar{\ell} \cdot p_2) 
        \frac{ B_{h}(\bar{\ell}, p_1,p_2) }
        {\bar{\ell}^2 (\bar{\ell} - \bar{q})^2}  
        \\
        &\times \Big[ P_{\gamma\delta\alpha\beta}   \bar{q}^{\gamma} p_{1}^{\delta} 
        \exp{\bigl(- i (a_1+a_2)* ( -\bar{\ell} + \bar{q} ) \bigr)}^{(\alpha}_{\ \lambda} p_{2}^{\beta)} p_{2}^{\lambda}
        \\
        &+ P_{\gamma\delta\alpha\beta}   p_{1}^{\gamma} \bar{q}^{\delta} 
        \exp{\bigl(- i (a_1+a_2)* ( -\bar{\ell} + \bar{q} ) \bigr)}^{(\alpha}_{\ \lambda} p_{2}^{\beta)} p_{2}^{\lambda}
        \Big]
    \end{aligned}
\end{equation}
and
\begin{equation}
    \widetilde{{\mathcal D}}^{(2)} =
    - \frac{i}{4\hbar}
    \int \frac{{\hat{d} ^4 \bar{\ell}}  
    \bar{\ell}^2}{\bar{\ell}^2 (\bar{\ell} - \bar{q})^2}
    B_{h}(\bar{\ell}, p_1,p_2) B_{h}(-\bar{\ell}+\bar{q}, p_1,p_2)
    \Big( \hat{\delta}^{\prime}( \bar{\ell} \cdot p_1) 
    \hat{\delta}( \bar{\ell} \cdot p_2 ) 
    - \hat{\delta}( \bar{\ell} \cdot p_1) 
    \hat{\delta}^{\prime}( \bar{\ell} \cdot p_2 ) \Big) .
\end{equation}
As above, $p_1^{\prime} = p_1 + \hbar \bar{q}$, $p_2^{\prime} = p_2 - \hbar \bar{q}$. When we collect all box-type terms, we see that all hyper-classical terms cancel out.

One important check is to verify whether the extra term $\Delta\mathscr{M}^{\mu(1)}_{\text{extra}}$ generates any hyper-classical contribution. In the exact classical kernel $\Delta\mathscr{M}^{\mu(1)}_{\text{extra}}$, the only part that can contribute to the hyper-classical sector is the term proportional to $\delta^\mu_{\nu}$, since the piece proportional to $\hbar p_1^\mu \bar{\ell}_\nu/m_1^2$ carries one additional explicit power of $\hbar$ and is therefore subleading. For the remaining part, using the product rule for the spin commutator, one finds
\begin{equation}
    \begin{aligned}
        \frac{1}{2}\left[s_1^\nu,B_h(\bar\ell)B_h(-\bar\ell+\bar q)\right]
        -\left[s_1^\nu,B_h(-\bar\ell+\bar q)\right]B_h(\bar\ell)
        =
        \frac{1}{2}&\left[s_1^\nu,B_h(\bar\ell)\right]B_h(-\bar\ell+\bar q)
        \\&-\frac{1}{2} B_h(\bar\ell)\left[s_1^\nu,B_h(-\bar\ell+\bar q)\right].
    \end{aligned}
\end{equation}
After the classical reduction, the tree kernels commute as ordinary functions, and the expression above is exactly odd under $\bar\ell\to \bar q-\bar\ell$, whereas the cut-box measure is invariant. Therefore $\Delta\mathscr{M}^{\mu(1)}_{\text{extra}}$ does not generate any hyper-classical contribution, to all orders in spin. Its first non-vanishing effect starts at classical order.

Our next and final task of this section is to compute the classical piece from the triangle contribution. We again follow ref.~\cite{Menezes:2022tcs} and employ leading-singularity (LS) techniques to compute the classical contribution from the triangle term. This method is convenient because it can isolate the classical part of the amplitude by focusing on triangle diagrams, following ref.~\cite{Cachazo:2017jef}, see also refs.~\cite{Guevara:2018wpp,Menezes:2022tcs}. For this we use the parametrization~\cite{Guevara:2017csg}
\begin{equation}
    \begin{gathered}
        p_{1}=|\eta]\langle\lambda|+|\lambda]\langle\eta| \qquad\qquad\qquad\qquad  -\frac{t}{m_1^{2}}=\frac{(x_{1}-1)^{2}}{x_{1}} \\ 
        p_{1}^{\prime}=x_{1}|\eta]\langle\lambda|+\frac{1}{x_{1}}|\lambda]\langle\eta|+|\lambda]\langle\lambda| \qquad        4-\frac{t}{m_1^{2}}=\frac{(x_{1}+1)^{2}}{x_{1}} \\
        \langle\lambda\eta\rangle=[\lambda\eta]=m_1
    \end{gathered}
\end{equation}
Similar expressions hold for $p_2, p_2^{\prime}$. In what follows we will identify the complex null vector 
$| \lambda\bigr] \langle \lambda |$ with the momentum mismatch $q$ when $x_1 \to 1$. The formula for one of these Leading Singularity contributions, called $LS^{(a)}_h$
\begin{equation}
    \begin{aligned}
        \textrm{LS}^{(a)}_{h} = \sum_{h_1,h_3 = \pm} 
        \oint&_{\Gamma_a} \frac{d^4 \ell}{(2\pi)^4 (\ell^2 - m_1^2)} \frac{1}{(\ell-p_1^{\prime})^2 (\ell-p_1)^2}
        \\
        &\times \Big\langle M_{S_{1}}(-1^{\prime},\ell_2,-\ell_{1}^{h_1})M_{S_{1}}(1,-\ell_2,-\ell_{3}^{h_3})M_{S_{2}}(2,\ell_{3}^{-h_3},\ell_{1}^{-h_1},-2^{\prime})\Big\rangle
    \end{aligned}
\end{equation}
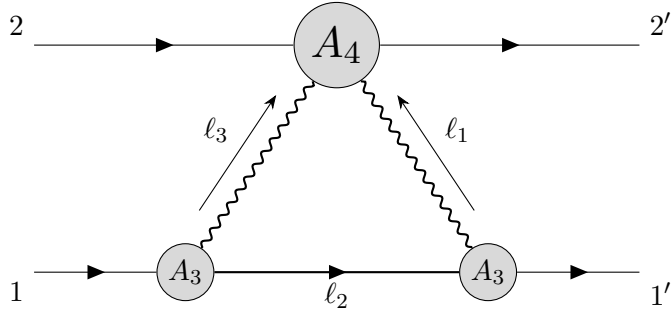
\begin{figure}[H]
    \centering
    \begin{tikzpicture}
        \begin{feynman}[every blob={/tikz/fill=gray!30,/tikz/inner sep=2pt}]
            \vertex[blob,scale=1.5] (a) at (0, 2){$A_4$};
            \vertex (p1) at (-4,2);
            \vertex (p2) at (4,2);
            
            \vertex[blob] (b) at (2,-1){$A_3$};
            \vertex at (4,-1) (p4);
            
            \vertex[blob] (c) at (-2,-1){$A_3$};
            \vertex at (-4,-1) (p3);
            
            \draw[fermion] (a)  -- (p2)node[above right]{$2'$}   ;
            \draw[fermion] (p1) node[above left]{$2$} -- (a)    ;
            \draw[fermion] (b)  -- (p4)node[below right]{$1'$}  ;
            \draw[fermion] (p3) node[below left]{$1$} --  (c)   ;
            
            \draw[photon, thick,momentum=$\ell_3$] (c) -- (a);
            \draw[photon, thick,momentum'=$\ell_1$] (b) -- (a);
            
            \draw[fermion,  thick] (c) --node[below] {$\ell_2$} (b);
        \end{feynman}
    \end{tikzpicture}
\end{figure}
and $LS_{h}^{(b)}$ 
\begin{equation}
    \begin{aligned}
        \textrm{LS}^{(b)}_{h} = \sum_{h_1,h_3 = \pm} 
        \oint&_{\Gamma_b} \frac{d^4 \ell}{(2\pi)^4 (\ell^2 - m_2^2)} \frac{1}{(\ell-p_2^{\prime})^2 (\ell-p_2)^2}
        \\
        &\times 
        \Big\langle M_{S_{2}}(-2^{\prime},\ell_4,\ell_{1}^{h_1})M_{S_{2}}(2,-\ell_4,\ell_{3}^{h_3})M_{S_{1}}(1,-\ell_{3}^{-h_3},-\ell_{1}^{-h_1},-1^{\prime})\Big\rangle
    \end{aligned}
\end{equation}
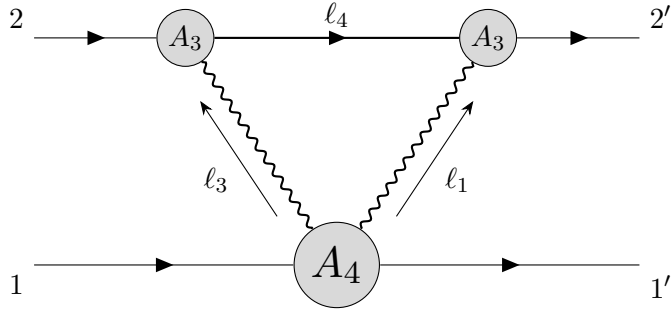
\begin{figure}[H]
    \centering
    \begin{tikzpicture}
        \begin{feynman}[every blob={/tikz/fill=gray!30,/tikz/inner sep=2pt}]
            \vertex[blob,scale=1.5] (a) at (0, -2){$A_4$};
            \vertex (p1) at (-4,-2);
            \vertex (p2) at (4,-2);
            
            \vertex[blob] (b) at (2,1){$A_3$};
            \vertex at (4,1) (p4);
            
            \vertex[blob] (c) at (-2,1){$A_3$};
            \vertex at (-4,1) (p3);
            
            \draw[fermion] (a) -- (p2)node[below right]{$1'$} ;
            \draw[fermion] (p1) node[below left]{$1$} -- (a);
            \draw[fermion] (b) -- (p4)node[above right]{$2'$} ;
            \draw[fermion] (p3) node[above left]{$2$} --  (c);
            
            \draw[photon, thick,momentum=$\ell_3$] (a) -- (c);
            \draw[photon, thick,momentum'=$\ell_1$] (a) -- (b);
            
            \draw[fermion,  thick] (c) --node[above] {$\ell_4$} (b);
        \end{feynman}
    \end{tikzpicture}
\end{figure}
The triangle topology, which is the one we are interested in, is obtained by choosing contours $\Gamma_a, \Gamma_b$ that compute residues at $z = 0$ or $z = \infty$. Focusing on the configuration $h_1 = ++$ and $h_3 = --$, we find that
\begin{equation}
    LS_{h}^{(a+-)}\!=\!\frac{x_{1}}{4m_1^{2}(x_{1}^{2}-1)}\frac{1}{2\pi i} \!\oint_{\Gamma_{a}}\!   \frac{dy}{y}\Big\langle M_{S_{1}}(-1^{\prime},\ell_2,-\ell_{1}^{+})M_{S_{1}}(1,-\ell_2,-\ell_{3}^{-})M_{S_{2}}(2,\ell_{3}^{+},\ell_{1}^{-},-2^{\prime})\Big\rangle
\end{equation}
and
\begin{equation}
    LS_{h}^{(b+-)}\!=\!\frac{x_{2}}{4m_2^{2}(x_{2}^{2}-1)}\frac{1}{2\pi i} \!\oint_{\Gamma_{b}} \!  \frac{dy}{y}\Big\langle M_{S_{2}}(-2^{\prime},\ell_4,\ell_{1}^{+})M_{S_{2}}(2,-\ell_4,\ell_{3}^{-})M_{S_{1}}(1,-\ell_{3}^{+},-\ell_{1}^{-},-1^{\prime})\Big\rangle
\end{equation}
The other opposite-helicity configurations $LS_{h}^{(a-+)}$ and $LS_{h}^{(b-+)}$ are related to the previous ones by a simple change of variables~\cite{Cachazo:2017jef,Guevara:2018wpp}. One can show that the equal helicity configurations, such as $LS_{h}^{(a++)}$ and $LS_{h}^{(b++)}$ have zero residue at both $0, \infty$~\cite{Cachazo:2017jef,Guevara:2018wpp,Menezes:2022tcs}, so they do not contribute to the triangle topology.

The LS integrands involve a product of three tree-level amplitudes inside these brackets of generalized expectation values (GEV) $\langle \cdots \rangle$. The evaluation of the term $\braket{M_1,M_2,M_3}$ requires careful treatment and, since these are spin-$S$ amplitudes, their direct algebraic multiplication is ill-defined. Instead, all open spin indices must be fully contracted. The precise prescription for this contraction is detailed in ref.~\cite{Guevara:2018wpp}, which mandates a proper normalization given by tensor products of the spin-$1$ polarization vectors
\begin{equation}
    m^{2S}\epsilon_{2,\mu_{1}\ldots\mu_{S}}\epsilon_{1}^{\mu_{1}\ldots\mu_{S}}=\lim_{S\rightarrow\infty}\exp{\mp i\frac{q^{\mu}\epsilon_{q}^{\nu-}S_{1\mu\nu}}{p_{1}\cdot\epsilon_{q}^{-}}\langle21\rangle^{2S}}
\end{equation}
for incoming ($-$) and outgoing ($+$) momenta $q$. Here, $\epsilon_1$ and $\epsilon_2$ denote the massive polarization tensors for the two spinning particles, and all $S$ of their indices are fully contracted. Consequently, the GEV serves as a prescription to systematically sew the three amplitudes, functioning equivalently to the double angle brackets employed previously.

Returning to the present calculation, the loop momenta for the leading singularity $LS_{h}^{(a)}$ are parametrized as follows
\begin{equation}
    \begin{aligned}
        \ell_{3 \dot{\alpha} \alpha }(y) =& | \ell_3 \bigr] \langle \ell_3 |
        \\
        =& \frac{1}{x_1 + 1} \left( | \eta \bigr] (x_1^2-1)y  + (1+ x_1 y) | \lambda \bigr] \right) 
        \frac{1}{x_1 + 1} \left( \langle \eta| (x_1^2-1) - \frac{1}{y} (1+ x_1 y)  \langle \lambda | \right)
        \\
        =& \left( -\frac{(x_1-1) (x_1 y+1)}{x_1+1} \right) | \eta\bigr] \langle \lambda | 
        + \left( \frac{(x_1-1) (x_1 y+1)}{x_1+1} \right) | \lambda\bigr] \langle \eta | 
        \\
        &+ \left( -\frac{(x_1 y+1)^2}{(x_1+1)^2 y} \right) | \lambda\bigr] \langle \lambda | 
        + y (x_1-1)^2 | \eta\bigr] \langle \eta | 
        \\
        \ell_{1 \dot{\alpha} \alpha}(y) =& | \ell_1 \bigr] \langle \ell_1 |
        \\
        =& \frac{1}{x_1 + 1} \left( - | \eta \bigr] x_1  (x_1^2-1)y + (1- x_1^2 y)  | \lambda \bigr] \right) 
        \frac{1}{x_1 + 1} \left( \langle \eta| \frac{(x_1^2-1)}{x_1} + \frac{(1-y)}{y}  \langle \lambda | \right) 
        \\
        =& \left( \frac{(x_1-1) x_1 (y-1)}{x_1+1} \right) | \eta\bigr] \langle \lambda | 
        + \left( -\frac{(x_1-1) \left(x_1^2 y-1\right)}{x_1 (x_1+1)}\right) | \lambda\bigr] \langle \eta | 
        \\
        &+ \left( \frac{(y-1) \left(x_1^2 y-1\right)}{(x_1+1)^2 y} \right) | \lambda\bigr] \langle \lambda | 
        - y (x_1-1)^2 | \eta\bigr] \langle \eta | .
    \end{aligned}
\end{equation}
An analogous parametrization applies to $LS_{h}^{(b)}$, differing only by an overall minus sign and the standard $1 \leftrightarrow 2$ exchange.

Substituting the tree-level amplitudes into the leading singularity formulation allows for the direct evaluation of the loop integrands. The relevant $3$-point amplitudes, particularly in their exponentiated form, were established previously. The $4$-point amplitude appearing in the aforementioned expressions corresponds to a spinning Compton amplitude. For the same-helicity configuration, this evaluates to~\cite{Johansson:2019dnu}
\begin{equation}
    M(1^S,2^{++}, 3^{++}, 4^S) = - i \mu^{4-2S} \frac{\kappa^2}{4 } \frac{\bigl[ 23 \bigr]^4 }{(p_2 + p_3)^2} \frac{ \langle {\mathbf 1} {\mathbf 4} \rangle^{2S} } { [(p_1 + p_2)^2 - \mu^2] [(p_1 + p_3)^2 - \mu^2] }.
\end{equation}
However, the opposite-helicity configuration is required for the one-loop computation. As is well known, this amplitude exhibits spurious, \emph{i.e.} unphysical, poles, necessitating a systematic procedure for their removal. In the context of BCFW recursion, these artifacts arise from boundary contributions at infinity. These are typically associated with contact terms, to which standard BCFW recursion relations and physical factorization channels are insensitive. The inclusion of such contact terms introduces ambiguities that must be taken care of throughout the calculation.

One candidate for the Compton amplitude that potentially represents Kerr Black Holes is the one calculated in refs.~\cite{Cangemi:2023bpe,Cangemi:2023ysz,Bohnenblust:2024hkw} which has the classical limit \footnote{There is also an additional contact term which presumably captures non-analytic contributions such as the ones found in the black-hole perturbation theory approach of refs.~\cite{Bautista:2021wfy,Bautista:2022wjf}. Since such terms are not well understood from the perspective of the massive higher-spin theory employed by ref.~\cite{Cangemi:2023bpe}, we choose not to include them in our analysis. However, we emphasize that, if one wants, such contact terms can be easily included in the subsequent computations.} 
\begin{equation}
        M(1^{S},2^{S},3^{--}, 4^{++}) = - i \frac{\kappa^2}{4 }   
        \frac{\bra{3}p_1\sket{4}^4}{{q}^2(p_1\cdot {q}_\perp)^2} \mathcal{M}_S
\end{equation}
where $\mathcal{M}_S$ is the spin-dependent part, namely
\begin{equation}
    \begin{aligned}
        \mathcal{M}_S & =    e^{{\bar{x}}}\cosh{{\bar{z}}} - {\bar{w}} e^{{\bar{x}}}\text{sinch} {\bar{z}} + \frac{{\bar{w}}^2-{\bar{z}}^2}{2}{E} \\
        & +({\bar{w}}^2 - {\bar{z}}^2)({\bar{x}}-{\bar{w}}){\tilde{E}} - \frac{({\bar{w}}^2-{\bar{z}}^2)^2}{2 \bar{\xi} }({\mathcal{E}} + \eta{\tilde{\mathcal{E}} }) 
    \end{aligned}
\end{equation}
where the entire functions $E, \tilde{E}, {\mathcal{E}}, \tilde{\mathcal{E}}$ are defined as
\begin{equation}
    \begin{aligned}
        E(\bar{x}, \bar{y}, \bar{z}) =& \frac{e^{\bar{y}}-e^{\bar{x}} \cosh \bar{z}+(\bar{x}-\bar{y}) e^{\bar{x}} \operatorname{sinhc} \bar{z}}{(\bar{x}-\bar{y})^2-\bar{z}^2}+(\bar{y} \rightarrow-\bar{y})
        \\
        \tilde{E}(\bar{x}, \bar{y}, \bar{z}) =& \frac{2 \bar{x} \cosh \bar{y}+\left(\bar{x}^2+\bar{y}^2-\bar{z}^2\right) \operatorname{sinhc} \bar{y}}{\left((\bar{x}-\bar{y})^2-\bar{z}^2\right)\left((\bar{x}+\bar{y})^2-\bar{z}^2\right)}+\binom{\bar{y} \leftrightarrow \bar{z}}{\bar{x} \rightarrow-\bar{x}} e^{\bar{x}},
    \end{aligned}
\end{equation}
and
\begin{equation}
    \begin{aligned}
        \mathcal{E}(\bar{x}, \bar{y}, \bar{z})= & \partial_{\bar{x}} \tilde{E}
        = \frac{e^{\bar{x}+\bar{z}}}{2 \bar{z}\left((\bar{x}+\bar{z})^2-\bar{y}^2\right)}-\frac{e^{\bar{x}+\bar{z}}(\bar{x}+\bar{z})}{\bar{z}\left((\bar{x}+\bar{z})^2-\bar{y}^2\right)^2} \\
        & +\frac{e^{-\bar{y}}(\bar{x}+\bar{y})}{\bar{y}\left((\bar{x}+\bar{y})^2-\bar{z}^2\right)^2}+\binom{\bar{y} \rightarrow-\bar{y}}{\bar{z} \rightarrow-\bar{z}} \\
        \tilde{\mathcal{E}}(\bar{x}, \bar{y}, \bar{z})= & \partial_{\bar{z}} \tilde{E}
        = \frac{e^{\bar{x}+\bar{z}}(\bar{z}-1)}{2 \bar{z}^2\left((\bar{x}+\bar{z})^2-\bar{y}^2\right)}-\frac{e^{\bar{x}+\bar{z}}(\bar{x}+\bar{z})}{\bar{z}\left((\bar{x}+\bar{z})^2-\bar{y}^2\right)^2} \\
        & -\frac{e^{-\bar{y}} \bar{z}}{\bar{y}\left((\bar{x}+\bar{y})^2-\bar{z}^2\right)^2}-\binom{\bar{y} \rightarrow-\bar{y}}{\bar{z} \rightarrow-\bar{z}} 
    \end{aligned}
\end{equation}
and $\eta = \pm 1$ corresponds to incoming/outgoing modes at the BH horizon. The definitions of the  parameters $\bar{w}, \bar{x}, \bar{y}, \bar{z}, \xi$ showing up in the above expressions are given by~\cite{Cangemi:2023bpe}

\begin{equation}
    \begin{gathered}
        \bar{x}_i = a_i \cdot (\ell_3-\ell_1) \qquad
        \bar{y}_i = a_i \cdot (\ell_3+\ell_1) \qquad
        \bar{z}_i = |a_i|v_i \cdot (\ell_3-\ell_1) \\
        \bar{w}_i = \frac{\bra{\ell_1}a_i\sket{\ell_3}}{\bra{\ell_1}v_i\sket{\ell_3}} v_i \cdot (\ell_3 - \ell_1) \qquad
        \bar{\xi}_i = \frac{\big(v_i \cdot (\ell_3 - \ell_1)\big)^2}{(\ell_1 + \ell_3)^2}
        = \frac{\bar{z}_i^2 - \bar{w}_i^2}{(\bar{w}_i-\bar{x}_i)^2 - \bar{y}_i^2}
         \label{156}
    \end{gathered}
\end{equation}
where $i=1,2$ and for $i=1$, associated with the topology $(b)$, we just need to remember that 
$\ell_{1,3}$ carry an overall minus sign in comparison with topology $(a)$. Observe that, in their approach, $a^{\mu} \sim 1/\hbar$, which is simply a rewording of the statement that $|a| \approx \hbar S/m$ is finite for $\hbar \to 0, S \to \infty$, where $S$ is the spin of the particle. To evaluate these observables in the appropriate classical regime, we use that
\begin{equation}
 \frac{x_1}{\left(x_1^2-1\right)} = \frac{m_1}{\sqrt{-t}} \frac{1}{\sqrt{4-t/m_1^2}}
 \label{4.42}
\end{equation} 
and since $t = \hbar^2 \bar{q}^2$, the integrand must be independent of $\hbar$ in order to produce a classical contribution to the angular impulse. So this indicates we need to employ the $x_1\to1$ limit~\cite{Menezes:2022tcs}, which yields the following asymptotic behaviors for the relevant spinors and momenta
\begin{equation}
    \begin{aligned}
        \lim_{x_1\to1} |\ell_3] &= \frac{1+y}{2}|\lambda]              \\ 
        \lim_{x_1\to1} \ell_3 &= -\frac{(y+1)^2}{4y}q                  
    \end{aligned}
    \qquad
    \begin{aligned}
        \lim_{x_1\to1} \langle\ell_1| &= \frac{1-y}{2y}\langle\lambda|  \\
        \lim_{x_1\to1} \ell_1 &= \frac{(y-1)^2}{4y}q
    \end{aligned}
\end{equation}
With these limits established, we can compute our LS integrals. After some algebra, we find that
\begin{equation}
    \begin{aligned}
\textrm{LS}^{(a+-)}_{h}  \biggl|_{x_1 \to 1} =&
- i \frac{\kappa^4}{2048 \hbar} 
\frac{m_1}{\sqrt{-\bar{q}^2}} 
\frac{1}{( s - m_1^2 - m_2^2 )^2 - 4 m_1^2 m_2^2}
 \frac{1}{2 \pi i} \oint_{\Gamma_a} \frac{dy}{y^3}
\frac{ (f(s, y))^4}
{(1-y^2)^2}
\\
&\times 
\biggl\langle
\frac{1}{m_1^{2S_1}}
\langle {\mathbf 1}^{\prime} |^{2S_1} \exp{- i \frac{\ell_{3}^{\mu} \epsilon_{\ell_3}^{\nu} J_{1\mu\nu}}
{p_1 \cdot \epsilon_{\ell_3}} }  | {\mathbf 1} \rangle^{2S_1}
M_{S_2}
\biggr\rangle
    \end{aligned}
\end{equation}
and
\begin{equation}
    \begin{aligned}
\textrm{LS}^{(b+-)}_{h}  \biggl|_{x_2 \to 1} =&
- i \frac{\kappa^4}{2048 \hbar} 
\frac{m_2}{\sqrt{-\bar{q}^2}} 
\frac{1}{( s - m_1^2 - m_2^2 )^2 - 4 m_1^2 m_2^2}
 \frac{1}{2 \pi i} \oint_{\Gamma_b} \frac{dy}{y^3}
\frac{ (f(s, y))^4}
{(1-y^2)^2}
\\
&\times 
\biggl\langle
\frac{1}{m_2^{2S_2}}
\langle {\mathbf 2}^{\prime} |^{2S_2} \exp{i \frac{\ell_{3}^{\mu} \epsilon_{\ell_3}^{\nu} J_{2\mu\nu}}
{p_2 \cdot \epsilon_{\ell_3}}}  | {\mathbf 2} \rangle^{2S_2}
M_{S_1}\biggr\rangle 
    \end{aligned}
\end{equation}
where 
\begin{equation}
f(s, y) \equiv  2y (s - m_1^2 - m_2^2) 
- {(1+y^2)}  \sqrt{( s - m_1^2 - m_2^2 )^2 - 4 m_1^2 m_2^2}    .
\end{equation}
Now it is clear that our integrands should not depend on $\hbar$, that is, we should integrate over the classical pieces of the integrand. So for instance we can use that~\cite{Guevara:2018wpp}
\begin{equation}
    \begin{aligned}
\Biggl \langle \frac{1}{m_1^{2S_1}}
\langle {\mathbf 1}^{\prime} |^{2S_1} \exp{-i \frac{\ell_{3}^{\mu} \epsilon_{\ell_3}^{\nu -} J_{1\mu\nu}}
{p_1 \cdot \epsilon^{-}_{\ell_3}}}  | {\mathbf 1} \rangle^{2S_1} \biggl|_{x_1 \to 1}
\Biggr \rangle
=& \exp{- i \frac{1+y^2}{2y} \frac{ q^{\mu} \epsilon_{q}^{\nu -} S_{1\mu\nu}}
{p_1 \cdot \epsilon^{-}_{q}}}
\\
=& \exp{\frac{1+y^2}{2y} \bar{q} \cdot a_1}
    \end{aligned}
\end{equation}
where we used that
$$
S^{\mu\nu} = \epsilon^{\mu\nu\alpha\beta} p_{\alpha} \frac{a_{\beta}}{\hbar},
\,\,\,   
a_{\lambda} = \frac{\hbar}{2 m^2} \epsilon_{\lambda\mu\nu\alpha} S^{\mu\nu} p^{\alpha}
$$
together with an evaluation of a Gram determinant for $3$-particle kinematics, which gives us
$$
a_1 \cdot \bar{q} = \pm i \frac{ q^{\mu} \epsilon_{q}^{\nu \pm} S_{1\mu\nu}}
{p_1 \cdot \epsilon^{\pm}_{q}}
\to 
\frac{ q^{\mu} \epsilon_{q}^{\nu -} J_{1\mu\nu}}
{p_1 \cdot \epsilon^{-}_{q}} = - 2 \frac{ q^{\mu} \epsilon_{q}^{\nu -} S_{1\mu\nu}}
{p_1 \cdot \epsilon^{-}_{q}} .
$$
We find a similar expression for the other topology. To complete our calculation, we need the $x_i \to 1$ limit of the spin-dependent part $M_{S_i}$ of the Compton amplitude. This amounts to taking the classical limit of the Compton, as presented above, and express it in terms of our parametrization. So now our task is to rewrite $\bar{w}_i, \bar{x}_i, \bar{y}_i, \bar{z}_i, \xi_i$ using our variables and then take $x_i \to 1$. Starting with the numerator of the spin variable $\bar{w}_2$, we substitute the asymptotic expressions directly into the spinor bracket, resulting in 
\begin{equation}
    \bra{\ell_1} a_2 \sket{\ell_3} \Big|_{x_1\to1} 
    \to \frac{1-y^2}{4y}a_2\cdot \bar{q}
\end{equation}
For the corresponding denominator, the calculation is more involved. By contracting the spinors with the four-velocity $v_2 = p_2/m_2$ and carefully expanding the terms before applying the limit, we find
\begin{equation}
    \begin{aligned}
        \bra{\ell_1} v_2 \sket{\ell_3} \Big|_{x_1\to1} 
        &= \frac{1}{m_2}\bra{\ell_1} p_2 \sket{\ell_3} \Big|_{x_1\to1} \\
        &= \frac{1}{4y   m_2}(x_1-1) f(s, y) .
         \label{140}
    \end{aligned}
\end{equation}
Next, we turn our attention to the velocity contraction $v_2 \cdot \left(\ell_3-\ell_1\right)$, 
$$
v_2 \cdot \left(\ell_3-\ell_1\right) = \frac{p_2}{m_2}\cdot \left(\ell_3-\ell_1\right) .
$$
In order to compute this, we first need to evaluate the momentum difference $\ell_3-\ell_1$. Expanding the definitions and gathering the relevant powers of $(x_1-1)$, we find, in the $x_1 \to 1$ limit:
\begin{equation}
    \begin{aligned}
        p_2 \cdot ( \ell_3-\ell_1 ) \bigg|_{x_1 \to 1}  =& \left(x_1-1\right) \left( - \frac{ \left(x_1 (2 y-1)+1\right)}{x_1+1}  \langle \lambda | p_2 | \eta\bigr] 
        + \frac{\left(x_1( 2 y x_1 + 1)-1\right)}{x_1 \left(x_1+1\right)}  \langle \eta | p_2 | \lambda\bigr] \right)
        \\
        &+ \frac{x_1 y \left(x_1 (1-2 y)-2\right)+y-2}{\left(x_1+1\right){}^2 y}  
        \langle \lambda | p_2 | \lambda\bigr] 
        + {\mathcal O}(\left(x_1-1\right){}^2 )
        \\
        =& \left(x_1-1\right) \left( - \frac{ \left(x_1 (2 y-1)+1\right)}{x_1+1}  \langle \lambda | p_2 | \eta\bigr] 
        + \frac{\left(x_1( 2 y x_1 + 1)-1\right)}{x_1 \left(x_1+1\right)}  \langle \eta | p_2 | \lambda\bigr] 
        \right. \\
        &+ \left. \frac{x_1 y \left(x_1 (1-2 y)-2\right)+y-2}{\left(x_1+1\right){}^2 y}  
        \Big( \langle \eta | p_2 |  \lambda \bigr] 
        - \langle \lambda | p_2 | \eta \bigr]   \Big) \right)
        \!+ {\mathcal O}(\left(x_1-1\right){}^2 )
    \end{aligned}
\end{equation}
where we used that
\begin{equation}
    2 (p_1^{\prime} - p_1) \cdot p_2 \biggl|_{x_1 \to 1} = 2 q \cdot p_2 \biggl|_{x_1 \to 1}
    = - (x_1-1) \Big( \langle \eta | p_2 |  \lambda \bigr] 
    - \langle \lambda | p_2 | \eta \bigr]   \Big) + \langle \lambda | p_2 |  \lambda \bigr] 
    = 0
\end{equation} 
after using the on-shell conditions $q \cdot p_2 \to 0$. Having computed both the numerator and the denominator, we can now assemble the full expression for the variable $\bar{w}_2$
\begin{equation}
    \begin{aligned}
        \bar{w}_2 \bigg|_{x_1 \to 1} &= 
        \frac{\bra{\ell_1}a_2\sket{\ell_3}}{\bra{\ell_1}v_2\sket{\ell_3}} v_2\cdot(\ell_3-\ell_1) 
        \bigg|_{x_1 \to 1}
        \\
        &=   \frac{h(s,y)}{ f(s,y)}     a_2\cdot \bar{q}
    \end{aligned}
\end{equation}
where
\begin{equation}
h(s,y) \equiv
\frac{(1-y^2)^2}{2y}  \sqrt{\big(s - m_1^2 - m_2^2\big)^2 - 4 m_1^2 m_2^2} .
\end{equation}
For the remaining variables $\bar{x}_2$, $\bar{y}_2$, $\bar{z}_2$, $\bar{\xi}_2$, we can use the above results to show that
\begin{equation}
    \begin{aligned}
        \bar{x}_2 \bigg|_{x_1 \to 1}  =& a_2 \cdot (\ell_3-\ell_1)\bigg| _{x_1 \to 1}
        \to\quad
        - \frac{1+y^2}{2 y}  a_2 \cdot \bar{q}
        \\
        \bar{y}_2 \bigg|_{x_1 \to 1}  =& a_2 \cdot (\ell_3+\ell_1)\bigg| _{x_1 \to 1}
        \to\quad
        - a_2 \cdot \bar{q}
        \\ 
        \bar{z}_2 \bigg|_{x_1 \to 1}  =& |a_2| v_2 \cdot (\ell_3-\ell_1)\bigg| _{x_1 \to 1}
        =  \frac{h(s,y)}{m_2 (1-y^2)} |a_2| \left(x_1-1\right) \bigg| _{x_1 \to 1}
        \to\quad \frac{h(s,y)}{m_2 (1-y^2)} |a_2| 
        \\
        \bar{\xi}_2 \bigg|_{x_1 \to 1}  =& 
        \frac{\bar{z}_2^2 - \bar{w}_2^2}{(\bar{w}_2-\bar{x}_2)^2 - \bar{y}_2^2} \bigg| _{x_1 \to 1}
        \to\quad\frac{ \left( \frac{h(s,y)}{m_2 (1-y^2)} \right)^2 |a_2|^2
        - \left( \frac{h(s,y)}{ f(s,y)} \right)^2  ( a_2\cdot \bar{q} )^2}
        {\left[ \left( \frac{1+y^2}{2 y} - \frac{h(s,y)}{ f(s,y)} \right)^2 - 1 \right] ( a_2 \cdot \bar{q} )^2}
    \end{aligned}
\end{equation}
where we used that
\begin{equation}
( \ell_3+\ell_1 ) \bigg|_{x_1 \to 1}  = (1-x_1) | \eta\bigr] \langle \lambda | 
+ \frac{(x_1-1)}{x_1} | \lambda\bigr] \langle \eta | 
- | \lambda\bigr] \langle \lambda | .
\end{equation}
We also used the fact that, due to the result~(\ref{4.42}), we take the interpretation that all quantities that scale as $1/\hbar$, such as $a_2^{\mu}$, should also scale as $1/(x_1-1)$ in our parametrization for the integrand, so we did the rescaling $|a_2| \to |a_2|/(x_1-1)$. Similar considerations can be done for the other topology. We hence find that
\begin{equation}
    \begin{aligned}
\textrm{LS}^{(a+-)}_{h}  \biggl|_{x_1 \to 1} =&
- i \frac{\kappa^4}{2048 \hbar} 
\frac{m_1}{\sqrt{-\bar{q}^2}} 
\frac{1}{( s - m_1^2 - m_2^2 )^2 - 4 m_1^2 m_2^2}
\\
&\times 
 \frac{1}{2 \pi i} \oint_{\Gamma_a} \frac{dy}{y^3}
\frac{ (f(s, y))^4}{(1-y^2)^2}
 \exp{ \frac{1+y^2}{2y} \bar{q} \cdot a_1 } M_{S_2}(y)
    \end{aligned}
\end{equation}
and
\begin{equation}
    \begin{aligned}
\textrm{LS}^{(b+-)}_{h}  \biggl|_{x_2 \to 1} =&
- i \frac{\kappa^4}{2048 \hbar} 
\frac{m_2}{\sqrt{-\bar{q}^2}} 
\frac{1}{( s - m_1^2 - m_2^2 )^2 - 4 m_1^2 m_2^2}
\\
&\times 
 \frac{1}{2 \pi i} \oint_{\Gamma_b} \frac{dy}{y^3}
\frac{ (f(s, y))^4}{(1-y^2)^2}
 \exp{ \frac{1+y^2}{2y} \bar{q} \cdot a_2 } M_{S_1}(y) .
    \end{aligned}
\end{equation}
Now we need to discuss the contours $\Gamma_{a,b}$. The triangle topology is associated with choosing a contour around zero or infinity, and such contours are associated with different opposite-helicity configurations in the computation of LS. To unify both descriptions, one resorts to the following change of variables\footnote{This map is known in the literature as the Joukowsky transform. It maps the unit circle in $y$ to the cut $(-1,1)$ in z. Inverting the relation, we find that $y=z\pm\sqrt{z^2-1}$. The two signs correspond to the two sheets across the branch cut. Henceforth, we choose the lower sign.}
$$
z = \frac{1+y^2}{2y} .
$$
In this way, both contours are mapped to $z = \infty$. Hence the full triangle leading singularity in the limit $x_1 \to 1$ reads
\begin{equation}
    \begin{aligned}
\textrm{LS}^{(a)}_{\triangle,h} \biggl|_{x_1 \to 1} =& - i \frac{\kappa^4}{256 \hbar}
\frac{m_1}{ \sqrt{-\bar{q}^2}}  
\frac{1}{ ( s - m_1^2 - m_2^2 )^2 - 4 m_1^2 m_2^2}
\\
&\times\frac{1}{2 \pi i} \oint_{\Gamma_{\infty}} dz
\frac{ ( g(s, z))^4}{(z^2-1)^{3/2}}
e^{z   \bar{q} \cdot a_1 }
M_{S_2}(z)
    \end{aligned}
\end{equation}
and
\begin{equation}
    \begin{aligned}
\textrm{LS}^{(b)}_{\triangle,h} \biggl|_{x_2 \to 1} =& - i \frac{\kappa^4}{256 \hbar}
\frac{m_2}{ \sqrt{-\bar{q}^2}}  
\frac{1}{ ( s - m_1^2 - m_2^2 )^2 - 4 m_1^2 m_2^2}
\\
&\times \frac{1}{2 \pi i} \oint_{\Gamma_{\infty}} dz
\frac{ (g(s, z))^4}{(z^2-1)^{3/2}}
e^{z   \bar{q} \cdot a_2 }
M_{S_1}(z) 
    \end{aligned}
\end{equation}
where
\begin{equation}
 g(s, z)\equiv (s - m_1^2 - m_2^2)
-  z \sqrt{( s - m_1^2 - m_2^2 )^2 - 4 m_1^2 m_2^2}    .
\end{equation}
Our parameters now have the following form\footnote{We emphasize that these expressions are related to topology (a). When going to topology (b), we must take into account an overall change of sign associated with the choice of the momenta $\ell_{1,3}$, see the discussion after their parametrization.}
\begin{equation}
    \begin{aligned}
\bar{w}_2(z) \bigg|_{x_1 \to 1}  =&   \frac{j(s,z)}{ g(s,z)}     a_2\cdot \bar{q},
\quad
j(s,z) \equiv \left(z^2-1\right)  \sqrt{\big(s - m_1^2 - m_2^2\big)^2 - 4 m_1^2 m_2^2} 
\\
 \bar{x}_2(z) \bigg|_{x_1 \to 1}  =& - z    a_2 \cdot \bar{q}
 \\
 \bar{y}_2(z) \bigg|_{x_1 \to 1}  =&  - a_2 \cdot \bar{q}
\\ 
\bar{z}_2(z) \bigg|_{x_1 \to 1}  =&  \frac{\kappa(s,z)}{m_2} |a_2| 
\quad
\kappa(s,z) \equiv \sqrt{z^2-1}  \sqrt{\big(s - m_1^2 - m_2^2\big)^2 - 4 m_1^2 m_2^2}
\\
\bar{\xi}_2(z) \bigg|_{x_1 \to 1}  =& \frac{ \left( \frac{\kappa(s,z)}{m_2} \right)^2 |a_2|^2
- \left( \frac{j(s,z)}{ g(s,z)} \right)^2  ( a_2\cdot \bar{q} )^2}
{\left[ \left( z - \frac{j(s,z)}{ g(s,z)} \right)^2 - 1 \right] ( a_2 \cdot \bar{q} )^2}
    \end{aligned}
\end{equation}
and similar to the other topology. Notice that, expressed in the $z$ variable, we have a branch-cut singularity at $z \in (-1,1)$. On the other hand, since the entire functions $E, \tilde{E}, {\mathcal{E}}, \tilde{\mathcal{E}}$ used to build the spinning part of the Compton amplitude are analytic everywhere on $\mathbb{C}^3$, we do not expect the LS integral to develop any essential singularity at finite $z$. This is to be contrasted with the problematic spinning Compton amplitude with unphysical poles at $p \cdot \varepsilon$, which generates an essential singularity at finite $z$, which in turn generates an additional pole when one calculates the scattering angle~\cite{Guevara:2018wpp}. 

However, the specific correct choice of contour in the spinning case is not so straightforward. The presence of spin fundamentally alters the analytic structure of the leading singularity integrands at infinity. For massive scalar particles, the integrand is a purely rational function, and consequently $z = +\infty$ corresponds to a simple pole with a well-defined residue. This allows the application of the standard residue theorem, including the contribution from infinity. However, for spinning particles, the integrand acquires exponential dependence on the loop variable through spin-dependent phases of the form $e^{z \bar{q} \cdot a_1}$ and similar terms within $M_{S_2}(z)$. These exponentials encode physical effects such as spin precession, Thomas precession in curved spacetime, and helicity structure. As a result, $z = +\infty$ becomes an \textit{essential singularity} rather than a pole, with no well-defined residue\footnote{This is not a feature of the above Joukowsky transform; indeed, one can easily show that the integral in the $y$ variable also has essential singularities at $y=0,+\infty$.}.

An interesting observation emerges when considering perturbative expansions around the spinless limit. If one expands the integrand in a Taylor series around $a_1 = a_2 = 0$, then at each finite order $n$, the integrand reduces to a polynomial in $z$ (truncating the exponential series), and $z = +\infty$ behaves as a pole of order $(n+1)$. However, the exact result, obtained by resumming all orders, recovers the exponential functions and thus the essential singularity. This demonstrates that spin effects are fundamentally non-perturbative in their impact on the analytic structure: the transition from a pole to an essential singularity occurs at infinite order in the spin expansion. The exponential factors are essential (pun intended!) for creating the essential singularity.

When extracting specific classical observables such as the scattering angle from the leading singularity, additional simplifications may occur. Through the classical limit ($\hbar \to 0$), stationary phase approximations, and projections onto specific spin configurations, the exponential phases can combine or cancel, potentially yielding a rational function. In such cases, the observable's integrand might have $z = +\infty$ as a genuine pole, allowing simpler residue-based contour prescriptions. This was exactly the case in ref.~\cite{Guevara:2018wpp}, in which their choice of contour was justified only after displaying the (unintegrated) expression for the scattering angle\footnote{Of course, in ref.~\cite{Guevara:2018wpp}, the situation is a bit simpler since the LS was calculated using the Compton with unphysical poles, so the result was valid only up to a certain order in a power-series expansion in $a_{1,2}$. So the correct procedure was indeed to envisage their result as already expanded around the spinless case and, as a result, it justifies calling $z = + \infty$ a pole instead of an essential singularity}. This highlights that the appropriate contour integration method depends not only on the leading singularity itself but also on the specific observable being computed.

Our calculation of the leading singularity for Kerr BH scattering to all orders in spin constitutes one of the main results of this paper. From it, the total classical one-loop triangle amplitude can be constructed by superposing the contributions from both kinematic channels, as explained in ref.~\cite{Cachazo:2017jef}. By summing the Leading Singularities of the $a$ and $b$ topologies, the amplitude is formally defined as
\begin{equation}
    M_{\triangle,\text{class}}^{(1)} = \frac{1}{4} \left( LS_{\triangle, h}^{(a)} 
    + LS_{\triangle,h}^{(b)} \right) .
\end{equation}
In addition to the angular impulse to be discussed in the next section, it is instructive to extract the corresponding non-relativistic gravitational potential from our amplitude. This provides an independent consistency check of our results and allows for a direct comparison with known post-Newtonian and effective field theory computations. For completeness, we present this derivation in Appendix~\ref{app:NR_LS}, where we obtain a representation of the potential valid to all orders in the spin variables at the integrand level, and verify that its expansion to linear order in spin reproduces the known spin-orbit interaction.

\section{Kerr black hole angular impulse}
\label{classical}
Having established the necessary mathematical framework and the relevant kinematic variables, we are now positioned to perform the explicit calculation of the angular impulse. This quantity is central to understanding the gravitational scattering of spinning black holes, as it encodes the total change in the internal angular momentum (spin) of the body during the interaction.

\subsection{LO angular impulse}
We begin our analysis with the computation of the LO angular impulse. To maintain consistency with recent developments in the scattering amplitudes program, we follow the methodology established in~\cite{Maybee:2019jus}. In this framework, the change in the spin vector is
\begin{equation}
    \begin{aligned}
        \langle \Delta s_1^{\mu,\text{LO}} \rangle =& i \kappa^2
        \llangle[\bigg]\hbar  \int \hat{d}^4\bar{q}   \hat{\delta}(2\bar{q}\cdot p_1 )
        \hat{\delta}(2\bar{q}\cdot p_2 )e^{-ib\cdot\bar{q}}\\
        &\times\left(\left[{s^\mu_1}{{\mathcal M}^{(0)}(\bar{q})}\right] - \frac{\hbar}{m_1^2}p_1^\mu\bar{q}_\nu s^\nu_1 {\mathcal M}^{(0)}(\bar{q})\right)\rrangle[\bigg]
        \\
        =&  -  \frac{i \kappa^2 m_1 m_2}{8}
        \int \hat{d}^4\bar{q}   \hat{\delta}(\bar{q}\cdot u_1 )
        \hat{\delta}(\bar{q}\cdot u_2 ) \frac{ e^{-ib\cdot\bar{q}}}{\bar{q}^2} 
        P_{\gamma\delta\alpha\beta}   u_{1}^{\gamma} u_{1}^{\delta} \\
        &\times  
        \left( i \epsilon^{\mu\nu\rho\sigma} u_{1\nu}s_{1\rho}\frac{\partial}{\partial s^\sigma_1}
         \exp{- i \left(\frac{s_1}{m_1}+\frac{s_2}{m_2}\right)* \bar{q} }^{(\alpha}_{\ \lambda}  \right. \\
        & - \left. \frac{1}{m_1} u_1^\mu\bar{q}_\nu s^\nu_1 
        \exp{- i \left(\frac{s_1}{m_1}+\frac{s_2}{m_2}\right)* \bar{q} }^{(\alpha}_{\ \lambda} \right) 
        u_{2}^{\beta)} u_{2}^{\lambda}
        \label{eq:LO_integral}         
    \end{aligned}
\end{equation}
where we have set $p_i \to m_i u_i $ and $a^{\mu}$ is now the classical ring radius. This is exactly what the double-angle brackets instruct us to do in the large quantum spin limit $S \to \infty$.

We now proceed to explicitly evaluate the integral to match the form presented in refs.~\cite{Maybee:2019jus,Vines:2017hyw}. Assuming, for simplicity, that black hole 2 2 has zero spin, we can rewrite the previous expression as
\begin{equation}
    \begin{aligned}
        \langle \Delta s_1^{\mu,\text{LO}} \rangle \!\!=& 
        m_1 m_2 \frac{\kappa^2}{16}  \sum_{k=0}^{2S_1} 
        \frac{1}{(2 S_1)^{k}} \binom{2S_1}{k}\!
         \biggl[ i k   \epsilon^{\mu\nu\rho\sigma} u_{1 \sigma} 
        \Big(  (2 \gamma^2 - 1) \bigl(1 + (-1)^{k}\bigr) I_{\nu}^{\ \mu_1 \ldots \mu_{k-1}} (b,u_1,u_2) 
        \\
        &- 2 i \gamma \epsilon_{\nu\kappa\alpha\beta} u_1^{\alpha} u_2^{\beta} 
        \bigl(1 - (-1)^{k}\bigr) I^{\kappa \mu_1 \ldots \mu_{k-1}}(b,u_1,u_2)  \Big) 
        a_{1 \mu_1} \cdots a_{1 \mu_{k-1}} a_{1 \rho} 
        \\
        &+ u_{1}^{\mu} 
        \Big( (2 \gamma^2 - 1) \bigl(1 + (-1)^{k}\bigr) I_{\nu}^{\ \mu_1 \ldots \mu_k}(b,u_1,u_2) 
        \\
        &- 2 i \gamma \epsilon_{\nu\kappa\alpha\beta} u_1^{\alpha} u_2^{\beta} 
        \bigl(1 - (-1)^{k}\bigr) I^{\kappa \mu_1 \ldots \mu_k}(b,u_1,u_2) \Big) 
        a^{\nu}_{1} a_{1 \mu_1} \cdots a_{1 \mu_k}  \biggr]
    \end{aligned}
\end{equation}
where we have employed elementary trigonometric identities, the binomial theorem, the standard definition of rapidity, and the relation~\cite{Arkani-Hamed:2019ymq}
$$
\bar{q}_{\nu} \sinh\phi = - i \epsilon_{\nu\kappa\alpha\beta} \bar{q}^{\kappa} u_1^{\alpha} u_2^{\beta}
$$
which holds on the support of the delta functions by virtue of a Schouten identity. We also use the integral definition from ref.~\cite{Maybee:2019jus}:
\begin{equation}
    I^{\mu_1 \ldots \mu_n}(b,u_1,u_2) = i \int  \frac{d^4 \bar{q}}{(2 \pi)^4} (2\pi)  \delta(\bar{q} \cdot u_1)
    (2\pi) \delta( \bar{q} \cdot u_2) \frac{e^{- i b \cdot \bar{q}}}{\bar{q}^2}
    \bar{q}^{\mu_1} \cdots \bar{q}^{\mu_n} .
\end{equation}
One obtains
\begin{equation}
    \begin{aligned}
        \langle \Delta s_1^{\mu,\text{LO}} \rangle =& m_1 m_2 \frac{\kappa^2}{16} \sum_{k=0}^{\infty} 
        \frac{1}{\Gamma(k+1)}
        \biggl[ i k   \epsilon^{\mu\nu\rho\sigma} u_{1 \sigma} 
        \Big( (2 \gamma^2 - 1) \bigl(1 + (-1)^{k}\bigr) I_{\nu}^{\ \mu_1 \ldots \mu_{k-1}}(b,u_1,u_2) 
        \\
        &- 2 i \gamma \epsilon_{\nu\kappa\alpha\beta} u_1^{\alpha} u_2^{\beta} 
        \bigl(1 - (-1)^{k}\bigr) I^{\kappa \mu_1 \ldots \mu_{k-1}}(b,u_1,u_2)  \Big) 
        a_{1 \mu_1}  \cdots  a_{1 \mu_{k-1}}  a_{1 \rho} 
        \\
        &+ u_{1}^{\mu} 
        \Big( (2 \gamma^2 - 1) \bigl(1 + (-1)^{k}\bigr) I_{\nu}^{\ \mu_1 \ldots \mu_k}(b,u_1,u_2) 
        \\
        &- 2 i \gamma \epsilon_{\nu\kappa\alpha\beta} u_1^{\alpha} u_2^{\beta} 
        \bigl(1 - (-1)^{k}\bigr) I^{\kappa \mu_1 \ldots \mu_k}(b,u_1,u_2) \Big) 
        a^{\nu}_{1}  a_{1 \mu_1} \cdots  a_{1 \mu_k}  \biggr]
        \\
        =& m_1 m_2 \frac{\kappa^2}{16} 
        \biggl[ u_{1}^{\mu} a^{\nu}_{1} 
        \Big( (2 \gamma^2 - 1) \eta_{\nu\kappa} \bigl( I^{\kappa}(b+i a_1,u_1,u_2) 
        + I^{\kappa}(b-i  a_1,u_1,u_2) \bigr) 
        \\
        &- 2 i \gamma \epsilon_{\nu\kappa\alpha\beta} u_1^{\alpha} u_2^{\beta} 
        \bigl( I^{\kappa}(b+i  a_1,u_1,u_2)  - I^{\kappa}(b-i a_1,u_1,u_2)  \bigr)  \Big) 
        \\
        &+ i   \epsilon^{\mu\nu\rho\sigma} u_{1 \sigma} a_{1 \rho} 
        \Big( (2 \gamma^2 - 1) \eta_{\nu\kappa}
        \bigl( I^{\kappa}(b+i  a_1,u_1,u_2) + I^{\kappa}(b-i a_1,u_1,u_2) \bigr) 
        \\
        &- 2 i \gamma \epsilon_{\nu\kappa\alpha\beta} u_1^{\alpha} u_2^{\beta} 
        \bigl( I^{\kappa}(b+i a_1,u_1,u_2)  - I^{\kappa}(b-i a_1,u_1,u_2)  \bigr) \Big)  
        \biggr]
        \\
        =& m_1 m_2 \frac{\kappa^2}{8} 
        \biggl\{ u_{1}^{\mu} a^{\nu}_{1}
        \textrm{Re}
        \Big[ \Big( (2 \gamma^2 - 1) \eta_{\nu\kappa} - 2 i \gamma \epsilon_{\nu\kappa\alpha\beta} u_1^{\alpha} u_2^{\beta} \Big) I^{\kappa}(b+i a_1,u_1,u_2)  \Big] 
        \\
        &+ \epsilon^{\mu\nu\rho\sigma} u_{1 \rho} a_{1 \sigma} 
        \textrm{Im}\Big[ \Big( (2 \gamma^2 - 1) \eta_{\nu\kappa}- 2 i \gamma \epsilon_{\nu\kappa\alpha\beta} u_1^{\alpha} u_2^{\beta} \Big) I^{\kappa}(b+i  a_1 ,u_1,u_2)  \Big]  
        \biggr\} .
    \end{aligned}
\end{equation}
It is evident that the terms $k=0,1,2$ in the series, up to order $\mathcal{O}(a^2)$, recover the results derived in ref.~\cite{Maybee:2019jus}. Indeed, this procedure completely reproduces the LO angular impulse obtained in ref.~\cite{Vines:2017hyw} for $a_2=0$, a direct consequence of the Newman-Janis shift. A straightforward calculation allows one to reproduce the entire result even for $a_2 \neq 0$. Physically, the real part describes the longitudinal shift in the spin vector along $u_1^\mu$, maintaining the covariant constraint $s_1 \cdot u_1 = 0$ after the impulse. On the other hand, the imaginary part accounts for the transverse precession of the spin vector.

\subsection{NLO angular impulse}

Having established the classical spin dynamics at leading order, we must now shift our focus toward the NLO corrections to the angular impulse, which require the one-loop scattering amplitude. Isolating the classical limit and assembling the components of that amplitude, we find that
\begin{equation}
    \langle \Delta s^{\mu,\text{NLO}}_{1} \rangle =
    \frac{i \kappa^4 m_1^2 m_2^2}{4} 
    \int  \hat{d}^4 \bar{q}   \hat{\delta}(\bar{q} \cdot u_1)
    \hat{\delta}(\bar{q} \cdot u_2) e^{- i b \cdot \bar{q}} 
    \Big( \mathscr{M}^{\mu(1)} + \Delta\mathscr{M}^{\mu(1)} \Big) 
    \label{eq:kmoc_nlo}
\end{equation}
where
\begin{equation}
    \begin{aligned}
        \mathscr{M}^{\mu(1)} =&  i \epsilon^{\mu\nu\rho\sigma} u_{1\nu}s_{1\rho}
        \frac{\partial}{\partial s^\sigma_1} \big( {\mathcal M}_b(\bar{q}) + {\mathcal M}_t(\bar{q}) \big)
        \\
        &- \frac{1}{m_1} u_{1}^{\mu} \bar{q}_{\nu} s^{\nu}_{1} 
        \big( {\mathcal M}_b(\bar{q}) + {\mathcal M}_t(\bar{q}) \big)
        \\
        &+ i \epsilon^{\mu\nu\rho\sigma} u_{1\nu}s_{1\rho}
        \frac{\partial}{\partial s^\sigma_1} {\mathcal M}_c(\bar{q})
        \\
        &+ \frac{i}{2 m_1} u_{1}^{\mu}  
        \epsilon^{\nu\tau\rho\sigma} u_{1\tau}s_{1\rho}\frac{\partial}{\partial s^\sigma_1} N_{\nu}(\bar{q})
        \\
        &- \frac{1}{m_1} u_{1}^{\mu}  s^{\nu}_{1} {\mathcal M}_{\nu}(\bar{q}) 
    \end{aligned}
\end{equation}
with
\begin{equation}
    \begin{aligned}
        {\mathcal M}_b(\bar{q}) \equiv&
        \frac{1}{2 m_1} 
        \int {\hat{d}^4 \bar{\ell}}  
        \frac{ \bar{\ell} \cdot (\bar{\ell} - \bar{q}) \hat{\delta}(\bar{\ell} \cdot u_2) }{\bar{\ell}^2 (\bar{\ell} - \bar{q})^2 
        ( \bar{\ell} \cdot u_1  + i\varepsilon )^2 } 
        B_{h}(\bar{\ell}, u_1,u_2) B_{h}(-\bar{\ell}+\bar{q}, u_1,u_2)
        \\
        &+
        \frac{1}{ 2 m_2} 
        \int {\hat{d}^4 \bar{\ell}}  
        \frac{ \bar{\ell} \cdot (\bar{\ell} - \bar{q}) \hat{\delta}(\bar{\ell} \cdot u_1) }
        {\bar{\ell}^2 (\bar{\ell} - \bar{q})^2 ( \bar{\ell} \cdot u_2  - i\varepsilon )^2 } 
        B_{h}(\bar{\ell}, u_1,u_2) B_{h}(-\bar{\ell}+\bar{q}, u_1,u_2)
        \\
        &+ \frac{i}{4 m_1 m_2} 
        \int {\hat{d}^4 \bar{\ell}}  
        ( 2 \bar{\ell} \cdot \bar{q} - \bar{\ell}^2 ) \frac{ B_{h}(\bar{\ell}, u_1,u_2) B_{h}(-\bar{\ell}+\bar{q}, u_1,u_2) }
        {\bar{\ell}^2 (\bar{\ell} - \bar{q})^2}
        \\
        &\times \Big( m_2 \hat{\delta}^{\prime}( \bar{\ell} \cdot u_1) 
        \hat{\delta}( \bar{\ell} \cdot u_2 ) 
        - m_1 \hat{\delta}( \bar{\ell} \cdot u_1) 
        \hat{\delta}^{\prime}( \bar{\ell} \cdot u_2 ) \Big) .
    \end{aligned}
\end{equation}
\begin{equation}
    \begin{aligned}
        {\mathcal M}_t(\bar{q})  \equiv& \frac{1}{256 m_2}
        \frac{1}{ \gamma^2 - 1}
        \frac{1}{ \sqrt{-\bar{q}^2}}  
        \oint_{\Gamma_{\infty}} \frac{dz}{2 \pi i}
        \frac{ (  \gamma -  z \sqrt{ \gamma^2 - 1} )^4}{(z^2-1)^{3/2}}
        e^{z   \bar{q} \cdot a_1 }
        M_{S_2}(z)
        \\
        &+ \frac{1}{256 m_1}
        \frac{1}{ \gamma^2 - 1}
        \frac{1}{ \sqrt{-\bar{q}^2}}  
        \oint_{\Gamma_{\infty}} \frac{dz}{2 \pi i}
        \frac{ (  \gamma -  z \sqrt{ \gamma^2 - 1} )^4}{(z^2-1)^{3/2}}
        e^{z   \bar{q} \cdot a_2 }
        M_{S_1}(z) 
    \end{aligned}
\end{equation}
\begin{equation}
    \begin{aligned}
        {\mathcal M}_c(\bar{q}) =& \frac{i}{8 m_2} 
        \int {\hat{d} ^4 \bar{\ell}}  \hat{\delta}( \bar{\ell} \cdot u_1) 
        \hat{\delta}( \bar{\ell} \cdot u_2) 
        \frac{ B_{h}(\bar{\ell}, u_1,u_2) }
        {\bar{\ell}^2 (\bar{\ell} - \bar{q})^2} 
        \\
        &\times \Big[ P_{\gamma\delta\alpha\beta}   u_{1}^{\gamma} u_{1}^{\delta} 
        \exp{\bigl(- i (a_1+a_2)* ( -\bar{\ell} + \bar{q} ) \bigr)}^{(\alpha}_{\ \lambda} \bar{q}^{\beta)} u_{2}^{\lambda}
        \\
        &+ P_{\gamma\delta\alpha\beta}   u_{1}^{\gamma} u_{1}^{\delta} 
        \exp{\bigl(- i (a_1+a_2)* ( -\bar{\ell} + \bar{q} ) \bigr)}^{(\alpha}_{\ \lambda} u_{2}^{\beta)} \bar{q}^{\lambda}
        \Big]
        \\
        &- \frac{i}{8 m_1} 
        \int {\hat{d} ^4 \bar{\ell}}  \hat{\delta}( \bar{\ell} \cdot u_1) 
        \hat{\delta}( \bar{\ell} \cdot u_2) 
        \frac{ B_{h}(\bar{\ell}, u_1,u_2) }
        {\bar{\ell}^2 (\bar{\ell} - \bar{q})^2}  
        \\
        &\times \Big[ P_{\gamma\delta\alpha\beta}   \bar{q}^{\gamma} u_{1}^{\delta} 
        \exp{\bigl(- i (a_1+a_2)* ( -\bar{\ell} + \bar{q} ) \bigr)}^{(\alpha}_{\ \lambda} u_{2}^{\beta)} 
        u_{2}^{\lambda}
        \\
        &+ P_{\gamma\delta\alpha\beta}   u_{1}^{\gamma} \bar{q}^{\delta} 
        \exp{\bigl(- i (a_1+a_2)* ( -\bar{\ell} + \bar{q} ) \bigr)}^{(\alpha}_{\ \lambda} u_{2}^{\beta)} 
        u_{2}^{\lambda}
        \Big]
        \\
        &- \frac{i}{4 m_1 m_2}
        \int {\hat{d} ^4 \bar{\ell}}  
        \bar{\ell}^2 \frac{1}{\bar{\ell}^2 (\bar{\ell} - \bar{q})^2}
        B_{h}(\bar{\ell}, u_1,u_2) B_{h}(-\bar{\ell}+\bar{q}, u_1,u_2)
        \\
        &\times\Big( m_2 \hat{\delta}^{\prime}( \bar{\ell} \cdot u_1) 
        \hat{\delta}( \bar{\ell} \cdot u_2 ) 
        - m_1 \hat{\delta}( \bar{\ell} \cdot u_1) 
        \hat{\delta}^{\prime}( \bar{\ell} \cdot u_2 ) \Big)
    \end{aligned}
\end{equation}
\begin{equation}
    N_{\nu}(\bar{q}) = - \frac{i}{2} 
    \int {\hat{d} ^4 \bar{\ell}}  \frac{\bar{q}_{\nu}}{\bar{\ell}^2 (\bar{\ell} - \bar{q})^2}
    \hat{\delta}( \bar{\ell} \cdot u_1 ) 
    \hat{\delta}( \bar{\ell} \cdot u_2 ) 
    B_{h}(\bar{\ell}, u_1,u_2) B_{h}(-\bar{\ell}+\bar{q}, u_1,u_2) 
\end{equation}
\begin{equation}
    \begin{aligned}
        {\mathcal M}_{\nu}(\bar{q}) =& 
        - \frac{i}{2 m_1 m_2} 
        \int {\hat{d} ^4 \bar{\ell}}  
        \bar{\ell} \cdot (\bar{\ell} - \bar{q}) \frac{\bar{\ell}_{\nu}}{\bar{\ell}^2 (\bar{\ell} - \bar{q})^2}
        B_{h}(\bar{\ell}, u_1,u_2) B_{h}(-\bar{\ell}+\bar{q}, u_1,u_2)
        \\
        &\times \Big( m_2 \hat{\delta}^{\prime}( \bar{\ell} \cdot u_1) 
        \hat{\delta}( \bar{\ell} \cdot u_2 ) 
        - m_1 \hat{\delta}( \bar{\ell} \cdot u_1) 
        \hat{\delta}^{\prime}( \bar{\ell} \cdot u_2 ) \Big)
        \\
        &- \frac{i}{4 m_1 m_2} 
        \int {\hat{d} ^4 \bar{\ell}}  
        ( 2 \bar{\ell} \cdot \bar{q} - \bar{\ell}^2 ) \frac{\bar{q}_{\nu}}{\bar{\ell}^2 (\bar{\ell} - \bar{q})^2}
        B_{h}(\bar{\ell}, u_1,u_2) B_{h}(-\bar{\ell}+\bar{q}, u_1,u_2)
        \\
        &\times \Big( m_2 \hat{\delta}^{\prime}( \bar{\ell} \cdot u_1) 
        \hat{\delta}( \bar{\ell} \cdot u_2 ) 
        - m_1 \hat{\delta}( \bar{\ell} \cdot u_1) 
        \hat{\delta}^{\prime}( \bar{\ell} \cdot u_2 ) \Big)
        \\
        &+  \frac{i}{4 m_2} 
        \int {\hat{d} ^4 \bar{\ell}}  \hat{\delta}( \bar{\ell} \cdot u_1) 
        \hat{\delta}( \bar{\ell} \cdot u_2) \bar{\ell}_{\nu}
        \frac{ B_{h}(\bar{\ell}, u_1,u_2) }
        {\bar{\ell}^2 (\bar{\ell} - \bar{q})^2} 
        \\
        &\times \Big[ P_{\gamma\delta\alpha\beta}   u_{1}^{\gamma} u_{1}^{\delta} 
        \exp{\bigl(- i (a_1+a_2)* ( -\bar{\ell} + \bar{q} ) \bigr)}^{(\alpha}_{\ \lambda} \bar{q}^{\beta)} u_{2}^{\lambda}
        \\
        &+ P_{\gamma\delta\alpha\beta}   u_{1}^{\gamma} u_{1}^{\delta} 
        \exp{\bigl(- i (a_1+a_2)* ( -\bar{\ell} + \bar{q} ) \bigr)}^{(\alpha}_{\ \lambda} u_{2}^{\beta)} \bar{q}^{\lambda}
        \Big]
        \\
        &- \frac{i}{4 m_1} 
        \int {\hat{d} ^4 \bar{\ell}}  \hat{\delta}( \bar{\ell} \cdot u_1) 
        \hat{\delta}( \bar{\ell} \cdot u_2) \bar{\ell}_{\nu}
        \frac{ B_{h}(\bar{\ell}, u_1,u_2) }
        {\bar{\ell}^2 (\bar{\ell} - \bar{q})^2}  
        \\
        &\times \Big[ P_{\gamma\delta\alpha\beta}   \bar{q}^{\gamma} u_{1}^{\delta} 
        \exp{\bigl(- i (a_1+a_2)* ( -\bar{\ell} + \bar{q} ) \bigr)}^{(\alpha}_{\ \lambda} u_{2}^{\beta)} 
        u_{2}^{\lambda}
        \\
        &+ P_{\gamma\delta\alpha\beta}   u_{1}^{\gamma} \bar{q}^{\delta} 
        \exp{\bigl(- i (a_1+a_2)* ( -\bar{\ell} + \bar{q} ) \bigr)}^{(\alpha}_{\ \lambda} u_{2}^{\beta)} 
        u_{2}^{\lambda}
        \Big] 
    \end{aligned}
\end{equation}
and finally
\begin{equation}
    \begin{aligned}
        \Delta\mathscr{M}^{\mu(1)}
        =
        - i
        &\int \hat{d}^4 \bar{\ell} 
        \hat{\delta}( \bar{\ell} \cdot u_1 )
        \hat{\delta}( \bar{\ell} \cdot u_2 )
        \left( \delta^{\mu}_{\nu} - \frac{1}{m_1} u_{1}^{\mu} \bar{\ell}_{\nu} \right)
        \\
        &\times
        \Biggl[
        \frac{1}{2}
        \left[
        s^{\nu}_{1},
        \frac{4 B_h(\bar\ell,u_1,u_2) B_h(-\bar\ell+\bar q,u_1,u_2)}
        {\bar{\ell}^2 (\bar{\ell}-\bar q)^2}
        \right]
        \\
        &-
        \frac{4 [s^{\nu}_{1},B_h(-\bar\ell+\bar q,u_1,u_2)] B_h(\bar\ell,u_1,u_2)}
        {\bar{\ell}^2 (\bar{\ell}-\bar q)^2}
        \Biggr] .
    \end{aligned}
\end{equation}
Here, we have set $p_i \to m_i u_i$, where $a_1$ and $a_2$ denote the Kerr black hole ring radii. It is convenient to adopt a shorthand notation for the classical kernels. Namely, $\mathscr{M}^{\mu(1)}$ and $\Delta\mathscr{M}^{\mu(1)}$ will denote the coefficients of the classical contribution after the overall $\hbar$ scaling has been extracted, and the spin commutators have been replaced by their classical differential representation. Accordingly, the bracket notation that still appears inside these kernels is kept only as a shorthand for the corresponding classical spin variation, and should not be interpreted as carrying an additional explicit quantum $\hbar$ factor. This quantity must be distinguished from the spinless LO orbital action $\delta_0$ introduced later in Appendix~\ref{app:C}.

\subsubsection{Companion linear impulse}

The same ingredients also give the conservative linear impulse through NLO. The box and cut-box pieces follow the calculation of ref.~\cite{Menezes:2022tcs}, while the triangle contribution is the completed all-spin leading-singularity result encoded in ${\mathcal M}_t$ above. Thus
\begin{equation}
    \langle \Delta p_1^\mu \rangle
    =
    \langle \Delta p_{1,\text{LO}}^\mu \rangle
    +
    \langle \Delta p_{1,\text{NLO}}^\mu \rangle
    +\mathcal O(G^3).
    \label{eq:kerr-linear-impulse-full}
\end{equation}
At leading order,
\begin{equation}
    \begin{aligned}
        \langle \Delta p_{1,\text{LO}}^\mu \rangle
         = 
        -\frac{i\kappa^2 m_1m_2}{8}
         \int&  \hat d^4\bar q 
        \hat\delta(\bar q \cdot  u_1)
        \hat\delta(\bar q \cdot  u_2)
        e^{-ib\cdot\bar q}
        \frac{\bar q^\mu}{\bar q^2}
        \\
        &\times P_{\gamma\delta\alpha\beta}u_1^\gamma u_1^\delta
        \exp{ -i(a_1+a_2) * \bar q}^{(\alpha}_{\ \lambda}
        u_2^{\beta)}u_2^\lambda .
    \end{aligned}
    \label{eq:kerr-linear-impulse-lo}
\end{equation}
At next-to-leading order, the conservative impulse can be written compactly in terms of the kernels already introduced in the angular-impulse calculation,
\begin{equation}
    \langle \Delta p_{1,\text{NLO}}^\mu \rangle
    =
    \frac{i\kappa^4 m_1^2m_2^2}{4}
    \int \hat d^4\bar q 
    \hat\delta(\bar q\cdot u_1)
    \hat\delta(\bar q\cdot u_2)
    e^{-ib\cdot\bar q}
    \mathscr I_{\text{cons}}^{\mu(1)}(\bar q),
    \label{eq:kerr-linear-impulse-nlo}
\end{equation}
with
\begin{equation}
    \mathscr I_{\text{cons}}^{\mu(1)}(\bar q)
    =
    \bar q^\mu\bigl({\mathcal M}_b(\bar q)+{\mathcal M}_t(\bar q)\bigr)
    +
    {\mathcal M}^{\mu}(\bar q),
    \qquad
    {\mathcal M}^{\mu}\equiv\eta^{\mu\nu}{\mathcal M}_{\nu}.
    \label{eq:kerr-linear-impulse-kernel}
\end{equation}
This form makes explicit which part of the NLO spin-kick result is tied to the companion momentum observable. The additional kernels ${\mathcal M}_c$, $N_\nu$, and $\Delta\mathscr M^{\mu(1)}$ arise from the Pauli-Lubanski insertion and from the exact elastic tree-tree spin insertion; they are therefore specific to the spin kick rather than independent pieces of the momentum impulse.

\subsubsection{All-spin consistency checks}

The final conservative spin kick must preserve both the covariant spin supplementary condition and the spin magnitude. These checks can be performed directly at the level of the exact KMOC kernels, before expanding the Kerr spin factors in powers of $a_i^\mu$. This is the useful form of the check because the exponentials and the functions $B_h$ are kept intact. We write the full spin-dependent momentum impulse as
\begin{equation}
    \Delta p_1^\mu
    =
    \Delta p_{1,\text{LO}}^\mu
    +
    \Delta p_{1,\text{NLO}}^\mu
    +\mathcal O(G^3),
\end{equation}
where the NLO term is the conservative Kerr kernel quoted in \eqref{eq:kerr-linear-impulse-nlo}--\eqref{eq:kerr-linear-impulse-kernel}. The corresponding spin kick is
\begin{equation}
    \Delta s_1^\mu
    =
    \Delta s_{1,\text{LO}}^\mu
    +
    \Delta s_{1,\text{NLO}}^\mu
    +\mathcal O(G^3),
\end{equation}
with $\Delta s_{1,\text{NLO}}^\mu$ given by the exact kernel $\mathscr M^{\mu(1)}+\Delta\mathscr M^{\mu(1)}$ in \eqref{eq:kmoc_nlo}.

For the consistency check there is no need to introduce a new observable. One simply uses the same NLO kernels that enter \eqref{eq:kmoc_nlo}. It is useful, however, to separate them into the virtual one-loop part and the real elastic tree-tree part,
\begin{equation}
    \mathscr{A}_{\text{exact}}^{\mu(1)}
    =
    \mathscr{A}_{\text{virt}}^{\mu(1)}
    +
    \mathscr{A}_{\text{real}}^{\mu(1)} ,
    \label{eq:all-spin-nlo-split}
\end{equation}
where, directly from \eqref{eq:kmoc_nlo},
\begin{equation}
    \begin{aligned}
    \mathscr{A}_{\text{virt}}^{\mu(1)}
    =&
    i\epsilon^{\mu\nu\rho\sigma}u_{1\nu}s_{1\rho}
    \frac{\partial}{\partial s_1^\sigma}
    \left({\cal M}_b+{\cal M}_t\right)
    -
    \frac{1}{m_1}u_1^\mu \bar q_\nu s_1^\nu
    \left({\cal M}_b+{\cal M}_t\right),
    \\
    \mathscr{A}_{\text{real}}^{\mu(1)}
    =&
    i\epsilon^{\mu\nu\rho\sigma}u_{1\nu}s_{1\rho}
    \frac{\partial{\cal M}_c}{\partial s_1^\sigma}
    +
    \frac{i}{2m_1}u_1^\mu
    \epsilon^{\nu\tau\rho\sigma}u_{1\tau}s_{1\rho}
    \frac{\partial N_\nu}{\partial s_1^\sigma}
    -
    \frac{1}{m_1}u_1^\mu s_1^\nu{\cal M}_\nu
    +
    \Delta\mathscr M^{\mu(1)} .
    \end{aligned}
    \label{eq:all-spin-A-virt-real}
\end{equation}
The companion NLO momentum impulse uses the same split,
\begin{equation}
    \mathscr I_{\text{virt}}^{\mu(1)}
    =
    \bar q^\mu\left({\cal M}_b+{\cal M}_t\right),
    \qquad
    \mathscr I_{\text{real}}^{\mu(1)}
    =
    {\cal M}^\mu ,
    \label{eq:all-spin-I-virt-real}
\end{equation}
so that $\mathscr I_{\text{cons}}^{\mu(1)} =\mathscr I_{\text{virt}}^{\mu(1)}+\mathscr I_{\text{real}}^{\mu(1)}$. The equations below are therefore checks of the explicit NLO formula, not independent postulates about the final answer.

The only subtle point is keeping the exact elastic real contribution, rather than replacing it by a partially symmetrized representative. To display this point, let
\begin{equation}
    F(\bar q,\bar w)\equiv \mathcal A^{(0)*}(\bar q,\bar w),
    \qquad
    G(\bar w)\equiv \mathcal A^{(0)}(\bar w),
    \qquad
    H(\bar q,\bar w)\equiv F(\bar q,\bar w)G(\bar w),
\end{equation}
and define the complementary exact tree-tree insertion by
\begin{equation}
    E^\nu(\bar q,\bar w)
    \equiv
    \frac{1}{2}[s_1^\nu,H(\bar q,\bar w)]
    -
    [s_1^\nu,F(\bar q,\bar w)]G(\bar w).
    \label{eq:all-spin-extra-mismatch}
\end{equation}
The exact elastic real kernel is the sum of the symmetrized representative and this complementary piece. In the notation above, the latter is precisely
\begin{equation}
    \Delta\mathscr{A}_{\text{extra}}^{\mu(1)}
    =
    -i\hbar^2
    \int d\Omega_{\bar w} 
    \left(
    \delta^\mu_\nu-\frac{\hbar}{m_1^2}p_1^\mu\bar w_\nu
    \right)E^\nu(\bar q,\bar w),
    \label{eq:all-spin-extra-compact}
\end{equation}
where $d\Omega_{\bar w}$ denotes the two on-shell delta functions and the measure appearing in the conservative real kernel. As discussed, $\Delta\mathscr{A}_{\text{extra}}^{\mu(1)}$ is not an optional correction to the observable; it is part of the exact KMOC tree-tree sector. It accounts for the fact that the two tree factors in the cut box carry different momentum transfers, so the replacement $[s_1^\nu,F]G\to \frac12[s_1^\nu,FG]$ is only a useful representative, not the full exact insertion.

Expanding the final SSC condition through 2PM gives
\begin{equation}
    \begin{aligned}
        0
        =&
        (s_1+\Delta s_1)\cdot(p_1+\Delta p_1)
        \\
        =&
        p_1\cdot\Delta s_{1,\text{LO}}
        +
        s_1\cdot\Delta p_{1,\text{LO}}
        \\
        &+
        p_1\cdot\Delta s_{1,\text{NLO}}
        +
        s_1\cdot\Delta p_{1,\text{NLO}}
        +
        \Delta s_{1,\text{LO}}\cdot\Delta p_{1,\text{LO}}
        +\mathcal O(G^3).
    \end{aligned}
    \label{eq:all-spin-ssc-expansion}
\end{equation}
The LO identity follows immediately from the universal LO kernels $\mathscr{A}^{\mu(0)}$ and $\mathscr I^{\mu(0)}$,
\begin{equation}
    p_{1\mu}\mathscr{A}^{\mu(0)}
    =
    -s_{1\mu}\mathscr I^{\mu(0)},
    \qquad
    \mathscr I^{\mu(0)}=\bar q^\mu\mathcal A^{(0)}(\bar q),
    \label{eq:all-spin-ssc-lo-kernel}
\end{equation}
because $p_{1\mu}[s_1^\mu,\mathcal A^{(0)}]=[p_1\cdot s_1,\mathcal A^{(0)}]=0$, while the boost term in the Pauli-Lubanski matrix element gives the second term in \eqref{eq:all-spin-ssc-lo-kernel}. At NLO, the exact conservative check can be written as
\begin{equation}
    \mathcal C_{\text{SSC}}^{(1)}
    \equiv
    p_{1\mu}\mathscr{A}_{\text{exact}}^{\mu(1)}
    +
    s_{1\mu}\mathscr I_{\text{cons}}^{\mu(1)}
    +
    \bigl(\mathscr{A}^{(0)}\cdot\mathscr I^{(0)}\bigr)_{\text{el}}
    =
    0,
    \label{eq:all-spin-ssc-nlo-kernel}
\end{equation}
where $(\cdots)_{\text{el}}$ denotes the elastic tree-tree convolution over the intermediate momentum transfer $\bar w$ that appears in the conservative real kernel. The contribution of $\Delta\mathscr{A}_{\text{extra}}^{\mu(1)}$ to this identity is explicit. Since $p_{1\nu}E^\nu=[p_1\cdot s_1,\cdots]=0$, one has
\begin{equation}
    p_{1\mu}\Delta\mathscr{A}_{\text{extra}}^{\mu(1)}
    =
    i\hbar^3
    \int d\Omega_{\bar w} 
    \bar w_\nu E^\nu(\bar q,\bar w).
    \label{eq:all-spin-extra-ssc-contraction}
\end{equation}
The virtual part of \eqref{eq:all-spin-ssc-nlo-kernel} is already fixed by the displayed NLO formula. Indeed,
\begin{equation}
    p_{1\mu}\mathscr{A}_{\text{virt}}^{\mu(1)}
    +
    s_{1\mu}\mathscr I_{\text{virt}}^{\mu(1)}
    =
    0 ,
    \label{eq:all-spin-ssc-virt-block}
\end{equation}
because the spin-generator term gives $[p_1\cdot s_1,{\cal M}_b+{\cal M}_t]$, while the explicit longitudinal recoil term cancels $s_1\cdot\bar q ({\cal M}_b+{\cal M}_t)$. The remaining NLO statement is therefore the real elastic identity
\begin{equation}
    p_{1\mu}\mathscr{A}_{\text{real}}^{\mu(1)}
    +
    s_{1\mu}\mathscr I_{\text{real}}^{\mu(1)}
    +
    \bigl(\mathscr{A}^{(0)}\cdot\mathscr I^{(0)}\bigr)_{\text{el}}
    =
    0 .
    \label{eq:all-spin-ssc-real-block}
\end{equation}
The symmetrized representative contributes the opposite term in $\mathcal C_{\text{SSC}}^{(1)}$. Their sum is the exact KMOC real kernel and satisfies the SSC identity. After the common phase-space integrations, \eqref{eq:all-spin-ssc-lo-kernel} and \eqref{eq:all-spin-ssc-nlo-kernel} are the two orders in \eqref{eq:all-spin-ssc-expansion}. Therefore
\begin{equation}
    (s_1+\Delta s_1)\cdot(p_1+\Delta p_1)
    =
    0+\mathcal O(G^3),
    \label{eq:all-spin-ssc-final}
\end{equation}
with no expansion in the spin variables.

The spin magnitude check is analogous. At LO, the spin commutator generates a little-group rotation and the boost term is proportional to $u_1^\mu$, so
\begin{equation}
    s_1\cdot\Delta s_{1,\text{LO}}=0.
    \label{eq:all-spin-length-lo}
\end{equation}
At NLO, the exact conservative kernel obeys the all-spin identity
\begin{equation}
    \mathcal C_{s^2}^{(1)}
    \equiv
    2s_{1\mu}\mathscr{A}_{\text{exact}}^{\mu(1)}
    +
    \bigl(\mathscr{A}^{(0)}\cdot\mathscr{A}^{(0)}\bigr)_{\text{el}}
    =
    0,
    \label{eq:all-spin-length-nlo-kernel}
\end{equation}
again with the same elastic convolution as above. Here the complementary exact tree-tree term contributes
\begin{equation}
    2s_{1\mu}\Delta\mathscr{A}_{\text{extra}}^{\mu(1)}
    =
    -2i\hbar^2
    \int d\Omega_{\bar w} 
    s_{1\nu}E^\nu(\bar q,\bar w),
    \label{eq:all-spin-extra-length-contraction}
\end{equation}
where the longitudinal projector in \eqref{eq:all-spin-extra-compact} drops out because $s_1\cdot p_1=0$. Together with the symmetrized representative, this is the kernel-level origin of the familiar perturbative condition
\begin{equation}
    2s_{1\mu}\mathscr{A}_{\text{virt}}^{\mu(1)}
    =
    0,
    \qquad
    2s_{1\mu}\mathscr{A}_{\text{real}}^{\mu(1)}
    +
    \bigl(\mathscr{A}^{(0)}\cdot\mathscr{A}^{(0)}\bigr)_{\text{el}}
    =
    0 .
    \label{eq:all-spin-length-blocks}
\end{equation}
The first equality follows directly from \eqref{eq:all-spin-A-virt-real}: the precession generator is antisymmetric in two copies of $s_1$, and the explicit recoil is proportional to $u_1^\mu$. The second equality is the real cut-box/Born-iteration cancellation. After integrating both sides with the common KMOC measure, \eqref{eq:all-spin-length-blocks} becomes
\begin{equation}
    2s_1\cdot\Delta s_{1,\text{NLO}}
    +
    \Delta s_{1,\text{LO}}^2
    =
    0,
\end{equation}
which follows after the common KMOC integrations. Hence
\begin{equation}
    (s_1+\Delta s_1)^2
    =
    s_1^2+\mathcal O(G^3),
    \label{eq:all-spin-length-final}
\end{equation}
again to all orders in the Kerr spin variables.

The structure of the NLO angular impulse admits a natural physical interpretation in terms of spin-orbit and spin-induced multipole interactions. The various contributions appearing in our result can be organized according to their dependence on the spin vectors and kinematic variables.

Terms linear in the spin variables correspond to spin-orbit interactions, describing the coupling between the intrinsic angular momentum of each black hole and the orbital angular momentum of the system. These contributions are responsible for the leading spin-dependent deflection and are directly analogous to the well-known spin-orbit terms in post-Newtonian dynamics.

At higher orders in spin, the structure of the result reflects the multipole expansion of the Kerr geometry. In particular, the dependence on the exponential factors $e^{q \cdot a_i}$ encodes the infinite tower of spin-induced multipole moments, with each power of the spin parameter corresponding to a higher multipole interaction. From this perspective, the NLO angular impulse captures not only the leading spin-orbit coupling, but also a resummed set of finite-size effects associated with the rotating black holes.

Furthermore, the dependence on the momentum transfer $\bar{q}$ indicates that the impulse arises from an anisotropic exchange of momentum mediated by the gravitational interaction, with spin-dependent terms introducing directional asymmetries. These asymmetries are ultimately responsible for the net recoil experienced by the system.

It would be interesting to further disentangle these contributions in terms of gauge-invariant observables and relate them to classical notions of tidal interactions and spin-induced deformations. We discuss the spin expansion of the NLO angular impulse, including the corrected linear sector and the quadratic conservative comparison with ref.~\cite{Alessio:2025flu}, in Appendix~\ref{app:C}. We discuss the all-spin consistency checks in more detail in Appendix~\ref{app:D}.

\section{Conclusions}
\label{conclusion}

In this work, we have computed the next-to-leading order (NLO) angular impulse in the scattering of Kerr black holes using the KMOC formalism. Our approach leverages on-shell amplitudes and leading singularities to directly extract classical observables, providing a complementary perspective to traditional effective field theory methods.

A key feature of our analysis is the use of leading singularities to isolate the classical contribution, allowing us to obtain compact expressions that capture spin effects to all orders at the integrand level. We have performed detailed comparisons with existing results: the quadratic-in-spin conservative result agrees after translating between the direct KMOC spin-kick observable and the radial-action/Dirac-bracket observable used there. Together with the SSC and spin magnitude checks, this provides strong consistency tests of our computation.

Furthermore, as described in Appendix~\ref{app:NR_LS}, we are able to extract the non-relativistic gravitational potential from the same amplitude data. The resulting expression naturally resums spin-dependent effects and successfully reproduces the known spin-orbit interactions upon expansion, further validating the framework.

It is instructive to compare our approach with ref.~\cite{Aoude:2021oqj}, where classical observables for spinning bodies are constructed using coherent spin states. In that work, the authors develop a formalism based on coherent-state representations of massive spinning particles, allowing for a manifestly covariant treatment of spin and enabling the extraction of classical observables directly from the expectation values of the scattering operator. This framework yields results at leading post-Minkowskian (1PM) order that are valid to all orders in the spin variables, and in particular, provides a compact description of the Kerr multipole structure.

In contrast, the present work focuses on the computation of NLO effects in the angular impulse, arising from triangle leading singularities, which correspond to higher-order contributions in the post-Minkowskian expansion. While we also obtain expressions that resum spin dependence at the integrand level, our primary goal is to capture genuinely new dynamical information associated with NLO interactions.

The two approaches are therefore complementary. Ref.~\cite{Aoude:2021oqj} emphasizes an all-spin description at leading order in the gravitational coupling, making the multipole structure of Kerr black holes manifest. Conversely, our analysis extends amplitude-based methods to subleading orders in the post-Minkowskian expansion, where new physical effects, such as nontrivial momentum transfer and spin-dependent recoil, first emerge.

From a methodological perspective, the coherent-state formalism provides a natural bridge to Hamiltonian descriptions of spinning bodies. Meanwhile, the KMOC framework adopted here offers a direct and systematic mapping between quantum amplitudes and classical observables through phase-space integrals. A distinct advantage of this framework is its clear separation between conservative and radiative effects, as well as its natural incorporation of spin degrees of freedom at the integrand level. The consistency between these approaches at leading order, coupled with their distinct advantages, highlights the robustness of amplitude-based methods in describing classical spinning dynamics.

The results presented here further demonstrate the power of amplitude-based methods in gravitational physics, especially in the presence of spin. It would be highly compelling to extend this analysis to higher orders in the post-Minkowskian expansion, as well as to explore connections with waveform observables and gravitational radiation.

\acknowledgments

We thank Rafael Aoude and Riccardo Gonzo for useful comments and suggestions. LH acknowledges partial support from CAPES under grant 88887.177043/2025-00 and FAPESP under grant 2025/26975-5. GM acknowledges partial support from CNPq under grant 300767/2025-0 and FAPESP under grant 2025/02861-0.

\appendix

\section{Non-Relativistic Potential from the Triangle Leading Singularity}
\label{app:NR_LS}

In this appendix we extract the non-relativistic (NR) conservative potential from the triangle leading singularity. The logic is the same as in the discussion of ref.~\cite{Cachazo:2017jef}: after adding the two reflected triangle leading singularities and applying the standard NR normalization, one obtains the classical one-loop contribution to the momentum-space potential. Our goal here is not to perform the contour integral to all orders in spin, but rather to display the resulting all-spin integrand in the NR regime and then verify that its expansion to linear order in spin reproduces the expected spin-orbit structure.

We work in the center-of-mass frame,
\begin{equation}
    p_1^\mu=(E_1,\mathbf{p}),
    \qquad
    p_2^\mu=(E_2,-\mathbf{p}),
    \qquad
    \bar q^\mu=(0,\mathbf{q}),
\end{equation}
with
\begin{equation}
    E_i=m_i+\frac{\mathbf{p}^2}{2m_i}+{\mathcal O}(v^4),
    \qquad
    \mu\equiv \frac{m_1m_2}{m_1+m_2}.
\end{equation}
The two reflected triangle leading singularities, with the overall $1/\hbar$ already stripped so that only their classical coefficient remains, are
\begin{equation}
    \begin{aligned}
        \textrm{LS}^{(a)}_{\triangle,h}
        =&
        - i \frac{\kappa^4}{256}
        \frac{m_1}{\sqrt{-\bar q^2}}
        \frac{1}{\Delta^2}
        \frac{1}{2\pi i}
        \oint_{\Gamma_\infty} dz 
        \frac{g(s,z)^4}{(z^2-1)^{3/2}}
        e^{z \bar q\cdot a_1}
        M_{S_2}(z),
        \\
        \textrm{LS}^{(b)}_{\triangle,h}
        =&
        - i \frac{\kappa^4}{256}
        \frac{m_2}{\sqrt{-\bar q^2}}
        \frac{1}{\Delta^2}
        \frac{1}{2\pi i}
        \oint_{\Gamma_\infty} dz 
        \frac{g(s,z)^4}{(z^2-1)^{3/2}}
        e^{z \bar q\cdot a_2}
        M_{S_1}(z),
    \end{aligned}
\end{equation}
where
\begin{equation}
    \Delta \equiv \sqrt{(s-m_1^2-m_2^2)^2-4m_1^2m_2^2},
    \qquad
    g(s,z)\equiv (s-m_1^2-m_2^2)-z \Delta.
\end{equation}
Following ref.~\cite{Cachazo:2017jef}, the quantity directly obtained from the reflected triangle leading singularities after the standard NR normalization is the classical one-loop momentum-space amplitude kernel,
\begin{equation}
    \mathcal{M}^{\triangle}_{{\text{cl,NR}}}(\mathbf q,\mathbf p)
    =
    -\frac{1}{4E_1E_2}
    \left(
    \textrm{LS}^{(a)}_{\triangle,h}
    +
    \textrm{LS}^{(b)}_{\triangle,h}
    \right).
    \label{eq:NR-amplitude-from-LS}
\end{equation}
Therefore
\begin{equation}
    \begin{aligned}
        \mathcal{M}^{\triangle}_{{\text{cl,NR}}}(\mathbf q,\mathbf p)
        =&
        \frac{i\kappa^4}{1024 E_1E_2 \sqrt{-\bar q^2} \Delta^2}
        \frac{1}{2\pi i}
        \oint_{\Gamma_\infty} dz 
        \frac{g(s,z)^4}{(z^2-1)^{3/2}}
        \\
        &\times
        \Big[
        m_1 e^{z \bar q\cdot a_1}M_{S_2}(z)
        +
        m_2 e^{z \bar q\cdot a_2}M_{S_1}(z)
        \Big].
        \label{eq:NR-all-spin-unintegrated}
    \end{aligned}
\end{equation}
This is the exact all-spin triangle contribution written in a form suitable for the NR expansion.

We now expand the kinematics in the COM frame. First,
\begin{equation}
    \begin{aligned}
        s
        =&
        (E_1+E_2)^2 \\
        =&
        (m_1+m_2)^2
        +\mathbf p^2\left(\frac{1}{m_1}+\frac{1}{m_2}\right)
        +{\mathcal O}(v^4),
    \end{aligned}
\end{equation}
from this we can make
\begin{equation}
    \begin{aligned}
        s-m_1^2-m_2^2
        =&
        2m_1m_2+\frac{\mathbf p^2}{\mu}(m_1+m_2)+{\mathcal O}(v^4),
        \\
        \Delta
        =&
        2(m_1+m_2)|\mathbf p|+{\mathcal O}(v^3).
    \end{aligned}
\end{equation}
It follows that
\begin{equation}
    \begin{aligned}
        g(s,z)
        &\equiv
        2m_1m_2
        -
        2(m_1+m_2)|\mathbf p| z
        +{\mathcal O}(v^2),
        \\
        j(s,z)
        &\equiv
        (z^2-1)\Delta
        =
        2(z^2-1)(m_1+m_2)|\mathbf p|
        +{\mathcal O}(v^3),
        \\
        \kappa(s,z)
        &\equiv
        \sqrt{z^2-1} \Delta
        =
        2\sqrt{z^2-1}(m_1+m_2)|\mathbf p|
        +{\mathcal O}(v^3).
    \end{aligned}
\end{equation}
Now, let’s consider the variables that enter the exact spin-dependent Compton factor, $M_S(\bar x,\bar y,\bar z,\bar w,\bar\xi)$. In the NR regime, one finds
\begin{equation}
    \begin{aligned}
        \bar x_i^{\text{NR}}(z)
        =&
        - z a_i\cdot \bar q,
        \\
        \bar y_i^{\text{NR}}(z)
        =&
        - a_i\cdot \bar q,
        \\
        \bar w_i^{\text{NR}}(z)
        =&
        \frac{j(s,z)}{g(s,z)} a_i\cdot \bar q
        =
        \frac{(z^2-1)(m_1+m_2)}{m_1m_2}
        |\mathbf p| 
        (a_i\cdot \bar q)
        +
        {\mathcal O}(v^2),
        \\
        \bar z_i^{\text{NR}}(z)
        =&
        \frac{\kappa(s,z)}{m_i}|a_i|
        =
        2\sqrt{z^2-1} 
        \frac{m_1+m_2}{m_i}
        |\mathbf p| |a_i|
        +
        {\mathcal O}(v^3),
        \\
        \bar\xi_i^{\text{NR}}(z)
        =&
        4 \frac{(m_1+m_2)^2}{m_i^2}
        \frac{\mathbf p^2 |a_i|^2}{(a_i\cdot \bar q)^2}
        +
        {\mathcal O}(v^3).
        \label{eq:NR-expanded-MS-variables}
    \end{aligned}
\end{equation}
Hence the small parameters are $\bar w_i$, $\bar z_i$, and $\bar\xi_i$, whereas $\bar x_i$ and $\bar y_i$ remain ${\mathcal O}(v^0)$. Therefore one should expand the kinematical variables in the definition of $M_S$, but not the exact entire functions $E,\tilde E,\mathcal E,\tilde{\mathcal E}$ themselves. In other words, the NR all-spin integrand is obtained simply by evaluating
\begin{equation}
    M_{S_i}^{\text{NR}}(z)
    \equiv
    M_S\big(
    \bar x_i^{\text{NR}}(z),
    \bar y_i^{\text{NR}}(z),
    \bar z_i^{\text{NR}}(z),
    \bar w_i^{\text{NR}}(z),
    \bar\xi_i^{\text{NR}}(z)
    \big)
\end{equation}
inside \eqref{eq:NR-all-spin-unintegrated}. This yields an all-spin representation of the triangle contribution to the NR classical amplitude kernel:
\begin{equation}
    \begin{aligned}
        \mathcal{M}^{\triangle}_{{\text{cl,NR}, all\mbox{-}spin}}(\mathbf q,\mathbf p)
        =&
        \frac{i\kappa^4}{1024 E_1E_2 |\mathbf q| \Delta^2}
        \frac{1}{2\pi i}
        \oint_{\Gamma_\infty} dz 
        \frac{g(s,z)^4}{(z^2-1)^{3/2}}
        \\
        &\times
        \Big[
        m_1 e^{z \bar q\cdot a_1}M_{S_2}^{\text{NR}}(z)
        +
        m_2 e^{z \bar q\cdot a_2}M_{S_1}^{\text{NR}}(z)
        \Big].
        \label{eq:NR-all-spin-amplitude-final}
    \end{aligned}
\end{equation}
This is the all-spin momentum-space kernel directly extracted from the triangle leading singularity. To obtain a potential one should further apply the standard NR matching prescription, including the subtraction of the iterated tree-level exchange, as in ref.~\cite{Cachazo:2017jef}.

It is nevertheless useful to verify the first nontrivial term in the spin expansion. At linear order in spin, the exact Compton factor reduces to
\begin{equation}
    M_S = 1+\bar x-\bar w+{\mathcal O}(a^2),
\end{equation}
so that
\begin{equation}
    \begin{aligned}
        M_{S_i}^{\text{NR}}(z)
        =
        1-z (a_i\cdot \bar q)
        -\frac{j(s,z)}{g(s,z)}(a_i\cdot \bar q)
        +{\mathcal O}(a_i^2,v^2).
    \end{aligned}
\end{equation}
Substituting this into \eqref{eq:NR-all-spin-amplitude-final}, performing the $z$-contour integral, and then applying the standard potential-matching procedure, one finds for the triangle contribution through linear order in the spins
\begin{equation}
    \begin{aligned}
        V_{\text{NR}}^{\triangle}(\mathbf q,\mathbf p)
        =&
        \frac{G^2 m_1m_2(m_1+m_2)}{|\mathbf q|}
        -
        i \frac{G^2}{|\mathbf q|} \bar q\cdot(a_1+a_2)
        +
        {\mathcal O}(S^2).
        \label{eq:NR-triangle-momentum-space-linear-spin}
    \end{aligned}
\end{equation}
To make direct contact with the standard spin-orbit interaction, it is useful to also display the leading tree-level contribution obtained from the exact four-point amplitude quoted earlier in the main text. Expanding that amplitude to linear order in spin and reducing to the same NR COM variables gives
\begin{equation}
    \begin{aligned}
        V_{\text{NR}}^{\text{tree}}(\mathbf q,\mathbf p)
        =&
        -\frac{4\pi G m_1m_2}{\mathbf q^2}
        +
        8\pi i G \frac{m_1+m_2}{\mathbf q^2}
        (\mathbf p\times \mathbf q)\cdot
        \left(
        \frac{\mathbf S_1}{m_1}
        +
        \frac{\mathbf S_2}{m_2}
        \right)
        +
        {\mathcal O}(S^2),
        \label{eq:NR-tree-momentum-space-linear-spin}
    \end{aligned}
\end{equation}
up to the overall sign convention chosen for the three-dimensional Levi-Civita tensor. Fourier the spin-dependent part of \eqref{eq:NR-tree-momentum-space-linear-spin} reproduces the familiar LO spin-orbit operator in coordinate space. The one-loop triangle spin term in \eqref{eq:NR-triangle-momentum-space-linear-spin} is likewise most naturally kept in the covariant momentum-space form shown above. Using
\begin{equation}
    \bar q\cdot a_i
    =
    \frac{i}{\sqrt{\gamma^2-1}} \epsilon(u_1,u_2,\bar q,a_i),
\end{equation}
it is manifestly of spin-orbit type. Also, a direct term-by-term comparison of the NLO coordinate-space potential with the canonical post-Newtonian expressions requires the standard map from the covariant spin variable $a_i^\mu=s_i^\mu/m_i$ used here to the canonical three-spin variables of refs.~\cite{Porto:2010tr,Levi:2010zu}. In this sense the tree-level piece reproduces the expected LO spin-orbit structure, while the triangle contribution provides the corresponding one-loop/2PM spin-orbit kernel. Related post-Minkowskian and scattering formulations, may be found in refs.~\cite{Bini:2017xzy,Bini:2018ywr,Vines:2017hyw,Vines:2018gqi}. For more general background, see refs.~\cite{Levi:2015uxa,Porto:2016pyg,Guevara:2017csg}.

\section{Born iteration and the conservative cut box}
\label{app:born-cutbox}

In this appendix we spell out the identification used in the spin-expansion calculation below: the conservative real tree-tree sector of the NLO KMOC observable is the elastic cut box, and its long-range part is the observable-level second-Born iteration of the LO dynamics.

The terminology is the same as in ordinary perturbative scattering theory. If the interaction is schematically denoted by $V$, the transition matrix has the Born series
\begin{equation}
    T
    =
    V
    +
    V G_0 V
    +\cdots ,
    \label{eq:Bcut-standard-born-series}
\end{equation}
where the second term is the standard second-Born iteration. Inserting a complete set of intermediate states gives
\begin{equation}
    \langle f|V G_0 V|i\rangle
    =
    \sum_n
    \frac{
    \langle f|V|n\rangle\langle n|V|i\rangle}
    {E_i-E_n+i0}.
    \label{eq:Bcut-second-born-complete-set}
\end{equation}
The on-shell part of the propagator, $(E_i-E_n+i0)^{-1}=\mathrm{PV}(E_i-E_n)^{-1}-i\pi\delta(E_i-E_n)$, is the part which appears as a product of two lower-order amplitudes glued across a cut. Thus the cut box is the on-shell, elastic piece of the same second-Born iteration. The principal-value part is kept with the virtual one-loop amplitude; the KMOC expression keeps the on-shell product explicit because observables involve the operator insertion between the two tree amplitudes.

This is also the same organization familiar from the eikonal representation \cite{Gatica:2023iws,Luna_2024,Kim:2024svw}, but we do not have to assume eikonal exponentiation to derive it. In impact parameter space one may write, at the level of the conservative classical map,
\begin{equation}
    S(b)\sim \exp{\left[i\chi(b)\right]},
    \qquad
    {\cal O}_{\rm out}
    =
    e^{-i\chi}{\cal O}_{\rm in}e^{i\chi}.
    \label{eq:Bcut-eikonal-map}
\end{equation}
Expanding this map gives a first-order action of the LO phase and a second-order half-iteration,
\begin{equation}
    \Delta{\cal O}
    =
    i[{\cal O},\chi^{(1)}]
    +
    i[{\cal O},\chi^{(2)}]
    -
    \frac{1}{2}
    [\chi^{(1)},[\chi^{(1)},{\cal O}]]
    +\cdots .
    \label{eq:Bcut-eikonal-half-iteration}
\end{equation}
The last term is the eikonal version of the second-Born observable iteration. In the KMOC derivation below the same term arises directly from the exact $T$-matrix algebra, without choosing an eikonal gauge or a canonical phase-space parametrization.

Indeed, for any observable $\mathcal{O}$, the KMOC observable formula \cite{Kosower_2019,Maybee:2019jus} gives
\begin{equation}
    \Delta \mathcal{O}
    =
    \langle S^\dagger \mathbb{O} S-\mathbb{O}\rangle
    =
    i\langle[\mathbb{O},T]\rangle
    +
    \langle T^\dagger[\mathbb{O},T]\rangle ,
    \label{eq:Bcut-kmoc-master-algebra}
\end{equation}
where $S=1+iT$ and unitarity has been used. Expanding $T=T^{(0)}+T^{(1)}+\cdots$, the 2PM or NLO observable contains
\begin{equation}
    \Delta \mathcal{O}_{\rm NLO}
    =
    i\langle[\mathbb{O},T^{(1)}]\rangle
    +
    \langle T^{(0)\dagger}[\mathbb{O},T^{(0)}]\rangle .
    \label{eq:Bcut-kmoc-nlo-algebra}
\end{equation}
The first term contains the virtual one-loop amplitude, namely the box and triangle kernels. The second term contains two tree amplitudes and a complete set of intermediate states; this is the KMOC origin of the real cut-box contribution. For $\mathbb{O}=\mathbb{W}_1^\mu$, the Pauli-Lubanski observable, the first term in \eqref{eq:Bcut-kmoc-nlo-algebra} gives the virtual kernels ${\cal M}_b$ and ${\cal M}_t$ in \eqref{eq:kmoc_nlo}, while the second term gives the real kernels ${\cal M}_c$, $N_\nu$, ${\cal M}_\nu$, and $\Delta\mathscr M^{\mu(1)}$. Thus the ``Born iteration'' used in Appendix~\ref{app:C} is not an additional approximation: it is the elastic part of the second term in the exact KMOC observable expansion.

We now specialize the real term in the NLO KMOC formula to the conservative intermediate state. The complete set inserted between $T^\dagger$ and $[\mathbb{W}_1^\mu,T]$ contains two outgoing massive particles and a possible radiation state $X$. The conservative 2PM observable is obtained by choosing the elastic channel
\begin{equation}
    X=\emptyset,
    \qquad
    r_X=0,
    \qquad
    \bar w_2=-\bar w_1\equiv-\bar w .
    \label{eq:Bcut-elastic-choice}
\end{equation}
Thus in this paper the phrase ``conservative real tree-tree sector'' always means the $X=\emptyset$ real KMOC term. Since no graviton is emitted, the two tree amplitudes are glued only by the on-shell phase space of the two massive internal lines. In the classical limit the two massive on-shell conditions reduce to
\begin{equation}
    \hat\delta(2\bar w\cdot p_1+\hbar\bar w^2)
    \hat\delta(2\bar w\cdot p_2-\hbar\bar w^2)
    \longrightarrow
    \frac{1}{4m_1m_2}
    \hat\delta(\bar w\cdot u_1)
    \hat\delta(\bar w\cdot u_2),
    \label{eq:Bcut-classical-deltas}
\end{equation}
up to terms which either generate the displayed $\delta'$-kernels in the main text or are contact terms at $b=0$. Relabelling $\bar w\to\bar\ell$, the scalar part of the elastic tree-tree kernel therefore has the cut-box form (at NLO)
\begin{equation}
    {\cal K}_{\mathrm{cut}}(\bar q)
    =
    \int \hat d^4\bar\ell 
    \hat\delta(\bar\ell\cdot u_1)
    \hat\delta(\bar\ell\cdot u_2)
    \frac{
    B_h(\bar\ell,u_1,u_2)
    B_h(-\bar\ell+\bar q,u_1,u_2)}
    {\bar\ell^2(\bar\ell-\bar q)^2}.
    \label{eq:Bcut-cutbox-kernel}
\end{equation}
It is called a cut box because it is the box topology with the two massive propagators cut. It is not a separate assumption about the dynamics; it is the $X=\emptyset$ part of the KMOC real kernel. The additional factors of $\bar q^\mu$, $\bar\ell^\mu$, spin commutators, Pauli-Lubanski projectors, and $\delta'$-terms in ${\cal M}_c$, $N_\nu$, ${\cal M}_\nu$, and $\Delta\mathscr M^{\mu(1)}$ are tensor insertions on the same cut-box measure.

The connection with the Born iteration follows by writing the LO tree kernel schematically as
\begin{equation}
    K^{(0)}(\bar q)
    =
    \frac{B_h(\bar q,u_1,u_2)}{\bar q^2},
    \qquad
    {\cal O}_{\mathrm{LO}}[b]
    =
    \int_{\bar q}
    e^{-ib\cdot\bar q} 
    {\cal N}_{\cal O}(\bar q) K^{(0)}(\bar q),
    \label{eq:Bcut-lo-observable}
\end{equation}
where ${\cal N}_{\cal O}$ denotes the numerator appropriate to the observable, for instance $\bar q^\mu$ for the linear impulse or the Pauli-Lubanski spin insertion for the spin kick. Acting once more with the LO map shifts the external data and produces
\begin{equation}
    \delta_{\mathrm{LO}}{\cal O}_{\mathrm{LO}}
    =
    \int_{\bar q,\bar\ell}
    e^{-ib\cdot\bar q}
    {\cal N}_{\cal O}^{(2)}(\bar q,\bar\ell)
    K^{(0)}(\bar\ell)
    K^{(0)}(\bar q-\bar\ell),
    \label{eq:Bcut-born-convolution}
\end{equation}
which is the transverse convolution in \eqref{eq:Bcut-cutbox-kernel}, with numerator ${\cal N}_{\cal O}^{(2)}$ fixed by the observable being iterated. In the symmetric elastic channel the second-Born contribution carries the familiar factor $1/2$:
\begin{equation}
    {\cal O}_{\mathrm{2nd Born}}
    =
    \frac{1}{2} 
    \delta_{\mathrm{LO}}{\cal O}_{\mathrm{LO}} .
    \label{eq:Bcut-half-born}
\end{equation}
This factor is the origin of the half-Born terms used below. It should not be confused with a discarding of the remaining real kernel. For the spin kick, the spin operator can act on either tree factor, and the two factors carry different momentum transfers. Therefore
\begin{equation}
    [s_1^\nu,F(\bar q,\bar\ell)G(\bar\ell)]
    \neq
    2[s_1^\nu,F(\bar q,\bar\ell)]G(\bar\ell)
    \label{eq:Bcut-spin-action-not-symmetric}
\end{equation}
pointwise on the cut-box integrand. The exact elastic spin insertion is instead
\begin{equation}
    E^\nu(\bar q,\bar\ell)
    =
    \frac{1}{2}[s_1^\nu,F(\bar q,\bar\ell)G(\bar\ell)]
    -
    [s_1^\nu,F(\bar q,\bar\ell)]G(\bar\ell),
    \label{eq:Bcut-extra-definition}
\end{equation}
with $F=\mathcal A^{(0)*}(\bar q,\bar\ell)$ and $G=\mathcal A^{(0)}(\bar\ell)$. This is the same complementary term written in \eqref{eq:all-spin-extra-mismatch}. Keeping it is what turns a convenient symmetrized representative into the exact $X=\emptyset$ cut-box kernel.

Finally, the cut-box/Born-iteration identification explains why the quadratic appendix below contains a push-forward of the LO spin kick. If $X^A$ denotes the external data on which the LO observable depends, the elastic second-Born map shifts this data by the LO impulse and LO spin map:
\begin{equation}
    X^A\longrightarrow X^A+\delta_{\mathrm{Born}}X^A .
    \label{eq:Bcut-data-shift}
\end{equation}
Expanding the LO spin kick on the shifted data gives
\begin{equation}
    \Delta s_{1,\mathrm{LO}}^\mu(X+\delta_{\mathrm{Born}}X)
    =
    \Delta s_{1,\mathrm{LO}}^\mu(X)
    +
    \delta_{\mathrm{Born}}X^A
    \frac{\partial}{\partial X^A}
    \Delta s_{1,\mathrm{LO}}^\mu(X)
    +\cdots .
    \label{eq:Bcut-push-forward-origin}
\end{equation}
Equation \eqref{eq:Bcut-push-forward-origin} is the source of the chain-rule operator used in Appendix~\ref{app:C}. The quantities $\delta_{\mathrm{Born}}X^A$ are derived from the LO KMOC impulse and spin kick; they are not fitted to the comparison with ref.~\cite{Alessio:2025flu}.

\section{Spin expansion of the NLO spin kick}
\label{app:C}
In this appendix we evaluate the NLO angular impulse by expanding it in powers of the spin vectors. Working order by order keeps the loop integrations tractable and separates the spin-orbit interactions from the higher spin-induced multipole effects. We proceed through the expansion one order at a time, isolating the gauge-invariant classical information at each order.
\subsection{Linear order in spin}
\label{app:B-linear-spin}

We begin our explicit derivation by isolating the linear-in-spin contributions to the NLO angular impulse. To systematically manage the complexity of the loop integrations, we organize the calculation into distinct topological sectors. Specifically, we partition the one-loop amplitude into its virtual box and triangle components, subsequently combining them with the exact conservative real tree-tree sector, namely the $X=\emptyset$ cut-box contribution derived in Appendix~\ref{app:born-cutbox}. This decomposition effectively isolates the infrared divergent pieces from the finite classical momentum transfers. Consequently, we must evaluate the following individual terms
\begin{equation}
    \text{Term 1: }    i \epsilon^{\mu\nu\rho\sigma} u_{1\nu}s_{1\rho}
    \frac{\partial}{\partial s^\sigma_1} \big( {\mathcal M}_b(\bar{q}) \big)
\end{equation}
\begin{equation}
    \text{Term 2: }   i \epsilon^{\mu\nu\rho\sigma} u_{1\nu}s_{1\rho}
    \frac{\partial}{\partial s^\sigma_1} \big(  {\mathcal M}_t(\bar{q}) \big)
\end{equation}
\begin{equation}
    \text{Term 3: }   - \frac{1}{m_1} u_{1}^{\mu} \bar{q}_{\nu} s^{\nu}_{1} 
    \big( {\mathcal M}_b(\bar{q})  \big)
\end{equation}
\begin{equation}
    \text{Term 4: }   - \frac{1}{m_1} u_{1}^{\mu} \bar{q}_{\nu} s^{\nu}_{1} 
    \big(  {\mathcal M}_t(\bar{q}) \big)
\end{equation}
\begin{equation}
    \text{Term 5: }  
    i \epsilon^{\mu\nu\rho\sigma} u_{1\nu}s_{1\rho}
    \frac{\partial}{\partial s^\sigma_1} {\mathcal M}_c(\bar{q})
    +
    \frac{i}{2 m_1} u_{1}^{\mu}
    \epsilon^{\nu\tau\rho\sigma} u_{1\tau}s_{1\rho}
    \frac{\partial}{\partial s^\sigma_1} N_{\nu}(\bar{q})
\end{equation}
\begin{equation}
    \text{Term 6: }   - \frac{1}{m_1} u_{1}^{\mu}  s^{\nu}_{1} {\mathcal M}_{\nu}(\bar{q}) 
\end{equation}
\begin{equation}
    \text{Term 7: }  
    \Delta\mathscr M^{\mu(1)}(\bar q)
\end{equation}

\subsubsection{Term 1}

We first direct our attention to the evaluation of Term $1$. The core of this calculation relies on the linear-in-spin expansion of the product $B_h(\bar\ell) B_h(-\bar\ell+\bar q)$. This quantity encodes the leading gravitational coupling to the spinning body. We begin by recalling the exact expression for the effective vertex function $B_h(q)$
\begin{equation}
    B_h(q) =
    \frac{1}{4}
    P_{\mu\nu\alpha\beta} 
    u_1^\mu u_1^\nu
    \left[
    \exp{\left(-i (a_1+a_2)*q\right)}
    \right]^{(\alpha}{}_{\lambda}
    u_2^{\beta)} u_2^\lambda,
\end{equation}
with the standard cross-product notation defined as
\begin{equation}
    (a*q)^\mu{}_\nu = \epsilon^\mu{}_{\nu\rho\sigma} a^\rho q^\sigma.
\end{equation}
To isolate the dipole interactions, we perform a Taylor expansion of the exponential operator up to linear order in the spin vectors. This yields
\begin{equation}
    \exp{\left(-i (a_1+a_2)*q\right)}
    =
    \mathbf{1}
    - i (a_1+a_2)*q
    + \mathcal{O}(a^2).
\end{equation}
Thus, its explicit tensor representation can be written as
\begin{equation}
    \left[\exp{(-i (a_1+a_2)*q)}\right]^{\alpha}{}_{\lambda}
    =
    \delta^\alpha_{\lambda}
    - i  \epsilon^\alpha{}_{\lambda\rho\sigma}
    (a_1+a_2)^\rho q^\sigma
    + \mathcal{O}(a^2)
\end{equation}
which immediately leads to the symmetrized structure
\begin{equation}
    \left[\exp{-i (a_1+a_2)*q}\right]^{(\alpha}{}_{\lambda}
    u_2^{\beta)} u_2^\lambda
    =
    u_2^\alpha u_2^\beta
    - \frac{i}{2}
    \Big(
    \epsilon^\alpha{}_{\lambda\rho\sigma}
    u_2^\beta
    +
    \epsilon^\beta{}_{\lambda\rho\sigma}
    u_2^\alpha
    \Big)
    u_2^\lambda
    (a_1+a_2)^\rho q^\sigma.
\end{equation}
By substituting this back into the vertex definition, we have
\begin{equation}
    B_h(q)
    =
    \frac{1}{4}
    P_{\mu\nu\alpha\beta} 
    u_1^\mu u_1^\nu
    u_2^\alpha u_2^\beta
    - \frac{i}{8}
    P_{\mu\nu\alpha\beta} 
    u_1^\mu u_1^\nu
    \Big(
    \epsilon^\alpha{}_{\lambda\rho\sigma}
    u_2^\beta
    +
    \epsilon^\beta{}_{\lambda\rho\sigma}
    u_2^\alpha
    \Big)
    u_2^\lambda
    (a_1+a_2)^\rho q^\sigma
    + \mathcal{O}(a^2).
\end{equation}
We define the purely scalar contribution as
\begin{equation}
    B_h^{(0)} =
    \frac{1}{4}
    P_{\mu\nu\alpha\beta} 
    u_1^\mu u_1^\nu
    u_2^\alpha u_2^\beta,
\end{equation}
and the linear-in-spin part as
\begin{equation}
    B_h^{(1)}(q)
    =
    - \frac{i}{8}
    P_{\mu\nu\alpha\beta} 
    u_1^\mu u_1^\nu
    \Big(
    \epsilon^\alpha{}_{\lambda\rho\sigma}
    u_2^\beta
    +
    \epsilon^\beta{}_{\lambda\rho\sigma}
    u_2^\alpha
    \Big)
    u_2^\lambda
    (a_1+a_2)^\rho q^\sigma.
\end{equation}
Expanding the product of the two vertices gives
\begin{equation}
    B_h(\bar\ell) B_h(-\bar\ell+\bar q)
    =
    B_h^{(0)} B_h^{(0)}
    +
    B_h^{(0)} B_h^{(1)}(-\bar\ell+\bar q)
    +
    B_h^{(1)}(\bar\ell) B_h^{(0)}
    + \mathcal{O}(a^2).
\end{equation}
Thus, the relevant linear-in-spin part yields to
\begin{equation}
    B_h(\bar\ell) B_h(-\bar\ell+\bar q)\Big|_{\text{linear}}
    =
    B_h^{(0)}
    \Big[
    B_h^{(1)}(\bar\ell)
    +
    B_h^{(1)}(-\bar\ell+\bar q)
    \Big].
\end{equation}
Since
\begin{equation}
    \begin{aligned}
        B_h^{(1)}(\bar\ell)
        &\propto
        (a_1+a_2)^\rho \bar\ell^\sigma,
        \\
        B_h^{(1)}(-\bar\ell+\bar q)
        &\propto
        (a_1+a_2)^\rho (-\bar\ell+\bar q)^\sigma
    \end{aligned}
\end{equation}
we deduce that the internal loop momentum $\bar\ell$ cancels out, leaving
\begin{equation}
    B_h^{(1)}(\bar\ell)
    +
    B_h^{(1)}(-\bar\ell+\bar q)
    \propto
    (a_1+a_2)^\rho \bar q^\sigma.
\end{equation}
Therefore, the total linear-in-spin contribution simplifies to
\begin{equation}
    B_h(\bar\ell) B_h(-\bar\ell+\bar q)\Big|_{\text{linear}}
    =
    - i  B_h^{(0)}  \mathcal{C}_{\gamma\delta} 
    (a_1+a_2)^\gamma \bar q^\delta,
\end{equation}
where we have defined the tensor
\begin{equation}
    \mathcal{C}_{\rho\sigma}
    =
    \frac{1}{8}
    P_{\mu\nu\alpha\beta} 
    u_1^\mu u_1^\nu
    \Big(
    \epsilon^\alpha{}_{\lambda\rho\sigma} u_2^\beta
    +
    \epsilon^\beta{}_{\lambda\rho\sigma} u_2^\alpha
    \Big)
    u_2^\lambda.
\end{equation}
Taking the relevant derivative with respect to the spin tensor yields
\begin{equation}
    \begin{aligned}
        i \epsilon^{\mu\nu\rho\sigma} u_{1\nu}s_{1\rho}
        \frac{\partial}{\partial s^\sigma_1} B_h(\bar\ell) B_h(-\bar\ell+\bar q)\Big|_{\text{linear}}
        =& B_h^{(0)}  \mathcal{C}_{\sigma\delta}
        \epsilon^{\mu\nu\rho\sigma} u_{1\nu} a_{1\rho} \bar q^\delta
    \end{aligned}
\end{equation}
This allows us to express the first term as a sum over specific loop integrals
\begin{equation}
    \begin{aligned}
        \text{Term 1} =& B_h^{(0)}  \mathcal{C}_{\sigma\delta}
        \epsilon^{\mu\nu\rho\sigma} u_{1\nu} a_{1\rho} \bar q^\delta
        ( I_1 + I_2 + I_3 )
        \\
        I_1 =&
        \frac{1}{2 m_1} 
        \int \hat{d}^4 \bar{\ell}  
        \frac{ \bar{\ell} \cdot (\bar{\ell} - \bar{q}) \hat{\delta}(\bar{\ell} \cdot u_2) }
        {\bar{\ell}^2 (\bar{\ell} - \bar{q})^2 
        ( \bar{\ell} \cdot u_1  + i\varepsilon )^2 }
        \\
        I_2 =& I_1(1 \leftrightarrow 2)
        \\
        I_3 =&
        \frac{i}{4 m_1 m_2} 
        \int \hat{d}^4 \bar{\ell}  
        ( 2 \bar{\ell} \cdot \bar{q} - \bar{\ell}^2 )
        \frac{1}{\bar{\ell}^2 (\bar{\ell} - \bar{q})^2}
        \Big(
        m_2 \hat{\delta}^{\prime}( \bar{\ell} \cdot u_1) 
        \hat{\delta}( \bar{\ell} \cdot u_2 ) 
        - m_1 \hat{\delta}( \bar{\ell} \cdot u_1) 
        \hat{\delta}^{\prime}( \bar{\ell} \cdot u_2 )
        \Big) .
    \end{aligned}
\end{equation}
The full physical contribution from Term $1$ is therefore given by
\begin{equation}
    \begin{aligned}
        \frac{i \kappa^4 m_1^2 m_2^2}{4} 
        \int  \hat{d}^4 \bar{q}   \hat{\delta}(\bar{q} \cdot u_1)
        \hat{\delta}(\bar{q} \cdot u_2) e^{- i b \cdot \bar{q}} \Big( \text{Term 1} \Big)
    \end{aligned}
\end{equation}
As established in ref.~\cite{Menezes:2022tcs}, the integrals $I_1$ and $I_2$ evaluate to zero due to the parity properties of the integrand under the support of the on-shell delta functions. To evaluate $I_3$, however, a more sophisticated approach is required. We can leverage the distributional nature of the delta function derivatives. Let us define a kernel function
\begin{equation}
    \begin{aligned}
        F(\bar{\ell},\bar{q}) \equiv 
        \frac{2 \bar{\ell}\cdot \bar{q} - \bar{\ell}^{2}}
        {\bar{\ell}^{2} (\bar{\ell}-\bar{q})^{2}},
    \end{aligned}
\end{equation}
such that the integral can be rewritten as
\begin{equation}
    \begin{aligned}
        I_3 =&
        \frac{i}{4 m_1 m_2} 
        \int \hat{d}^4 \bar{\ell}  
        F(\bar{\ell},\bar{q})
        \Big(
        m_2 \hat{\delta}^{\prime}( \bar{\ell} \cdot u_1) 
        \hat{\delta}( \bar{\ell} \cdot u_2 ) 
        - m_1 \hat{\delta}( \bar{\ell} \cdot u_1) 
        \hat{\delta}^{\prime}( \bar{\ell} \cdot u_2 )
        \Big) .
    \end{aligned}
\end{equation}
To effectively perform the integration by parts and exploit the kinematic constraints, we construct a specialized basis of orthogonal four-velocities. We introduce the auxiliary vectors
\begin{equation}
    \begin{aligned}
        \tilde{u}_1^\mu =& \frac{u_1^\mu - \gamma u_2^\mu}{\gamma^2-1},
        \\
        \tilde{u}_2^\mu =& \frac{u_2^\mu - \gamma u_1^\mu}{\gamma^2-1},
    \end{aligned}
\end{equation}
which, by construction, satisfy the orthogonality relations
\begin{equation}
    \begin{aligned}
        \tilde{u}_1\cdot u_1 = -1, \qquad \tilde{u}_1\cdot u_2 = 0,
        \\
        \tilde{u}_2\cdot u_2 = -1, \qquad \tilde{u}_2\cdot u_1 = 0.
    \end{aligned}
\end{equation}
Because of these relations, the derivatives of the delta functions can be mapped to derivatives along this new basis
\begin{equation}
    \begin{aligned}
        \hat{\delta}^{\prime}(\bar{\ell}\cdot u_1) =&
        - \tilde{u}_1\cdot \frac{\partial}{\partial \bar{\ell}}
         \hat{\delta}(\bar{\ell}\cdot u_1),
        \\
        \hat{\delta}^{\prime}(\bar{\ell}\cdot u_2) =&
        - \tilde{u}_2\cdot \frac{\partial}{\partial \bar{\ell}}
         \hat{\delta}(\bar{\ell}\cdot u_2).
    \end{aligned}
\end{equation}
Substituting this into the first term of $I_3$ and integrating by parts yields
\begin{equation}
    \begin{aligned}
        \int \hat{d}^4 \bar{\ell} 
        F(\bar{\ell},\bar{q})
        \hat{\delta}^{\prime}( \bar{\ell} \cdot u_1) 
        \hat{\delta}( \bar{\ell} \cdot u_2 )
        =&
        - \int \hat{d}^4 \bar{\ell} 
        F(\bar{\ell},\bar{q})
        \left(
        \tilde{u}_1\cdot \frac{\partial}{\partial \bar{\ell}}
        \hat{\delta}(\bar{\ell}\cdot u_1)
        \right)
        \hat{\delta}( \bar{\ell} \cdot u_2 )
        \\
        =&
        \int \hat{d}^4 \bar{\ell} 
        \hat{\delta}( \bar{\ell} \cdot u_1) 
        \hat{\delta}( \bar{\ell} \cdot u_2 )
        \left(
        \tilde{u}_1\cdot \frac{\partial}{\partial \bar{\ell}}
        F(\bar{\ell},\bar{q})
        \right),
    \end{aligned}
\end{equation}
where the boundary terms vanish, and we utilized the identity
\begin{equation}
    \begin{aligned}
        \tilde{u}_1\cdot \frac{\partial}{\partial \bar{\ell}}
        \hat{\delta}(\bar{\ell}\cdot u_2)
        =
        (\tilde{u}_1\cdot u_2)\hat{\delta}^{\prime}(\bar{\ell}\cdot u_2)=0.
    \end{aligned}
\end{equation}
Similarly, applying the exact same procedure to the second term provides
\begin{equation}
    \begin{aligned}
        &
        \int \hat{d}^4 \bar{\ell} 
        F(\bar{\ell},\bar{q})
        \hat{\delta}( \bar{\ell} \cdot u_1) 
        \hat{\delta}^{\prime}( \bar{\ell} \cdot u_2 ) =
        \int \hat{d}^4 \bar{\ell} 
        \hat{\delta}( \bar{\ell} \cdot u_1) 
        \hat{\delta}( \bar{\ell} \cdot u_2 )
        \left(
        \tilde{u}_2\cdot \frac{\partial}{\partial \bar{\ell}}
        F(\bar{\ell},\bar{q})
        \right).
    \end{aligned}
\end{equation}
Hence, the full integral $I_3$ becomes
\begin{equation}
    \begin{aligned}
        I_3 =&
        \frac{i}{4 m_1 m_2} 
        \int \hat{d}^4 \bar{\ell} 
        \hat{\delta}( \bar{\ell} \cdot u_1) 
        \hat{\delta}( \bar{\ell} \cdot u_2 )\times
        \left(
        m_2 \tilde{u}_1\cdot \frac{\partial}{\partial \bar{\ell}}
        -
        m_1 \tilde{u}_2\cdot \frac{\partial}{\partial \bar{\ell}}
        \right)
        F(\bar{\ell},\bar{q}) .
    \end{aligned}
\end{equation}
Thus, on the support of the delta functions, the following condition holds
\begin{equation}
    \begin{aligned}
        \bar{\ell}\cdot u_1 = \bar{\ell}\cdot u_2 = 0
        \qquad \implies \qquad
        \bar{\ell}\cdot \tilde{u}_1 = \bar{\ell}\cdot \tilde{u}_2 = 0.
    \end{aligned}
\end{equation}
Moreover, the full physical expression for the observable is constrained by the outer support $\hat{\delta}(\bar{q}\cdot u_1)\hat{\delta}(\bar{q}\cdot u_2)$, ensuring that
\begin{equation}
    \begin{aligned}
        \bar{q}\cdot u_1 = \bar{q}\cdot u_2 = 0
        \qquad \implies \qquad
        \bar{q}\cdot \tilde{u}_1 = \bar{q}\cdot \tilde{u}_2 = 0.
    \end{aligned}
\end{equation}
Combining these constraints implies
\begin{equation}
    \begin{aligned}
        (\bar{\ell}-\bar{q})\cdot \tilde{u}_1 = 0,
        \qquad
        (\bar{\ell}-\bar{q})\cdot \tilde{u}_2 = 0.
    \end{aligned}
\end{equation}
Since the kernel $F(\bar{\ell},\bar{q})$ depends on $\bar{\ell}$ exclusively through the Lorentz scalars $\bar{\ell}^{2}$, $(\bar{\ell}-\bar{q})^{2}$, and $\bar{\ell}\cdot \bar{q}$, its derivatives along the directions $\tilde{u}_1$ and $\tilde{u}_2$ vanish on the support. We can explicitly verify this
\begin{equation}
    \begin{aligned}
        \tilde{u}_i\cdot \frac{\partial}{\partial \bar{\ell}} \bar{\ell}^{2}
        =& 2 \bar{\ell}\cdot \tilde{u}_i = 0,
        \\
        \tilde{u}_i\cdot \frac{\partial}{\partial \bar{\ell}} (\bar{\ell}-\bar{q})^{2}
        =& 2 (\bar{\ell}-\bar{q})\cdot \tilde{u}_i = 0,
        \\
        \tilde{u}_i\cdot \frac{\partial}{\partial \bar{\ell}} (\bar{\ell}\cdot \bar{q})
        =& \bar{q}\cdot \tilde{u}_i = 0,
    \end{aligned}
\end{equation}
for $i=1,2$. Hence,
\begin{equation}
    \begin{aligned}
        \tilde{u}_1\cdot \frac{\partial}{\partial \bar{\ell}} F(\bar{\ell},\bar{q}) = 0,
        \qquad
        \tilde{u}_2\cdot \frac{\partial}{\partial \bar{\ell}} F(\bar{\ell},\bar{q}) = 0,
    \end{aligned}
\end{equation}
which implies that the third integral vanishes
\begin{equation}
    I_3 = 0.
\end{equation}
Combining this conclusion with the earlier results from ref.~\cite{Menezes:2022tcs}, we conclusively find that all integral components of this term are identically zero
\begin{equation}
    I_1 = I_2 = I_3 = 0.
\end{equation}

\subsubsection{Term 2}
Turning our attention to the triangle contribution, we must extract the specific components that will actively participate in the dynamics. We express the amplitude up to linear order in the relevant variables as follows
\begin{equation}
    \begin{aligned}
        {\mathcal M}_t(\bar{q}) \bigg|_{\begin{smallmatrix}\text{up to}\\\text{linear order}\end{smallmatrix}}   =&  \frac{1}{256 m_2}
        \frac{1}{ \gamma^2 - 1}
        \frac{1}{ \sqrt{-\bar{q}^2}}  
        \oint_{\Gamma_{\infty}} \frac{dz}{2 \pi i}
        \frac{ (  \gamma -  z \sqrt{ \gamma^2 - 1} )^4}{(z^2-1)^{3/2}}
        ( 1 + z   \bar{q} \cdot a_1 )
        \\
        &+ \frac{1}{256 m_1}
        \frac{1}{ \gamma^2 - 1}
        \frac{1}{ \sqrt{-\bar{q}^2}}  
        \oint_{\Gamma_{\infty}} \frac{dz}{2 \pi i}
        \frac{ (  \gamma -  z \sqrt{ \gamma^2 - 1} )^4}{(z^2-1)^{3/2}}
        ( 1 + \bar{w}_1(z) - \bar{x}_1(z)  )
        \\
        \bar{w}_1(z) =&  \frac{ \left(z^2-1\right)  \sqrt{  \gamma^2 - 1}  }
        {  \gamma-  z \sqrt{  \gamma^2 - 1} }     a_1\cdot \bar{q},
        \\
        \bar{x}_1(z)  =& - z    a_1 \cdot \bar{q}
    \end{aligned}
\end{equation}
where we have retained only these terms, anticipating that the spin derivative with respect to $s_1$ acts non-trivially upon them. This formulation naturally introduces the following contour integrals over the complex parameter $z$
\begin{equation}
    \begin{aligned}
        J_1 =&  \oint_{\Gamma_{\infty}} \frac{dz}{2 \pi i}
        \frac{ (  \gamma -  z \sqrt{ \gamma^2 - 1} )^4}{(z^2-1)^{3/2}} z
        \\
        J_2 =&  \oint_{\Gamma_{\infty}} \frac{dz}{2 \pi i}
        \frac{ (  \gamma -  z \sqrt{ \gamma^2 - 1} )^4}{(z^2-1)^{3/2}}
        \frac{ \left(z^2-1\right)  \sqrt{  \gamma^2 - 1}  }
        {  \gamma-  z \sqrt{  \gamma^2 - 1} }
    \end{aligned}
\end{equation}
We evaluate the contour integrals $J_{1,2}$ by defining $\Gamma_{\infty}$ such that it encloses only the contribution from the physical pole situated at infinity ($z=\infty$). It is also convenient to define an auxiliary integral
\begin{equation}
    \begin{aligned}
        J_3 =& \oint_{\Gamma_{\infty}} \frac{dz}{2 \pi i}
        \frac{ (  \gamma -  z \sqrt{ \gamma^2 - 1} )^4}{(z^2-1)^{3/2}},
    \end{aligned}
\end{equation}
which will prove useful in our subsequent algebraic simplifications. To simplify the notation, we define
\begin{equation}
    s \equiv \sqrt{\gamma^2-1} = \gamma \beta,
\end{equation}
To extract the residues at infinity, we expand the integrands using the standard Laurent series representations for large $z$
\begin{equation}
    \begin{aligned}
        \sqrt{z^2-1} =& z \sqrt{1-\frac{1}{z^2}},
        \\
        \frac{1}{(z^2-1)^{3/2}} =& \frac{1}{z^3}\left(1+\frac{3}{2z^2}+{\mathcal O}(z^{-4})\right),
        \\
        \frac{1}{\sqrt{z^2-1}} =& \frac{1}{z}\left(1+\frac{1}{2z^2}+{\mathcal O}(z^{-4})\right),
    \end{aligned}
\end{equation}
The residue theorem for a contour at infinity dictates that the integral evaluates to
\begin{equation}
    \oint_{\Gamma_{\infty}} \frac{dz}{2\pi i}  f(z) = - [z^{-1}]  f(z),
\end{equation}
where $[z^{-1}] f(z)$ denotes the coefficient of the $1/z$ term in the Laurent expansion of $f(z)$ around $z=\infty$.

Evaluating the integrand for $J_3$, we find
\begin{equation}
    \begin{aligned}
        \frac{(\gamma-zs)^4}{(z^2-1)^{3/2}}
        =&
        \left(
        s^4 z^4 - 4 \gamma s^3 z^3 + 6 \gamma^2 s^2 z^2 - 4 \gamma^3 s z + \gamma^4
        \right)
        \frac{1}{z^3}\left(1+\frac{3}{2z^2}+{\mathcal O}(z^{-4})\right)
        \\
        =&
        s^4 z - 4 \gamma s^3 + \frac{1}{z}\left( 6 \gamma^2 s^2 + \frac{3}{2}s^4 \right)
        + {\mathcal O}(z^{-2}) .
    \end{aligned}
\end{equation}
Isolating the relevant coefficient, we obtain
\begin{equation}
    \begin{aligned}
        J_3 =& - \left( 6 \gamma^2 s^2 + \frac{3}{2}s^4 \right)
        \\
        =&
        -\frac{3}{2}(\gamma^2-1)(5\gamma^2-1).
    \end{aligned}
\end{equation}
By a similar procedure for $J_1$, the extra factor of $z$ modifies the expansion to
\begin{equation}
    \begin{aligned}
        z \frac{(\gamma-zs)^4}{(z^2-1)^{3/2}}
        =&
        \left(
        s^4 z^2 - 4 \gamma s^3 z + 6 \gamma^2 s^2 - \frac{4 \gamma^3 s}{z} + {\mathcal O}(z^{-2})
        \right)
        \left(1+\frac{3}{2z^2}+{\mathcal O}(z^{-4})\right)
        \\
        =&
        s^4 z^2 - 4 \gamma s^3 z + 6 \gamma^2 s^2
        -\frac{1}{z}\left( 4 \gamma^3 s + 6 \gamma s^3 \right)
        + {\mathcal O}(z^{-2}) .
    \end{aligned}
\end{equation}
Therefore, extracting the residue yields
\begin{equation}
    \begin{aligned}
        J_1 =& 4 \gamma^3 s + 6 \gamma s^3
        \\
        =&
        2 \gamma \sqrt{\gamma^2-1} (5\gamma^2-3).
    \end{aligned}
\end{equation}
To evaluate $J_2$, rather than performing a direct and tedious Laurent expansion, we exploit a total derivative identity. We see that
\begin{equation}
    \begin{aligned}
        \frac{d}{dz}\left(
        \frac{(\gamma-zs)^4}{\sqrt{z^2-1}}
        \right)
        =&
        - z \frac{(\gamma-zs)^4}{(z^2-1)^{3/2}}
        - 4s \frac{(\gamma-zs)^3}{\sqrt{z^2-1}} .
    \end{aligned}
\end{equation}
Since the integral of a total derivative along a closed contour identically vanishes, we establish a relationship between the integrals
\begin{equation}
    \begin{aligned}
        0 =&
        \oint_{\Gamma_{\infty}} \frac{dz}{2\pi i} 
        \frac{d}{dz}\left(
        \frac{(\gamma-zs)^4}{\sqrt{z^2-1}}
        \right)
        \\
        =&
        - J_1 - 4 J_2 .
    \end{aligned}
\end{equation}
This allows us to solve directly for $J_2$
\begin{equation}
    \begin{aligned}
        J_2 = -\frac{J_1}{4}
        = - \frac{1}{2}\gamma \sqrt{\gamma^2-1} (5\gamma^2-3).
    \end{aligned}
\end{equation}
Collecting these contour integral results, we have
\begin{equation}
    \begin{aligned}
        J_1 =& 2 \gamma \sqrt{\gamma^2-1} (5\gamma^2-3),
        \\
        J_2 =& - \frac{1}{2}\gamma \sqrt{\gamma^2-1} (5\gamma^2-3),
        \\
        J_3 =& -\frac{3}{2}(\gamma^2-1)(5\gamma^2-1).
    \end{aligned}
\end{equation}
In particular, the linear combination required for our amplitude evaluates to
\begin{equation}
    J_1+J_2 = \frac{3}{2}\gamma \sqrt{\gamma^2-1} (5\gamma^2-3).
\end{equation}
With the integration performed, we now focus on evaluating the physical observable specified in Term $2$
\begin{equation}
    {\text{Term}} 2
    =
    i\epsilon^{\mu\nu\rho\sigma}u_{1\nu}s_{1\rho}
    \frac{\partial}{\partial s_1^\sigma}
    {\mathcal M}_t(\bar q).
\end{equation}
An important point in this derivation is that, within the triangle sector, the spin-linear factor $\bar q\cdot a_1$ must be understood through its on-shell representative. Specifically, on the support of the triangle kinematics, the following identity holds
\begin{equation}
    i\epsilon_{\mu\nu\rho\sigma}
    p_1^\mu p_2^\nu \bar q^\rho a_1^\sigma
    =
    m_1 m_2 \sqrt{\gamma^2-1} 
    (\bar q\cdot a_1).
\end{equation}
Substituting the velocity definitions $p_a^\mu=m_1u_1^\mu$ and $p_b^\mu=m_2u_2^\mu$, this gives the expression
\begin{equation}
    \bar q\cdot a_1
    =
    \frac{i}{\sqrt{\gamma^2-1}} 
    \epsilon(u_1,u_2,\bar q,a_1)
    \label{eq:B2-onshell-qdot-a}
\end{equation}
up to the global convention for the orientation of the epsilon tensor.

The sequence of mathematical operations is non-commutative in this context. One must first substitute the on-shell equality \eqref{eq:B2-onshell-qdot-a}, and only afterwards compute the spin derivative. If one erroneously differentiates the representative $\bar q\cdot a_1$ directly, one obtains the dual structure $i\epsilon^{\mu\nu\rho\sigma}u_{1\nu}a_{1\rho}\bar q_\sigma$, which fails to project onto the correct in-plane basis appropriate for describing the final conservative observable.

Proceeding carefully, the linear-in-$a_1$ part of the triangle extraction yields
\begin{equation}
    {\mathcal M}_t(\bar q)\Big|_{a_1}
     \longrightarrow 
    \frac{\gamma(5\gamma^2-3)}{256\sqrt{\gamma^2-1}}
    \left(
    \frac{2}{m_2}
    +
    \frac{3}{2m_1}
    \right)
    \frac{\bar q\cdot a_1}{\sqrt{-\bar q^2}} .
\end{equation}
Using the on-shell condition \eqref{eq:B2-onshell-qdot-a}, this expression transforms into
\begin{equation}
    {\mathcal M}_t(\bar q)\Big|_{a_1}
     \longrightarrow 
    \frac{i\gamma(5\gamma^2-3)}{256(\gamma^2-1)}
    \left(
    \frac{2}{m_2}
    +
    \frac{3}{2m_1}
    \right)
    \frac{\epsilon(u_1,u_2,\bar q,a_1)}{\sqrt{-\bar q^2}} .
    \label{eq:B2-Mt-linear-onshell}
\end{equation}
We are now positioned to act with the spin commutator. Noting the rescaling $a_1^\mu=s_1^\mu/m_1$, we perform the differentiation
\begin{equation}
    \begin{aligned}
        &i\epsilon^{\mu\nu\rho\sigma}u_{1\nu}s_{1\rho}
        \frac{\partial}{\partial s_1^\sigma}
        \left(\bar q\cdot a_1\right)_{\text{on\ shell}}
        \\
        &\qquad =
        i\epsilon^{\mu\nu\rho\sigma}u_{1\nu}s_{1\rho}
        \frac{\partial}{\partial s_1^\sigma}
        \left[
        \frac{i}{\sqrt{\gamma^2-1}}
        \epsilon(u_1,u_2,\bar q,a_1)
        \right]
        \\
        &\qquad =
        \frac{1}{\sqrt{\gamma^2-1}}
        \left[
        \bar q^\mu(u_2\cdot a_1)
        -
        u_2^\mu(\bar q\cdot a_1)
        +
        \gamma u_1^\mu(\bar q\cdot a_1)
        \right].
        \label{eq:B2-spin-derivative-onshell}
    \end{aligned}
\end{equation}
Therefore, the complete expression in momentum space becomes
\begin{align}
    {\text{Term}} 2
    =&
    \frac{\gamma(5\gamma^2-3)}{256(\gamma^2-1)}
    \left(
    \frac{2}{m_2}
    +
    \frac{3}{2m_1}
    \right)
    \\
    &\times
    \frac{
    \bar q^\mu(u_2\cdot a_1)
    -
    u_2^\mu(\bar q\cdot a_1)
    +
    \gamma u_1^\mu(\bar q\cdot a_1)
    }
    {\sqrt{-\bar q^2}} .
    \label{eq:B2-term2-momentum-space}
\end{align}
Finally, we transform this result back to impact parameter space by evaluating the integral
\begin{equation}
    i\frac{\kappa^4m_1^2m_2^2}{4}
    \int \hat d^4\bar q 
    \hat\delta(\bar q\cdot u_1)
    \hat\delta(\bar q\cdot u_2)
    e^{-ib\cdot\bar q}
    \left({\text{Term}} 2\right).
\end{equation}
The relevant master Fourier transform connecting momentum space to the impact parameter $b$ is given by
\begin{equation}
    \int \hat d^4\bar q 
    \hat\delta(\bar q\cdot u_1)
    \hat\delta(\bar q\cdot u_2)
    e^{-ib\cdot\bar q}
    \frac{\bar q^\alpha}{\sqrt{-\bar q^2}}
    =
    \frac{i}{2\pi\sqrt{\gamma^2-1}}
    \frac{b^\alpha}{b^3}.
    \label{eq:B2-master-q-over-sqrt}
\end{equation}
Using \eqref{eq:B2-master-q-over-sqrt}, we obtain
\begin{equation}
    \begin{aligned}
        &i\frac{\kappa^4m_1^2m_2^2}{4}
        \int \hat d^4\bar q 
        \hat\delta(\bar q\cdot u_1)
        \hat\delta(\bar q\cdot u_2)
        e^{-ib\cdot\bar q}
        \left({\text{Term}} 2\right)
        \\
        &\qquad =
        \frac{\kappa^4m_1m_2(4m_1+3m_2)}{4096\pi}
        \frac{\gamma(5\gamma^2-3)}{(\gamma^2-1)^{3/2}}
        \\
        &\qquad\quad \times
        \frac{
        b^\mu(u_2\cdot a_1)
        -
        u_2^\mu(b\cdot a_1)
        +
        \gamma u_1^\mu(b\cdot a_1)
        }{b^3}.
    \end{aligned}
\end{equation}
Equivalently, expressing the final result in terms of the proper spin variable $s_1^\mu=m_1a_1^\mu$, the variation gives
\begin{equation}
    \begin{aligned}
        \Delta s^\mu_{(2)}
        = 
        \frac{\kappa^4m_2(4m_1+3m_2)}{4096\pi}
        \frac{\gamma(5\gamma^2-3)}{(\gamma^2-1)^{3/2}}
        \frac{
        b^\mu(u_2\cdot s_1)
        -
        u_2^\mu(b\cdot s_1)
        +
        \gamma u_1^\mu(b\cdot s_1)
        }{b^3}.
        \label{eq:B2-term2-final-onshell}
    \end{aligned}
\end{equation}
This derived form resides in the physical in-plane basis defined by the set
\begin{equation}
    \{u_1^\mu,u_2^\mu,b^\mu\},
\end{equation}
and depends only on the well-defined scalar projections
\begin{equation}
    b\cdot s_1,
    \qquad
    u_2\cdot s_1.
\end{equation}

\subsubsection{Term 3}

Now looking at term $3$, we observe that this contribution is already explicitly linear in the spin variable $a_1$. Because the spin-dependent prefactor does not rely on the internal loop momentum $\bar{\ell}$, it factors cleanly out of the virtual integration. The expression simplifies to
\begin{equation}
    \text{Term 3} = - B_h^{(0)} B_h^{(0)}
    u_{1}^{\mu} \bar{q}_{\nu} a^{\nu}_{1} 
    ( I_1 + I_2 + I_3 )
\end{equation}
To obtain the corresponding physical observable in impact parameter space, we must evaluate the Fourier integral over the momentum transfer $\bar{q}$
\begin{equation}
    \begin{aligned}
        \frac{i \kappa^4 m_1^2 m_2^2}{4} 
        \int  \hat{d}^4 \bar{q}   \hat{\delta}(\bar{q} \cdot u_1)
        \hat{\delta}(\bar{q} \cdot u_2) e^{- i b \cdot \bar{q}} \Big( \text{Term 3} \Big)
    \end{aligned}
\end{equation}
However, from our analysis of term $1$, we recall that the scalar loop integrals $I_1$, $I_2$, and $I_3$ vanish identically due to the exact parity properties and kinematic constraints imposed by the on-shell delta functions. Thus, this integral yields a vanishing contribution

\subsubsection{Term 4}

Similarly, the contribution arising from term $4$ is intrinsically linear in the spin. By extracting the relevant kinematic components from the triangle sector and isolating the integration over the loop variables, we can write
\begin{equation}
    \text{Term 4} = 
    - u_{1}^{\mu} \bar{q}_{\nu} a^{\nu}_{1} \left[ \frac{1}{256 m_2}
    \frac{1}{ \gamma^2 - 1}
    + \frac{1}{256 m_1}
    \frac{1}{ \gamma^2 - 1}
    \right] \frac{J_3}{ \sqrt{-\bar{q}^2}} . 
\end{equation}
Substituting the contour integral $J_3 = -\frac{3}{2}(\gamma^2-1)(5\gamma^2-1)$ derived in the previous section, this expression yields
\begin{equation}
    \text{Term 4}
    =
    \frac{3}{512}(5\gamma^2-1)
    \left(
    \frac{1}{m_1}+\frac{1}{m_2}
    \right)
    u_1^\mu \frac{\bar{q}\cdot a_1}{\sqrt{-\bar{q}^2}}.
\end{equation}
We now proceed to transform this result into impact parameter space by computing the integral
\begin{equation}
    \frac{i \kappa^4 m_1^2 m_2^2}{4} 
    \int  \hat{d}^4 \bar{q}   \hat{\delta}(\bar{q} \cdot u_1)
    \hat{\delta}(\bar{q} \cdot u_2) e^{- i b \cdot \bar{q}} \Big( \text{Term 4} \Big)
\end{equation}
To do so, we employ the master Fourier transform established in our earlier derivations
\begin{equation}
    \int \hat{d}^4 \bar{q} 
    \hat{\delta}( \bar{q} \cdot u_1 )\hat{\delta}( \bar{q} \cdot u_2 ) 
    e^{- i b \cdot \bar{q}}
    \frac{\bar{q}^{\nu}}{\sqrt{-\bar{q}^2}}
    =
    \frac{i}{2\pi \sinh\phi}\frac{b^{\nu}}{b^3}.
\end{equation}
Contracting this tensorial identity directly with the specific spin vector $a_{1\nu}$ yields
\begin{equation}
    \int \hat{d}^4 \bar{q} 
    \hat{\delta}( \bar{q} \cdot u_1 )\hat{\delta}( \bar{q} \cdot u_2 ) 
    e^{- i b \cdot \bar{q}}
    \frac{\bar{q}\cdot a_1}{\sqrt{-\bar{q}^2}}
    =
    \frac{i}{2\pi \sinh\phi}\frac{a_1\cdot b}{b^3}.
\end{equation}
Inserting this evaluated integral back into the full expression, we assemble the components
\begin{equation}
    \begin{aligned}
        \frac{i \kappa^4 m_1^2 m_2^2}{4} 
        \int  \hat{d}^4 \bar{q}   \hat{\delta}(\bar{q} \cdot u_1)
        \hat{\delta}(\bar{q} \cdot u_2) &e^{- i b \cdot \bar{q}} \Big( \text{Term 4} \Big)
        \\
        =&
        \frac{i \kappa^4 m_1^2 m_2^2}{4}
        \left[
        \frac{3}{512}(5\gamma^2-1)
        \left(
        \frac{1}{m_1}+\frac{1}{m_2}
        \right)
        u_1^\mu
        \right] 
        \frac{i}{2\pi \sinh\phi}\frac{a_1\cdot b}{b^3}
        \\
        =&
        - \frac{3 \kappa^4 m_1^2 m_2^2}{4096\pi}
        \frac{5\gamma^2-1}{\sqrt{\gamma^2-1}}
        \left(
        \frac{1}{m_1}+\frac{1}{m_2}
        \right)
        u_1^\mu \frac{a_1\cdot b}{b^3}.
    \end{aligned}
\end{equation}
Finally, we arrive at the final expression for this contribution
\begin{equation}
    \begin{aligned}
        \frac{i \kappa^4 m_1^2 m_2^2}{4} 
        \int  \hat{d}^4 \bar{q}   \hat{\delta}(\bar{q} \cdot u_1)
        \hat{\delta}(\bar{q} \cdot u_2) e^{- i b \cdot \bar{q}} \Big( \text{Term 4} \Big)
        =&
        - \frac{3 \kappa^4 m_1 m_2 (m_1+m_2)}{4096\pi} \\
        &\times\frac{5\gamma^2-1}{\sqrt{\gamma^2-1}}
        u_1^\mu \frac{a_1\cdot b}{b^3}.
    \end{aligned}
\end{equation}

\subsubsection{Term 6}

Focusing on term $6$, we observe that it is already explicitly linear in the spin. Thus, to isolate the overall linear-in-spin contribution of this term, we require only the spinless component of the kernel ${\mathcal M}_\nu(\bar q)$. By applying the leading-order vertex replacement $B_h(\bar\ell)B_h(-\bar\ell+\bar q)\to (B_h^{(0)})^2$, we deduce that the second term within ${\mathcal M}_\nu$ is proportional to the scalar integral $I_3$, which we have previously demonstrated to vanish identically. Furthermore, the remaining two terms reduce to purely spinless contact kernels, which vanish under the support of the external kinematic constraints imposed by $\hat\delta(\bar q\cdot u_1)\hat\delta(\bar q\cdot u_2)$. Hence, only the first term survives this, leaving us with
\begin{equation}
    {\mathcal M}_\nu^{(0)}(\bar q)
    =
    -\frac{i(B_h^{(0)})^2}{2m_1m_2}
    \int \hat d^4\bar\ell 
    \frac{\bar\ell\cdot(\bar\ell-\bar q) \bar\ell_\nu}{\bar\ell^2(\bar\ell-\bar q)^2}
    \Big(
    m_2\hat\delta'(\bar\ell\cdot u_1)\hat\delta(\bar\ell\cdot u_2)
    -
    m_1\hat\delta(\bar\ell\cdot u_1)\hat\delta'(\bar\ell\cdot u_2)
    \Big).
\end{equation}
To evaluate this expression, we apply the $n=1$ version of the integration lemma introduced previously. This enables us to directly map the derivatives of the delta functions onto the auxiliary orthogonal basis vectors
\begin{equation}
    \begin{aligned}
        \int \frac{\hat d^4\bar\ell}{\bar\ell^2(\bar\ell-\bar q)^2}
        \hat\delta'(\bar\ell\cdot u_1)\hat\delta(\bar\ell\cdot u_2) 
        \bar\ell\cdot(\bar\ell-\bar q) \bar\ell_\nu
        =&
        \tilde u_{1\nu} J(\bar q),
        \\
        \int \frac{\hat d^4\bar\ell}{\bar\ell^2(\bar\ell-\bar q)^2}
        \hat\delta(\bar\ell\cdot u_1)\hat\delta'(\bar\ell\cdot u_2) 
        \bar\ell\cdot(\bar\ell-\bar q) \bar\ell_\nu
        =&
        \tilde u_{2\nu} J(\bar q),
    \end{aligned}
\end{equation}
where we have defined the master scalar integral $J(\bar q)$ as
\begin{equation}
    J(\bar q)\equiv
    \int \hat d^4\bar\ell 
    \hat\delta(\bar\ell\cdot u_1)\hat\delta(\bar\ell\cdot u_2)
    \frac{\bar\ell\cdot(\bar\ell-\bar q)}{\bar\ell^2(\bar\ell-\bar q)^2}.
\end{equation}
Substituting these relations back into the expression for the kernel, we obtain the following form
\begin{equation}
    {\mathcal M}_\nu^{(0)}(\bar q)
    =
    -\frac{i(B_h^{(0)})^2}{2m_1m_2}
    \big(
    m_2\tilde u_{1\nu}-m_1\tilde u_{2\nu}
    \big) J(\bar q).
\end{equation}
Hence, by contracting this with the kinematic prefactor, the full expression for Term $6$ becomes
\begin{equation}
    \begin{aligned}
        \text{Term 6}
        =&
        -\frac{1}{m_1}u_1^\mu s_1^\nu {\mathcal M}_\nu^{(0)}(\bar q)
        \\
        =&
        \frac{i(B_h^{(0)})^2}{2m_1^2m_2}
        u_1^\mu s_1^\nu
        \big(
        m_2\tilde u_{1\nu}-m_1\tilde u_{2\nu}
        \big) J(\bar q).
    \end{aligned}
\end{equation}
To proceed, we incorporate the spin constraint $a_1\cdot u_1=0$ and the relation $s_1^\nu=m_1 a_1^\nu$. Recalling the explicit definitions of our orthogonal basis vectors
\begin{equation}
    \begin{aligned}
        \tilde u_1^\mu=\frac{u_1^\mu-\gamma u_2^\mu}{\gamma^2-1},
        \qquad
        \tilde u_2^\mu=\frac{u_2^\mu-\gamma u_1^\mu}{\gamma^2-1},
    \end{aligned}
\end{equation}
we compute the exact projections of the spin vector onto this basis
\begin{equation}
    \begin{aligned}
        a_1\cdot\tilde u_1
        =
        -\frac{\gamma}{\gamma^2-1}(a_1\cdot u_2),
        \qquad
        a_1\cdot\tilde u_2
        =
        \frac{1}{\gamma^2-1}(a_1\cdot u_2).
    \end{aligned}
\end{equation}
Inserting these projections directly into our expression significantly simplifies the scalar coefficients, yielding
\begin{equation}
    \text{Term 6}
    =
    -\frac{i(B_h^{(0)})^2}{2m_1m_2}
    \frac{\gamma m_2+m_1}{\gamma^2-1}
    u_1^\mu (a_1\cdot u_2) J(\bar q).
\end{equation}
Next, we must transform this momentum-space result into impact parameter space. This requires evaluating the full Fourier transform associated with the master scalar integral, which we define as
\begin{equation}
    {\mathcal J}(b)\equiv
    \int \hat d^4\bar q 
    \hat\delta(\bar q\cdot u_1)\hat\delta(\bar q\cdot u_2)
    e^{-ib\cdot\bar q} J(\bar q).
\end{equation}
To resolve this integral analytically, we employ an variable shift $\bar q\to \bar q+\bar\ell$. Utilizing the fact that $\bar\ell\cdot u_i=0$ on the support of the inner delta functions, the exponential factorizes. This operation decouples the integration, giving
\begin{equation}
    \begin{aligned}
        {\mathcal J}(b)
        =&
        -\int \hat d^4\bar q 
        \hat\delta(\bar q\cdot u_1)\hat\delta(\bar q\cdot u_2)
        \frac{e^{-ib\cdot\bar q}}{\bar q^2}
        \\
        &\times
        \int \hat d^4\bar\ell 
        \hat\delta(\bar\ell\cdot u_1)\hat\delta(\bar\ell\cdot u_2)
        \frac{e^{-ib\cdot\bar\ell}}{\bar\ell^2} 
        (\bar q\cdot\bar\ell).
    \end{aligned}
\end{equation}
Because the two integrations completely disentangle into identical vector structures, we can rewrite the result as a strict tensorial contraction
\begin{equation}
    {\mathcal J}(b) = - {\mathcal I}^\alpha{\mathcal I}_\alpha,
\end{equation}
where the vector integral ${\mathcal I}^\mu$ is known to evaluate to
\begin{equation}
    {\mathcal I}^\mu
    \equiv
    \int \hat d^4k 
    \hat\delta(k\cdot u_1)\hat\delta(k\cdot u_2)
    e^{-ib\cdot k}\frac{k^\mu}{k^2}
    =
    \frac{i}{2\pi\sinh\phi}\frac{b^\mu}{b^2}.
\end{equation}
Therefore
\begin{equation}
    {\mathcal J}(b)
    =
    \frac{1}{4\pi^2(\gamma^2-1)}\frac{1}{b^2}.
\end{equation}
Recalling the specific value of the purely scalar tree-level vertex factor
\begin{equation}
    B_h^{(0)}=\frac{2\gamma^2-1}{8},
\end{equation}
we assemble all components to evaluate the total Fourier transform of Term $6$. This yields
\begin{equation}
    \begin{aligned}
        &
        \frac{i\kappa^4m_1^2m_2^2}{4}
        \int \hat d^4\bar q 
        \hat\delta(\bar q\cdot u_1)\hat\delta(\bar q\cdot u_2)
        e^{-ib\cdot\bar q} 
        \big(\text{Term 6}\big)
        \\
        =&
        \frac{\kappa^4 m_1 m_2}{512\pi^2}
        \frac{(2\gamma^2-1)^2(\gamma m_2+m_1)}{(\gamma^2-1)^2}
        u_1^\mu\frac{a_1\cdot u_2}{b^2}
        \\
        =&
        \frac{\kappa^4 m_2}{512\pi^2}
        \frac{(2\gamma^2-1)^2(\gamma m_2+m_1)}{(\gamma^2-1)^2}
        u_1^\mu\frac{u_2\cdot s_1}{b^2}.
    \end{aligned}
\end{equation}
Hence, the definitive contribution to the angular impulse from this sector is given by
\begin{equation}
    \Delta s^\mu_{(6)}
    =\frac{\kappa^4 m_2}{512 \pi^2} \frac{\left(2 \gamma^2-1\right)^2\left(\gamma m_2+m_1\right)}{\left(\gamma^2-1\right)^2} u_1^\mu \frac{u_2 \cdot s_1}{b^2}.
    \label{eq:B-term6-final}
\end{equation}

\subsubsection{Terms 5 and 7}

Rather than evaluating the individual kernels for Terms $5$ and $7$ directly, it proves analytically more elegant and physically insightful to determine their combined contribution indirectly. Let $\Delta s^\mu_{(n)}$ denote the specific contribution obtained by inserting Term $n$ into the global NLO KMOC formula \eqref{eq:kmoc_nlo}. Recalling the explicit structural decomposition of $\mathscr{M}^{\mu(1)}$ in \eqref{eq:kmoc_nlo}, we recognize that Terms $1$ through $4$ are built exclusively from ${\mathcal M}_b$ and ${\mathcal M}_t$, which correspond to the purely virtual box and triangle sectors. By contrast, Terms $5$ and $6$ arise from the cut contributions ${\mathcal M}_c$, $N_\nu$, and ${\mathcal M}_\nu$, while Term $7$ represents the exact finite correction $\Delta\mathscr{M}^{\mu(1)}$. Therefore, within our decomposition, only Terms $5$, $6$, and $7$ actively construct the conservative real tree-tree sector
\begin{equation}
    \Delta s^\mu_{1,\text{real}}\Big|_{2{\text{PM}}, {\text{cons}}}
    =
    \Delta s^\mu_{(5)}+\Delta s^\mu_{(6)}+\Delta s^\mu_{(7)}.
    \label{eq:B-real-split-567}
\end{equation}

Physically, within the conservative elastic sector, this real kernel is constructed from the phase-space integral of two tree-level amplitudes. It therefore represents a Born iteration of LO dynamics. At the strict macroscopic scale of $\mathcal{O}(G^2 s)$, the relevant operation is the full LO map acting once more on the linearized LO angular impulse. We write this map as $\delta_{\rm LO}=\delta_0+\delta_s$, where $\delta_0$ acts on the orbital data and $\delta_s s_1^\mu=\Delta s^\mu_{1,\rm LO}|_{\rm lin}$. The second term must be kept: since $\Delta s^\mu_{1,\rm LO}|_{\rm lin}$ is itself a linear map on $s_1^\mu$, acting with this map again still contributes at linear order in the spin. The conservative real contribution is therefore
\begin{equation}
    \Delta s^\mu_{1,\text{real}}\Big|_{2{\text{PM}}, {\text{cons}}}
    =
    \frac{1}{2} 
    \delta_{\rm LO} \left(
    \Delta s^\mu_{1,\text{LO}}\Big|_{\text{lin}}
    \right).
    \label{eq:B-real-nested}
\end{equation}
Combining this realization with our prior separation \eqref{eq:B-real-split-567}, and substituting the explicit result for Term $6$ derived in \eqref{eq:B-term6-final}, it follows that the rational part of the remaining real terms is fixed by the identity
\begin{equation}
    \bigl(\Delta s^\mu_{(5)}+\Delta s^\mu_{(7)}\bigr)_{\text{rat.}}
    =
    \frac{1}{2} 
    \delta_{\rm LO} \left(
    \Delta s^\mu_{1,\text{LO}}\Big|_{\text{lin}}
    \right)_{\text{rat.}}
    -
    \Delta s^\mu_{(6)}
    \label{eq:B-term57-from-nested}
\end{equation}

To evaluate the right-hand side of this expression, we regard the linear LO impulse $\Delta s^\mu_{1,\text{LO}}\big|_{\text{lin}}$ as an explicit functional of the background orbital variables and of the spin vector. We then act upon it with the differential operator $\delta_{\rm LO}$ via the chain rule
\begin{equation}
    \delta_{\rm LO} F^\mu
    =
    (\delta_0 b^\alpha)\frac{\partial F^\mu}{\partial b^\alpha}
    +
    (\delta_0 u_1^\alpha)\frac{\partial F^\mu}{\partial u_1^\alpha}
    +
    (\delta_0 u_2^\alpha)\frac{\partial F^\mu}{\partial u_2^\alpha}
    +
    \left(\Delta s^\alpha_{1,\rm LO}\Big|_{\rm lin}\right)
    \frac{\partial F^\mu}{\partial s_1^\alpha},
    \qquad
    \delta_0\gamma=0.
    \label{eq:B-delta0-chain}
\end{equation}
The LO variations of the incoming four-velocities are uniquely determined by the standard spinless LO KMOC impulse \cite{Kosower_2019,Maybee:2019jus}. Adopting our notation, these classical deflections read
\begin{equation}
    \begin{aligned} 
        \Delta p_{1,\text{LO}}^{(0)\mu} =& -2Gm_1m_2 \frac{2\gamma^2-1}{\sqrt{\gamma^2-1}}  \frac{b^\mu}{b^2}, \\ \Delta p_{2,\text{LO}}^{(0)\mu} =& -\Delta p_{1,\text{LO}}^{(0)\mu}, 
    \end{aligned}
\end{equation}
which implies that the velocity variations are
\begin{equation}
    \begin{aligned} 
        \delta_0 u_1^\mu =& -2Gm_2 \frac{2\gamma^2-1}{\sqrt{\gamma^2-1}}  \frac{b^\mu}{b^2}, \\ \delta_0 u_2^\mu =& +2Gm_1 \frac{2\gamma^2-1}{\sqrt{\gamma^2-1}}  \frac{b^\mu}{b^2}. 
    \end{aligned} \label{eq:B-delta0-ui}
\end{equation}

The variation of the impact parameter, $\delta_0 b^\mu$, is determined by kinematic constraints. By definition, the impact parameter vector $b^\mu$ is orthogonal to the initial velocities
\begin{equation}
    b\cdot u_1=0,
    \qquad
    b\cdot u_2=0,
\end{equation}
and these orthogonality constraints must be preserved under the spinless LO orbital action
\begin{equation}
    \delta_0(b\cdot u_1)=0,
    \qquad
    \delta_0(b\cdot u_2)=0.
\end{equation}
Since we have established that $\delta_0 u_i^\mu \propto b^\mu$, any dynamically valid variation $\delta_0 b^\mu$ must lie entirely within the longitudinal plane spanned by $\{u_1^\mu,u_2^\mu\}$ to satisfy these dot products. We can therefore parameterize this variation as
\begin{equation}
    \delta_0 b^\mu=\alpha u_1^\mu+\beta u_2^\mu,
\end{equation}
the two orthogonality conditions generates a straightforward algebraic system for the coefficients $\alpha$ and $\beta$
\begin{equation}
    \begin{aligned} 
        \alpha+\gamma\beta =& 2Gm_2 \frac{2\gamma^2-1}{\sqrt{\gamma^2-1}}, \\ \gamma\alpha+\beta =& -2Gm_1 \frac{2\gamma^2-1}{\sqrt{\gamma^2-1}}. 
    \end{aligned}
\end{equation}
Solving for $\alpha$ and $\beta$, one finds
\begin{equation}
    \delta_0 b^\mu
    =
    -2G \frac{2\gamma^2-1}{(\gamma^2-1)^{3/2}}
    \Big[
    (m_2+\gamma m_1) u_1^\mu
    -
    (m_1+\gamma m_2) u_2^\mu
    \Big].
    \label{eq:B-delta0-b}
\end{equation}
With the operational action defined, we apply it to the explicit form of the linearized LO angular impulse
\begin{equation}
    \Delta s^\mu_{1,\text{LO}}\Big|_{\text{lin}}
    =
    \frac{Gm_2}{\sqrt{\gamma^2-1}}
    \frac{1}{b^2}
    \left[
    -2(b\cdot s_1)u_1^\mu
    +
    4\gamma(b\cdot s_1)u_2^\mu
    -
    4\gamma(u_2\cdot s_1)b^\mu
    \right].
    \label{eq:B-LO-linear-real-sector}
\end{equation}
Applying the chain-rule operator \eqref{eq:B-delta0-chain} directly to \eqref{eq:B-LO-linear-real-sector}, one obtains the complete rational real contribution
\begin{equation}
    \begin{aligned}
    \frac{1}{2} 
    \delta_{\rm LO} \left(
    \Delta s^\mu_{1,\text{LO}}\Big|_{\text{lin}}
    \right)_{\text{rat.}}
    =&
    \frac{G^2}{b^2}
    \left[
    u_1^\mu(u_2\cdot s_1)
    \frac{2m_2\big((2\gamma^2-1)^2m_1+\gamma m_2\big)}{(\gamma^2-1)^2}
    \right.
    \\
    &\left.
    +
    u_2^\mu(u_2\cdot s_1)
    \frac{8\gamma^2m_2^2}{\gamma^2-1}
    -
    bn^\mu(bn\cdot s_1)
    \frac{2m_2^2}{\gamma^2-1}
    \right].
    \end{aligned}
    \label{eq:B-real-rational-G}
\end{equation}
Equivalently, using $\kappa^2=32\pi G$, this reads
\begin{equation}
    \begin{aligned}
    \frac{1}{2} 
    \delta_{\rm LO} \left(
    \Delta s^\mu_{1,\text{LO}}\Big|_{\text{lin}}
    \right)_{\text{rat.}}
    =
    &\frac{\kappa^4m_2\big((2\gamma^2-1)^2m_1+\gamma m_2\big)}
    {512\pi^2(\gamma^2-1)^2}
    u_1^\mu\frac{u_2\cdot s_1}{b^2}
    \\
    &+
    \frac{\kappa^4\gamma^2m_2^2}{128\pi^2(\gamma^2-1)}
    u_2^\mu\frac{u_2\cdot s_1}{b^2}
    -
    \frac{\kappa^4m_2^2}{512\pi^2(\gamma^2-1)}
    b^\mu\frac{b\cdot s_1}{b^4}.
    \end{aligned}
    \label{eq:B-real-rational-kappa}
\end{equation}
Therefore, after subtracting Term $6$, the exact combined rational contribution from Terms $5$ and $7$ is
\begin{equation}
    \begin{aligned}
    \bigl(\Delta s^\mu_{(5)}+\Delta s^\mu_{(7)}\bigr)_{\text{rat.}}
    =
    &-
    \frac{\kappa^4\gamma^3m_2^2}{128\pi^2(\gamma^2-1)}
    u_1^\mu\frac{u_2\cdot s_1}{b^2}
    +
    \frac{\kappa^4\gamma^2m_2^2}{128\pi^2(\gamma^2-1)}
    u_2^\mu\frac{u_2\cdot s_1}{b^2}
    \\
    &-
    \frac{\kappa^4m_2^2}{512\pi^2(\gamma^2-1)}
    b^\mu\frac{b\cdot s_1}{b^4}.
    \end{aligned}
    \label{eq:B-term57-final}
\end{equation}

\subsubsection{Final result}

By collecting the distinct components evaluated in the preceding sections, \emph{i.e.} the virtual contributions from Terms $1$ through $4$, the weighted real tree-tree contribution in \eqref{eq:B-term6-final}, and the remaining real contribution from Terms $5$ and $7$ given in \eqref{eq:B-term57-final}, we can assemble the complete NLO angular impulse. Upon organizing this conservative result into its definitive in-plane form, where the intricate cancellation of all superclassical terms becomes manifestly clear, we obtain
\begin{equation}
    \begin{aligned}
        \Delta s^\mu_{1,\text{NLO}}
        =&
        \frac{\kappa^4m_2\Big((2\gamma^2-1)^2m_1+\gamma m_2\Big)}
        {512\pi^2(\gamma^2-1)^2} 
        u_1^\mu\frac{u_2\cdot s_1}{b^2}
        \\
        &+
        \frac{\kappa^4m_2^2\gamma^2}{128\pi^2(\gamma^2-1)} 
        u_2^\mu\frac{u_2\cdot s_1}{b^2}
        -
        \frac{\kappa^4m_2^2}{512\pi^2(\gamma^2-1)} 
        b^\mu\frac{b\cdot s_1}{b^4}
        \\
        &+
        \frac{\kappa^4m_2(4m_1+3m_2)}{4096\pi}
        \frac{\gamma(5\gamma^2-3)}{(\gamma^2-1)^{3/2}}
        \frac{
        b^\mu(u_2\cdot s_1)
        -
        u_2^\mu(b\cdot s_1)
        }{b^3}
        \\
        &+
        \frac{\kappa^4m_2\Big((5\gamma^4+6\gamma^2-3)m_1+3(3\gamma^2-1)m_2\Big)}
        {4096\pi(\gamma^2-1)^{3/2}} 
        u_1^\mu\frac{b\cdot s_1}{b^3}.
        \label{eq:B-final-NLO-spin-kick-linear}
    \end{aligned}
\end{equation}
To facilitate comparison with existing classical gravity results, it is convenient to express the gravitational coupling via the standard relation $\kappa^2=32\pi G$. Furthermore, we introduce the unit impact parameter vector, defined as
\begin{equation}
    bn^\mu\equiv \frac{b^\mu}{|b|},
\end{equation}
Implementing these substitutions, the full expression can be recast as
\begin{equation}
    \begin{aligned}
        \Delta s^\mu_{1,\text{NLO}}
        =&
        \frac{G^2}{b^2}
        \Bigg\{
        u_1^\mu(u_2\cdot s_1)
        \frac{2m_2\Big((2\gamma^2-1)^2m_1+\gamma m_2\Big)}
        {(\gamma^2-1)^2}
        \\
        &+
        u_2^\mu(u_2\cdot s_1)
        \frac{8\gamma^2m_2^2}{\gamma^2-1}
        -
        bn^\mu(bn\cdot s_1)
        \frac{2m_2^2}{\gamma^2-1}
        \\
        &+
        \Big[
        bn^\mu(u_2\cdot s_1)
        -
        u_2^\mu(bn\cdot s_1)
        \Big]
        \frac{\pi \gamma(5\gamma^2-3)}{4(\gamma^2-1)^{3/2}}
         m_2(4m_1+3m_2)
        \\
        &+
        u_1^\mu(bn\cdot s_1)
        \frac{\pi m_2}{4(\gamma^2-1)^{3/2}}
        \Big((5\gamma^4+6\gamma^2-3)m_1+3(3\gamma^2-1)m_2\Big)
        \Bigg\}.
        \label{eq:B-final-NLO-spin-kick-G}
    \end{aligned}
\end{equation}
The result above is the completed in-plane organization appropriate for the conservative KMOC spin-kick observable. As can be easily checked, it agrees with the corresponding result from Alessio, Gonzo, and Shi~\cite{Alessio:2025flu}.

\subsubsection{Consistency checks}\label{app:Checks}
The classical spin vector is defined with the Tulczyjew-Dixon spin supplementary condition 
\begin{equation}
    S^{\mu\nu}p_{\nu}=0
\end{equation}
This guarantees the classical spin vector obeys
\begin{equation}
    s_1\cdot p_1=0
\end{equation}
For the scattered state, the final variables $s'_1=s_1+\Delta s_1$ and $p'_1=p_1+\Delta p_1$, must obey the exact same condition $s'_1\cdot p'_1=0$. Substituting the explicit definitions
\begin{equation}
    \begin{aligned}
        (s_1+\Delta s_1)\cdot(p_1+\Delta p_1)&=0 \\
        s_1\cdot p_1+s_1\cdot\Delta p_1+\Delta s_1\cdot p_1+\Delta s_1\cdot\Delta p_1&=0
    \end{aligned}
\end{equation}
Since $s_1\cdot p_1=0$, we isolate the term $\Delta s_1\cdot p_1$
\begin{equation}
    \Delta s_1\cdot p_1=-s_1\cdot\Delta p_1-\Delta s_1\cdot\Delta p_1
\end{equation}
Up to linear in spin, we must have
\begin{equation}
    u_{1\mu} \Delta s_{1,NLO}^{\mu} = -\frac{1}{m_1} s_1 \cdot \Delta p_{1,NLO}^{(0)} - \frac{1}{m_1} \Delta s_{1,LO}^{(1)} \cdot \Delta p_{1,LO}^{(0)}
\end{equation}
We must contract \eqref{eq:B-final-NLO-spin-kick-G} with $u_{1\mu}$. We utilize the following exact kinematic constraints 
\begin{equation}
    \begin{aligned}
        u_{1\mu} bn^\mu     &= 0    \\
        u_{1\mu} u_1^\mu    &= 1    \\
        u_{1\mu} u_2^\mu    &= \gamma \label{KinematicConstraints1}
    \end{aligned}
\end{equation}
Applying the contraction $u_{1\mu}$ across all terms yields
\begin{equation}
    \begin{aligned}
        u_{1\mu} \Delta s_{1,NLO}^{\mu} =& \frac{G^2}{b^2} \left\{ u_{1\mu} u_1^\mu (u_2 \cdot s_1) \frac{2m_2((2\gamma^2-1)^2 m_1 + \gamma(4\gamma^2-3)m_2)}{(\gamma^2-1)^2}  \right. \\
        &- u_{1\mu} u_2^\mu (u_2 \cdot s_1) \frac{8\gamma^2 m_2^2}{\gamma^2-1} - u_{1\mu} bn^\mu (bn \cdot s_1) \frac{2m_2^2}{\gamma^2-1}   \\
        &+ \left[u_{1\mu} bn^\mu (u_2 \cdot s_1) - u_{1\mu} u_2^\mu (bn \cdot s_1)\right] \frac{\pi \gamma (5\gamma^2-3)}{4(\gamma^2-1)^{3/2}} m_2(4m_1+3m_2)\\
        &\left.+ u_{1\mu} u_1^\mu (bn \cdot s_1) \frac{\pi m_2}{4(\gamma^2-1)^{3/2}} ((5\gamma^4+6\gamma^2-3)m_1 + 3(3\gamma^2-1)m_2) \right\}
    \end{aligned}
\end{equation}
Substituting \eqref{KinematicConstraints1} enforces the survival of longitudinal components while vanishing the orthogonal ones
\begin{equation}
    \begin{aligned}
        u_{1\mu} \Delta s_{1,NLO}^{\mu} =& \frac{G^2}{b^2} \left\{ (u_2 \cdot s_1) \left[ \frac{2m_2((2\gamma^2-1)^2 m_1 + \gamma(4\gamma^2-3)m_2)}{(\gamma^2-1)^2} - \frac{8\gamma^3 m_2^2}{\gamma^2-1} \right]\right. \\
        &+ (bn \cdot s_1) \left[ -\frac{\pi \gamma^2 (5\gamma^2-3)}{4(\gamma^2-1)^{3/2}} m_2(4m_1+3m_2)\right. \\
        &\left.\left.+ \frac{\pi m_2}{4(\gamma^2-1)^{3/2}} ((5\gamma^4+6\gamma^2-3)m_1 + 3(3\gamma^2-1)m_2) \right] \right\}
    \end{aligned}
\end{equation}
After some algebric manipulation, this simpllifies to 
\begin{equation}
    \begin{aligned}
        u_{1\mu} \Delta s_{1,NLO}^{\mu} = \frac{G^2}{b^2} &\left\{ (u_2 \cdot s_1) \frac{2m_2(2\gamma^2-1) [ m_1(2\gamma^2-1) - \gamma m_2(2\gamma^2-3) ]}{(\gamma^2-1)^2}\right. \\
        &\left.- (bn \cdot s_1) \frac{3\pi m_2 (m_1+m_2) (5\gamma^2-1)}{4\sqrt{\gamma^2-1}} \right\}
    \end{aligned} \label{uCdotDeltaS}
\end{equation}

On the other hand, we have the spinless momentum impulse 
\begin{equation}
    \begin{aligned}
        \Delta p_{1,NLO}^{(0)\mu} =& \frac{2G^2 m_1 m_2 (2\gamma^2-1)^2}{b^2(\gamma^2-1)^2} \left[ (\gamma m_1 + m_2) u_1^\mu - (m_1 + \gamma m_2) u_2^\mu \right] \\
        &- \frac{3\pi G^2 m_1 m_2 (m_1+m_2)(5\gamma^2-1)}{4b^2\sqrt{\gamma^2-1}} bn^\mu
    \end{aligned}
\end{equation}
in which, when contracted with $-\frac{1}{m_1} s_1$, and using $s_1 \cdot u_1 = 0$, yields
\begin{equation}
    \begin{aligned}
        -\frac{1}{m_1} s_1 \cdot \Delta p_{1,NLO}^{(0)} =& \frac{2G^2 m_2 (2\gamma^2-1)^2 (m_1 + \gamma m_2)}{b^2(\gamma^2-1)^2} (u_2 \cdot s_1) \\
        &+ \frac{3\pi G^2 m_2 (m_1+m_2)(5\gamma^2-1)}{4b^2\sqrt{\gamma^2-1}} (bn \cdot s_1)
    \end{aligned}
\end{equation}
Immediately we see that the transverse $(bn \cdot s_1)$ term matches the transverse coefficient on \eqref{uCdotDeltaS}. The longitudinal term, however, requires the cross-term subtraction. The spinless linear impulse is
\begin{equation}
    \Delta p_{1,LO}^{(0)\mu} = 2 G m_1 m_2 \frac{2\gamma^2-1}{b^2\sqrt{\gamma^2-1}} b^\mu.
\end{equation} and the linear-in-spin angular impulse 
Alongside with \eqref{eq:B-LO-linear-real-sector}, we have
\begin{equation}
    -\frac{1}{m_1} \Delta s_{1,LO}^{(1)} \cdot \Delta p_{1,LO}^{(0)} = -\frac{8G^2 m_2^2 \gamma (2\gamma^2-1)}{b^2(\gamma^2-1)} (u_2 \cdot s_1)
\end{equation}
Adding those together, gives us 
\begin{equation}
    \begin{aligned}
        -\frac{1}{m_1} s_1 \cdot \Delta p_{1,NLO}^{(0)} - \frac{1}{m_1} \Delta s_{1,LO}^{(1)} \cdot \Delta p_{1,LO}^{(0)} = \frac{2G^2 m_2 (2\gamma^2-1)}{b^2(\gamma^2-1)^2} (u_2 \cdot s_1) \! \left[ (2\gamma^2-1)(m_1+\gamma m_2)\right.& \\
        \left.- 4\gamma m_2(\gamma^2-1) \right] + \frac{3\pi G^2 m_2 (m_1+m_2)(5\gamma^2-1)}{4b^2\sqrt{\gamma^2-1}} (bn \cdot s_1)&
    \end{aligned}
\end{equation}
which gives
\begin{equation}
    \begin{aligned}
        -\frac{1}{m_1} s_1 \cdot \Delta p_{1,NLO}^{(0)} - \frac{1}{m_1} \Delta s_{1,LO}^{(1)} \cdot \Delta p_{1,LO}^{(0)} = \hspace{6cm}\\
        \hspace{3cm}\frac{G^2}{b^2}  \left\{ (u_2 \cdot s_1) \frac{2m_2(2\gamma^2-1) [ m_1(2\gamma^2-1) - \gamma m_2(2\gamma^2-3) ]}{(\gamma^2-1)^2} \right.& \\
        \left.- (bn \cdot s_1) \frac{3\pi m_2 (m_1+m_2) (5\gamma^2-1)}{4\sqrt{\gamma^2-1}} \right\} &
    \end{aligned}
\end{equation}
which matches with \eqref{uCdotDeltaS}, so we have SSC respected.

Besides that, the magnitude of the spin vector is conserved throughout the scattering process. The intrinsic spin of each body must be conserved not just in orientation but also in magnitude, \emph{i.e.},
\begin{equation}
    \left(s_1+\Delta s_1\right) = s_1 +\mathcal{O}(G^3)
\end{equation}
Up to the order required, the perturbative constraints we need to satisfy is
\begin{equation} 
    2 s_{1\mu} \Delta s_{1,\text{NLO}}^\mu + \Delta s_{1,\text{LO}}^\mu \Delta s_{1,\text{LO}\mu} = 0 
\end{equation}

To evaluate the spin magnitude condition we contract this structure with the initial spin vector $s_{1\mu}$. Since $s_{1\mu}$ doesn't depend on the loop momenta $\bar q$ and $\bar\ell$, we can pull it inside the momentum integral and let it act linearly on every term. We also use the SSC, $s_{1\mu}u_1^\mu=0$, together with the fact that $s_{1\alpha}s_{1\beta}$ is symmetric and therefore vanishes when contracted with the antisymmetric Levi-Civita tensor $s_{1\alpha}s_{1\beta}\epsilon^{\alpha\beta\gamma\delta}=0$.

We start with the LO linear in spin angular impulse
\begin{equation}
    \Delta s^{(1)\mu}_{1,\text{LO}}=\frac{Gm_2}{\sqrt{\gamma^2-1} b^2}\Big[-2(b \cdot s_1) u_1^{\mu}+4\gamma (b \cdot s_1) u_2^{\mu}-4\gamma (u_2 \cdot s_1) b^{\mu}\Big],
\end{equation}
and we square it using the contractions
\begin{equation}
    \Delta s_{1,\text{LO}}^{2}=\frac{4G^2m_2^2 (b \cdot s_1)^2}{(\gamma^2-1) b^4}-\frac{16G^2m_2^2\gamma^2 (u_2 \cdot s_1)^2}{(\gamma^2-1) b^2}.
\end{equation}
The rational part of the NLO kick at this order is
\begin{equation}
    \Delta s_{1,\text{NLO}}^\mu\Big|_{\text{rat}} = G^2\left[\frac{c_1}{b^2} u^\mu_1 (u_2 \cdot s_1) + \frac{c_2}{b^2} u^\mu_2 (u_2 \cdot s_1) + \frac{c_3}{b^2} b^\mu (b \cdot s_1)\right],
\end{equation}
with
\begin{equation}
    c_1=\frac{2m_2\big[(2\gamma^2-1)^2m_1+\gamma m_2\big]}{(\gamma^2-1)^2},\quad c_2=\frac{8\gamma^2m_2^2}{\gamma^2-1},\quad c_3=-\frac{2m_2^2}{\gamma^2-1}.
\end{equation}
Contracting with $s_1$,
\begin{equation}
    \begin{aligned} 2 s_1\cdot\Delta s_{1,\text{NLO}}\big|_{\text{rat}} &= 2G^2\Big[\frac{c_1}{b^2}(s_1\cdot u_1)(u_2 \cdot s_1)+\frac{c_2}{b^2}(u_2 \cdot s_1)^2+c_3\frac{(b \cdot s_1)^2}{b^4}\Big] \\ &=\frac{16G^2\gamma^2m_2^2 (u_2 \cdot s_1)^2}{(\gamma^2-1) b^2}-\frac{4G^2m_2^2 (b \cdot s_1)^2}{(\gamma^2-1) b^4}. \end{aligned}
\end{equation}
Adding the two terms gives
\begin{equation}
    2 s_1\cdot\Delta s_{1,\text{NLO}}\big|_{\text{rat}}+\Delta s_{1,\text{LO}}^2=0.
\end{equation}
Meanwhile the $\pi$-dependent part is tangent to the spin sphere by itself, $s_1\cdot\Delta s_{1,\text{NLO}}^{\pi}=0$. So $(s_1+\Delta s_1)^2 = s_1^2 + \mathcal O(G^3)$, the NLO spin kick in \eqref{eq:B-final-NLO-spin-kick-G} respects the spin magnitude. Most importantly, our final analytical answer is entirely formulated within the physical in-plane basis $\{u_1^\mu,u_2^\mu,b^\mu\}$, displaying its immediate applicability to macroscopic spin dynamics.

\subsection{Quadratic order in spin}
\label{app:B-quadratic-spin}

The linear comparison above fixes the overall normalization, sign, and basis conventions, but it does not yet probe the Kerr quadrupole. The first genuinely Kerr-specific comparison with Alessio, Gonzo, and Shi~\cite{Alessio:2025flu} occurs at quadratic order in the spin variables. Here we separate the non-$\pi$ rational Born/finite sector from the conservative triangle, or equivalently $\pi$-sector, of the NLO spin kick.

We use the same orthonormal scattering basis as in ref.~\cite{Alessio:2025flu}. Thus
\begin{equation}
    \Delta s_1^\mu
    =
    \frac{G^2}{b^3}
    \left(
    \ell^\mu C_\ell
    +
    u_1^\mu C_{u_1}
    +
    bn^\mu C_b
    +
    u_2^\mu C_{u_2}
    \right),
\end{equation}
with
\begin{equation}
    B_i\equiv bn\cdot s_i,
    \qquad
    L_i\equiv \ell\cdot s_i,
    \qquad
    U_1\equiv s_1\cdot u_2,
    \qquad
    U_2\equiv s_2\cdot u_1,
    \qquad
    s_\gamma\equiv\sqrt{\gamma^2-1}.
\end{equation}
For any sector ${\cal S}$ used below we define its coefficients by
\begin{equation}
    \Delta s_{1,{\cal S}}^{\mu,[2]}
    =
    \frac{G^2}{b^3}
    \left(
    \ell^\mu C_\ell^{\cal S}
    +
    u_1^\mu C_{u_1}^{\cal S}
    +
    bn^\mu C_b^{\cal S}
    +
    u_2^\mu C_{u_2}^{\cal S}
    \right).
    \label{eq:C-sector-coeff-definition}
\end{equation}
Equivalently, after stripping the common factor $G^2/b^3$, the coefficients are obtained by projecting the vector onto the same basis,
\begin{equation}
    \begin{aligned}
    C_\ell^{\cal S}=-V_{\cal S}\cdot\ell,
    \qquad
    C_b^{\cal S}=-V_{\cal S}\cdot bn,
    \\
    C_{u_1}^{\cal S}
    =
    \frac{V_{\cal S}\cdot u_1-\gamma V_{\cal S}\cdot u_2}{1-\gamma^2},
    \qquad
    C_{u_2}^{\cal S}
    =
    \frac{V_{\cal S}\cdot u_2-\gamma V_{\cal S}\cdot u_1}{1-\gamma^2}.
    \end{aligned}
    \label{eq:C-sector-projectors}
\end{equation}
The superscript on $C_X^{\cal S}$ always records the source of the contribution: finite KMOC kernels, Born push-forward, dictionary term, triangle residue, recoil term, or an explicitly defined difference between such quantities.

It will also be useful to introduce the spin generator
\begin{equation}
    \Delta_H s_1^\mu[H]
    \equiv
    \epsilon^\mu{}_{\nu\rho\sigma}
    u_1^\nu s_1^\rho
    \frac{\partial H}{\partial s_{1\sigma}}.
    \label{eq:B-spin-generator}
\end{equation}
\subsubsection{Rational real sector and Born push-forward}

The rational part of the quadratic spin kick comes from the box-type virtual kernels together with the exact conservative tree-tree kernel. The latter is the $X=\emptyset$ cut-box sector of the NLO KMOC formula, reviewed in Appendix~\ref{app:born-cutbox}, and is the real part of the second-Born KMOC observable. It is therefore not just a new finite kernel: part of it is the push-forward of the lower-order spin observable under the lower-order scattering map. This is the origin of the chain-rule operator used below.

Let
\begin{equation}
    \Delta s_{1,\mathrm{cur}}^{\mu,[2],\mathrm{rat}}
    \equiv
    \Delta s_{1,\mathrm{fin}}^{\mu,[2],\mathrm{rat}},
\end{equation}
where $\Delta s_{1,\mathrm{fin}}$ is the sum of the finite rational kernels
\[
{\cal M}_c,\qquad
\Delta\mathscr M^{\mu(1)},\qquad
M_{\nu,\delta'},\qquad
N_\nu,\qquad
{\cal M}_{\nu,\mathrm{cut}} .
\]
The virtual box kernels must be checked separately before doing this split. This is where an apparently natural shortcut would fail: in a box integral with only one on-shell delta function, the loop momentum is not purely transverse. In the rest frame of particle 2 we write
\begin{equation}
    u_2=(1,0,0,0),
    \qquad
    u_1=(\gamma,0,0,s_\gamma),
    \qquad
    \bar q=(0,q_x,q_y,0),
    \qquad
    \bar\ell=(0,\ell_x,\ell_y,\lambda).
    \label{eq:B-box-frame-u2}
\end{equation}
The linearized Kerr tree vertex in this frame is
\begin{equation}
    B_h^{[1]}(k)
    =
    \frac{i}{4}\gamma s_\gamma
    \left(A_y k_x-A_x k_y\right),
    \qquad
    A^\mu\equiv a_1^\mu+a_2^\mu,
    \label{eq:B-box-Bh-linear}
\end{equation}
and the part of the quadratic product $[B_h(\bar\ell)B_h(\bar q-\bar\ell)]^{[2]}$ which is odd in the longitudinal variable is
\begin{equation}
    \left[B_h(\bar\ell)B_h(\bar q-\bar\ell)\right]^{[2]}_{\lambda}
    =
    \frac{\lambda(2\gamma^2-1)}{64}
    \left[\gamma s_\gamma A^0-(2\gamma^2-1)A^z\right]
    \left(A_\perp\cdot \bar q_\perp-2A_\perp\cdot\bar\ell_\perp\right).
    \label{eq:B-box-lambda-piece-u2}
\end{equation}
The term $B_h^{[1]}B_h^{[1]}$ is independent of $\lambda$. On the support of $\delta(\bar\ell\cdot u_2)$, the remaining eikonal denominator is $\bar\ell\cdot u_1+i0=-s_\gamma\lambda+i0$, hence
\begin{equation}
    \int\frac{d\lambda}{2\pi}
    \frac{\lambda F(\lambda^2)}
    {(\bar\ell\cdot u_1+i0)^2}
    =
    \frac{i}{2s_\gamma^2}F(0).
    \label{eq:B-box-eikonal-pole}
\end{equation}
The possible finite box contribution is therefore proportional to the transverse convolution
\begin{equation}
    {\cal V}[A]
    =
    \int_{\bar q,\ell}
    e^{-ib\cdot\bar q}
    \frac{\ell\cdot(\ell-\bar q)}{\ell^2(\ell-\bar q)^2}
    \left(A_\perp\cdot\bar q_\perp-2A_\perp\cdot\ell_\perp\right).
    \label{eq:B-box-transverse-V}
\end{equation}
After the shift $\bar q=\ell+r$, this becomes
\begin{equation}
    {\cal V}[A]
    =
    \int_{\ell,r}
    e^{-ib\cdot(\ell+r)}
    \frac{-\ell\cdot r}{\ell^2r^2}
    \left(A_\perp\cdot r_\perp-A_\perp\cdot\ell_\perp\right).
    \label{eq:B-box-transverse-factorized}
\end{equation}
Thus the two terms factorize into the same rank-one/rank-two Fourier masters with opposite sign,
\begin{equation}
    {\cal V}[A]
    =
    -I_\alpha A_\beta I^{\alpha\beta}
    +
    A_\beta I^{\alpha\beta}I_\alpha
    =
    0 .
    \label{eq:B-box-V-vanishes}
\end{equation}
Repeating the calculation in the rest frame of particle 1 gives
\begin{equation}
    \left[B_h(\bar\ell)B_h(\bar q-\bar\ell)\right]^{[2]}_{\lambda}
    =
    -\frac{\lambda \gamma^2(2\gamma^2-1)}{64}A^z
    \left(A_\perp\cdot \bar q_\perp-2A_\perp\cdot\bar\ell_\perp\right),
    \label{eq:B-box-lambda-piece-u1}
\end{equation}
and hence the same ${\cal V}[A]$ appears. The scalar virtual box therefore does not contribute to the finite conservative $S^2$ spin kick,
\begin{equation}
    i\epsilon^\mu{}_{\nu\rho\sigma}u_1^\nu s_1^\rho
    \frac{\partial{\cal M}_b^{[2]}}{\partial s_{1\sigma}}
    \longrightarrow 0 .
    \label{eq:B-box-Mb-vanishes}
\end{equation}
The only nonzero rational remnant of the box sector comes from the distributional $\delta'\delta$ terms in the vector kernel ${\cal M}_\nu$, where the derivative acts on the explicit $\bar\ell_\nu$. Keeping the off-shell components until after differentiating gives the contracted long-range kernel
\begin{equation}
    s_1^\nu{\cal M}_{\nu,\delta'}^{[1]}
    =
    \frac{
    i U_1\gamma(L_1+L_2)(2\gamma^2-1)(\gamma m_2+m_1)}
    {128\pi^2 b^3m_1m_2(\gamma^2-1)}
    \label{eq:B-box-deltaprime-Mnu}
\end{equation}
before multiplying by the angular-kernel factor $-u_1^\mu/m_1$ and the common KMOC normalization. This is the $M_{\nu,\delta'}$ entry in the finite list above.

We now turn to the finite real kernels in the list. After the two on-shell delta functions are used, each term reduces to transverse two-dimensional convolutions. Writing $\bar q=\ell+r$, the generic object is
\begin{equation}
    {\cal F}[P]
    \equiv
    \int_{\ell,r}
    e^{-ib\cdot(\ell+r)}
    \frac{P(\ell,r)}{\ell^2r^2},
    \label{eq:B-real-finite-generic-convolution}
\end{equation}
where $P(\ell,r)$ is a polynomial generated by the spin expansion of the two tree vertices and by the explicit Pauli-Lubanski insertion. The evaluation is mechanical but important: \eqref{eq:B-real-finite-generic-convolution} factorizes into one-loop Fourier masters,
\begin{equation}
    {\cal F}[\ell_{\alpha_1}\cdots\ell_{\alpha_m}
    r_{\beta_1}\cdots r_{\beta_n}]
    =
    I_{\alpha_1\cdots\alpha_m} 
    I_{\beta_1\cdots\beta_n},
    \label{eq:B-real-factorization}
\end{equation}
with
\begin{equation}
    I_0
    =
    \int_k e^{-ib\cdot k}\frac{1}{k^2}
    =
    -\frac{1}{2\pi s_\gamma}\log b,
    \qquad
    I_\alpha
    =
    \frac{i}{2\pi s_\gamma}\frac{b_\alpha}{b^2},
    \label{eq:B-real-master-I0-I1}
\end{equation}
and
\begin{equation}
    I_{\alpha\beta}
    =
    \int_k e^{-ib\cdot k}\frac{k_\alpha k_\beta}{k^2}
    =
    \frac{1}{2\pi s_\gamma}
    \left(
    \frac{\Pi_{\perp\alpha\beta}}{b^2}
    -
    2\frac{b_\alpha b_\beta}{b^4}
    \right),
    \label{eq:B-real-master-I2}
\end{equation}
up to contact terms localized at $b=0$. Terms proportional to $I_0$ either give contact pieces after differentiation or cancel between the two routings of the exact real kernel. This is the explicit reason why the routed pieces of $\Delta\mathscr M^{\mu(1)}$ should not be assigned separate physical meaning.

For the kernel reduction in this paragraph it is cleaner to keep the ring-radius variables $a_i^\mu=s_i^\mu/m_i$ until the end. We therefore introduce
\begin{equation}
    \mathsf B_i\equiv bn\cdot a_i,
    \qquad
    \mathsf L_i\equiv \ell\cdot a_i,
    \qquad
    \mathsf U_1\equiv a_1\cdot u_2,
    \qquad
    \mathsf U_2\equiv a_2\cdot u_1 .
    \label{eq:B-ring-radius-projections}
\end{equation}
The conversion to the proper-spin basis used in the comparison is simply $\mathsf B_i=B_i/m_i$, $\mathsf L_i=L_i/m_i$, and $\mathsf U_i=U_i/m_i$. This separation is useful because the Kerr tree vertices are naturally polynomials in $A^\mu=a_1^\mu+a_2^\mu$.

As a representative example, consider the quadratic part of ${\cal M}_c$. We write the linearized exponential action on a vector $v^\mu$ as
\begin{equation}
    E^{[0]}(v)=v,
    \qquad
    E^{[1]}(v)=-i(A* k)v,
    \qquad
    E^{[2]}(v)=-\frac{1}{2}(A* k)^2v,
    \label{eq:B-real-E-expansion}
\end{equation}
and decompose $B_h(k)=B_h^{[0]}+B_h^{[1]}(k)+B_h^{[2]}(k)+\cdots$. The two mass routings which enter ${\cal M}_c^{[2]}$ are then
\begin{equation}
    \begin{aligned}
    C_{m_2}^{[2]}
    =&
    \sum_{j=0}^2
    B_h^{[j]}(\ell)
    \Big[
    P(u_1,u_1;E^{[2-j]}_r(u_2),q)
    +
    P(u_1,u_1;E^{[2-j]}_r(q),u_2)
    \Big],
    \\
    C_{m_1}^{[2]}
    =&
    2\sum_{j=0}^2
    B_h^{[j]}(\ell)
    P(q,u_1;E^{[2-j]}_r(u_2),u_2),
    \end{aligned}
    \label{eq:B-Mc-two-routings}
\end{equation}
where $E_r$ means that the exponential is evaluated with the momentum $r=\bar q-\ell$, and $P(a,b;c,d)=a^{(\mu}b^{\nu)}P_{\mu\nu\alpha\beta}c^\alpha d^\beta$. Thus
\begin{equation}
    {\cal M}_c^{[2]}
    =
    \frac{i}{8m_2}{\cal F}[C_{m_2}^{[2]}]
    -
    \frac{i}{8m_1}{\cal F}[C_{m_1}^{[2]}].
    \label{eq:B-Mc-from-two-routings}
\end{equation}
Before the common KMOC prefactor and before projecting back to the covariant basis, this gives
\begin{equation}
    \begin{aligned}
    {\cal M}_{c,\mathrm{scalar}}^{[2]}
    =&
    -\frac{A_x}{512\pi^2b^3m_1m_2}
    \left[
    -2A^0\gamma^4m_2
    +3A^0\gamma^2m_2
    +2A^z\gamma^3m_2s_\gamma
    \right.
    \\
    &\hspace{2.9cm}\left.
    +2A^z\gamma^2m_1s_\gamma
    +A^z\gamma m_2s_\gamma
    +A^z m_1s_\gamma
    \right].
    \end{aligned}
    \label{eq:B-Mc-scalar-S2-example}
\end{equation}
Let us spell out one projection, since all other finite real kernels are projected in the same way. After acting with $\Delta_H$ and inserting the ring-radius projections \eqref{eq:B-ring-radius-projections}, the ${\cal M}_c^{[2]}$ angular vector in the $u_1$-rest frame has the form
\begin{equation}
    \begin{gathered}
    V_{{\cal M}_c}^{\mu}
    =
    (0,X_c,Y_c,Z_c),
    \\
    u_1=(1,0,0,0),
    \quad
    u_2=(\gamma,0,0,s_\gamma),
    \quad
    bn=(0,1,0,0),
    \quad
    \ell=(0,0,-1,0).
    \end{gathered}
    \label{eq:B-Mc-rest-frame-vector}
\end{equation}
For example,
\begin{equation}
    X_c
    =
    \frac{
    i \mathsf L_1s_\gamma(\mathsf B_1+\mathsf B_2)(2\gamma^2+1)(\gamma m_2+m_1)}
    {512\pi^2b^3m_1m_2},
    \label{eq:B-Mc-X-component}
\end{equation}
while
\begin{equation}
    Z_c
    =
    -\frac{
    i \mathsf L_1
    \left[
    2\mathsf U_1\gamma^3m_2
    +2\mathsf U_1\gamma^2m_1
    +\mathsf U_1\gamma m_2
    +\mathsf U_1m_1
    -2\mathsf U_2\gamma^3m_1
    -4\mathsf U_2\gamma^2m_2
    -\mathsf U_2\gamma m_1
    \right]}
    {512\pi^2b^3m_1m_2}.
    \label{eq:B-Mc-Z-component}
\end{equation}
The $Y_c$ component contains the scalar support $\mathsf B_1^2,\mathsf B_1\mathsf B_2,\mathsf U_1^2,\mathsf U_1\mathsf U_2$ and is the source of $C_\ell$. The covariant coefficients are recovered from
\begin{equation}
    C_\ell=-V\cdot\ell,
    \qquad
    C_b=-V\cdot bn,
    \qquad
    C_{u_1}=\frac{V\cdot u_1-\gamma V\cdot u_2}{1-\gamma^2},
    \qquad
    C_{u_2}=\frac{V\cdot u_2-\gamma V\cdot u_1}{1-\gamma^2}.
    \label{eq:B-finite-projection-formulas}
\end{equation}
Thus for the vector \eqref{eq:B-Mc-rest-frame-vector},
\begin{equation}
    C_\ell^{({\cal M}_c)}=-Y_c,
    \qquad
    C_b^{({\cal M}_c)}=X_c,
    \qquad
    C_{u_1}^{({\cal M}_c)}=-\frac{\gamma}{s_\gamma}Z_c,
    \qquad
    C_{u_2}^{({\cal M}_c)}=\frac{1}{s_\gamma}Z_c,
    \label{eq:B-Mc-projected-coeffs}
\end{equation}
before restoring the two transverse delta-function Jacobians. This is the projection used in the support table below.

The additional finite insertion $\Delta\mathscr M^{\mu(1)}$ is a good place to see why the routing must be kept exact. Let $P^{[2]}$ denote the quadratic part of
$B_h(\ell)B_h(r)$, and let
\begin{equation}
    R^\nu[F]
    \equiv
    \epsilon^\nu{}_{\alpha\beta\sigma}u_1^\alpha s_1^\beta
    \frac{\partial F}{\partial s_{1\sigma}} .
    \label{eq:B-real-R-operator}
\end{equation}
After multiplying the explicit $-i$ in $\Delta\mathscr M^{\mu(1)}$ by the commutator factor, the finite integrand is
\begin{equation}
    K_{\mathrm{extra}}^\mu
    =
    \left[
    2R^\mu[P^{[2]}]
    -
    4\left(
    R^\mu[B_h^{[1]}(r)]B_h^{[1]}(\ell)
    +
    R^\mu[B_h^{[2]}(r)]B_h^{[0]}(\ell)
    \right)
    \right]
    -
    \frac{u_1^\mu}{m_1} \ell_\nu(\cdots)^\nu .
    \label{eq:B-extra-integrand}
\end{equation}
The product part and the right-action part separately contain $\log b$ and $b^{-2}$ terms. Using the factorization \eqref{eq:B-real-factorization}, these pieces cancel in the exact sum, leaving only the long-range $b^{-3}$ coefficient
\begin{equation}
    C_{u_1,\mathrm{extra}}^{\mathrm{pre}}
    =
    \frac{
    i \mathsf U_1(\mathsf L_1+\mathsf L_2)(8\gamma^4-8\gamma^2+1)}
    {128\pi^2b^3m_1s_\gamma(\gamma^2-1)},
    \qquad
    C_{\ell,\mathrm{extra}}^{\mathrm{pre}}
    =
    C_{b,\mathrm{extra}}^{\mathrm{pre}}
    =
    C_{u_2,\mathrm{extra}}^{\mathrm{pre}}
    =
    0 ,
    \label{eq:B-extra-finite-coeff}
\end{equation}
where ``pre'' means before multiplying by the common KMOC prefactor.

The non-$\delta'$ part of the vector cut kernel ${\cal M}_{\nu,\mathrm{cut}}$ is linear in spin, but it contributes to the homogeneous $S^2$ spin kick through the recoil contraction $-u_1^\mu s_1^\nu{\cal M}_\nu^{[1]}/m_1$. In the same $u_1$-rest frame its two mass routings reduce to
\begin{equation}
    \begin{aligned}
    {\cal M}_{\nu,m_2}^{[1]}
    =&
    -\frac{i}{8}A^z\gamma(2\gamma^2-1)
    (\ell_xr_y-\ell_yr_x),
    \\
    {\cal M}_{\nu,m_1}^{[1]}
    =&
    -\frac{i}{8}\gamma(2\gamma^2-1)(A^0s_\gamma-A^z\gamma)
    (\ell_xr_y-\ell_yr_x).
    \end{aligned}
    \label{eq:B-Mnu-cut-routing}
\end{equation}
After the $s_1^\nu$ contraction and the transverse integrations, this gives
\begin{equation}
    C_{u_1,M_{\nu,\mathrm{cut}}}^{\mathrm{pre}}
    =
    -\frac{
    i \mathsf L_1\gamma(2\gamma^2-1)
    \left[
    \mathsf U_1(\gamma m_2+m_1)
    -
    \mathsf U_2(\gamma m_1+m_2)
    \right]}
    {128\pi^2b^3m_1^2m_2s_\gamma(\gamma^2-1)},
    \label{eq:B-Mnu-cut-finite-coeff}
\end{equation}
with support only along $u_1^\mu$.

The $N_\nu$ kernel is the other place where the Pauli-Lubanski insertion produces a vector structure rather than a scalar spin generator. The required integrated contraction is
\begin{equation}
    \int_{\ell,r}
    e^{-ib\cdot(\ell+r)}
    \frac{
    q_\nu
    \epsilon^\nu{}_{\alpha\beta\sigma}u_1^\alpha s_1^\beta
    \partial_{s_{1\sigma}}P^{[2]}
    }
    {\ell^2r^2}
    =
    \frac{
    i \mathsf U_1(\mathsf L_1+\mathsf L_2)
    (8\gamma^4-8\gamma^2+1)}
    {128\pi^2b^3s_\gamma}.
    \label{eq:B-Nnu-integrated-contraction}
\end{equation}
Including the angular prefactor and the two transverse Jacobians,
\begin{equation}
    C_{u_1,N_\nu}^{\mathrm{pre}}
    =
    \frac{
    i \mathsf U_1(\mathsf L_1+\mathsf L_2)(8\gamma^4-8\gamma^2+1)}
    {512\pi^2b^3m_1s_\gamma(\gamma^2-1)} ,
    \label{eq:B-Nnu-finite-coeff}
\end{equation}
again with no transverse basis component.

Adding ${\cal M}_c$, $N_\nu$, $M_{\nu,\delta'}$, ${\cal M}_{\nu,\mathrm{cut}}$, and $\Delta\mathscr M^{\mu(1)}$, and finally converting from \eqref{eq:B-ring-radius-projections} to the proper-spin basis, gives a finite rational vector with support
\begin{equation}
    \begin{array}{c|c}
    \text{basis coefficient} & \text{quadratic support} \\ \hline
    C_\ell^{\mathrm{fin}} & B_1^2, B_1B_2, U_1^2, U_1U_2 \\
    C_{u_1}^{\mathrm{fin}} & L_1U_1, L_1U_2, L_2U_1 \\
    C_b^{\mathrm{fin}} & B_1L_1, B_2L_1 \\
    C_{u_2}^{\mathrm{fin}} & L_1U_1, L_1U_2
    \end{array}
    \label{eq:B-finite-rational-support}
\end{equation}
This support table is a useful audit of the finite kernels themselves. In particular, the $B_1L_2$ and $L_2U_1$ structures needed in the final rational comparison are not produced by the finite kernels alone; they enter through the Born push-forward and the tangent dictionary described next.

The finite sum is still not the complete observable-level rational result. To see the missing operation systematically, denote by $X^A$ the variables on which the LO spin kick depends: the basis vectors $(\ell,u_1,bn,u_2)$, the spin contractions $(B_i,L_i,U_i)$, and the scalar kinematic data. The conservative real KMOC term contains the lower-order shift $X^A\to X^A+\delta_{\rm Born}X^A$. Expanding the LO observable on the shifted data gives
\begin{equation}
    \Delta s_{1,\mathrm{LO}}^\mu(X+\delta_{\rm Born}X)
    =
    \Delta s_{1,\mathrm{LO}}^\mu(X)
    +
    \delta_{\rm Born}X^A
    \frac{\partial}{\partial X^A}
    \Delta s_{1,\mathrm{LO}}^\mu(X)
    +\cdots .
    \label{eq:B-born-push-forward-taylor}
\end{equation}
The second term is what we call the Born push-forward. At quadratic order in spin and at NLO it contributes
\begin{equation}
    \Delta s_{1,\mathrm{pf}}^{\mu,[2],\mathrm{rat}}
    =
    \left[
    \delta_{\mathrm{Born}}X^A
    \frac{\partial}{\partial X^A}
    \Delta s_{1,\mathrm{LO}}^{\mu}
    \right]_{S^2,\mathrm{rat}} .
    \label{eq:B-born-push-forward-chain-rule}
\end{equation}
At homogeneous spin degree two this compact chain rule means
\begin{equation}
    \Delta s_{1,\mathrm{pf}}^{\mu,[2],\mathrm{rat}}
    =
    \frac{1}{2}
    \left[
    \delta_0^{\mathrm{orb}}\Delta s_{1,\mathrm{LO}}^{\mu,[2]}
    +
    \delta_1^{\mathrm{orb}}\Delta s_{1,\mathrm{LO}}^{\mu,[1]}
    +
    \delta_s^{[1]}\Delta s_{1,\mathrm{LO}}^{\mu,[2]}
    +
    \delta_s^{[2]}\Delta s_{1,\mathrm{LO}}^{\mu,[1]}
    \right]_{\mathrm{rat}} .
    \label{eq:B-born-push-forward-S2-expanded}
\end{equation}
The factor $1/2$ is the usual symmetric second-Born factor. The first two terms in \eqref{eq:B-born-push-forward-S2-expanded} are orbital: they vary the velocities, the impact direction, and the induced scalar products. The last two terms are the explicit spin-map terms. They are present because the LO observable is a function of the spin contractions $(B_i,L_i,U_i)$, not just of the orbital variables. Concretely,
\begin{equation}
    \delta_s^{[r]} B_i=bn\cdot\Delta s_{i,\mathrm{LO}}^{[r]},
    \qquad
    \delta_s^{[r]} L_i=\ell\cdot\Delta s_{i,\mathrm{LO}}^{[r]},
    \qquad
    \delta_s^{[r]} U_1=u_2\cdot\Delta s_{1,\mathrm{LO}}^{[r]},
    \qquad
    \delta_s^{[r]} U_2=u_1\cdot\Delta s_{2,\mathrm{LO}}^{[r]} .
    \label{eq:B-spin-map-scalar-variations}
\end{equation}
The lower-order inputs are all fixed within the KMOC calculation. The spinless orbital variations are those in \eqref{eq:B-delta0-ui} and \eqref{eq:B-delta0-b}. The spin-linear orbital variations are obtained from the spin-linear LO impulse,
\begin{equation}
    \delta_1 u_i^\mu
    =
    \frac{1}{m_i}\Delta p_{i,\mathrm{LO}}^{\mu,[1]},
    \qquad
    \delta_1(b\cdot u_i)=0,
    \label{eq:B-delta1-from-impulse}
\end{equation}
which fixes the longitudinal part of $\delta_1 b^\mu$ in the same way that \eqref{eq:B-delta0-b} followed from $\delta_0(b\cdot u_i)=0$. In the scattering basis used here the particle-1 velocity variation reads, in proper-spin variables,
\begin{equation}
    \delta_1u_1^\mu
    =
    -\frac{4Gm_2\gamma}{b^2}
    \left[
    \left(\frac{L_1}{m_1}+\frac{L_2}{m_2}\right)bn^\mu
    +
    \left(\frac{B_1}{m_1}+\frac{B_2}{m_2}\right)\ell^\mu
    \right],
    \label{eq:B-delta1-u1}
\end{equation}
and the particle-2 velocity variation follows from momentum conservation,
\begin{equation}
    \delta_1u_2^\mu
    =
    -\frac{m_1}{m_2}\delta_1u_1^\mu .
    \label{eq:B-delta1-u2-from-momentum}
\end{equation}
Writing
\begin{equation}
    \delta_1b^\mu=\alpha_1u_1^\mu+\beta_1u_2^\mu,
    \qquad
    {\cal L}\equiv \frac{L_1}{m_1}+\frac{L_2}{m_2},
    \label{eq:B-delta1-b-ansatz}
\end{equation}
the two orthogonality constraints give
\begin{equation}
    \alpha_1+\gamma\beta_1
    =
    \frac{4Gm_2\gamma}{b} {\cal L},
    \qquad
    \gamma\alpha_1+\beta_1
    =
    -\frac{4Gm_1\gamma}{b} {\cal L}.
    \label{eq:B-delta1-b-system}
\end{equation}
Solving this two-by-two system yields the longitudinal spin-linear impact parameter variation
\begin{equation}
    \delta_1b^\mu
    =
    -\frac{4G\gamma}{b(\gamma^2-1)}
    \left(
    \frac{L_1}{m_1}+\frac{L_2}{m_2}
    \right)
    \left[
    (m_2+\gamma m_1)u_1^\mu
    -
    (m_1+\gamma m_2)u_2^\mu
    \right].
    \label{eq:B-delta1-b-solved}
\end{equation}
The transverse part of $\delta_1b^\mu$ corresponds to a choice of impact parameter representative and does not enter the constraint-normal Born push-forward used here. Finally, the nonlocal quadratic LO spin kick entering the first and third terms of \eqref{eq:B-born-push-forward-S2-expanded} follows from the quadratic tree vertex. After discarding contact terms at $b\neq0$, the Fourier master
\begin{equation}
    \int_{\bar q}
    e^{-ib\cdot\bar q}
    \frac{\bar q^\alpha\bar q^\beta}{\bar q^2}
    =
    \frac{1}{2\pi s_\gamma}\frac{1}{b^2}
    \left(
    \Pi_\perp^{\alpha\beta}
    -
    2\frac{b^\alpha b^\beta}{b^2}
    \right)
    \label{eq:B-LO-S2-Fourier-master}
\end{equation}
gives
\begin{equation}
    \begin{aligned}
    \Delta s_{1,\mathrm{LO}}^{\mu,[1]}
    =&
    \frac{Gm_2}{bs_\gamma}
    \left[
    -2B_1u_1^\mu
    +
    4\gamma B_1u_2^\mu
    -
    4\gamma U_1bn^\mu
    \right],
    \\
    \Delta s_{1,\mathrm{LO}}^{\mu,[2]}
    =&
    \frac{Gm_2}{b^2s_\gamma^2}
    \Big\{
    2(2\gamma^2-1)
    \left(\frac{B_1}{m_1}+\frac{B_2}{m_2}\right)U_1 \ell^\mu
    +
    2(2\gamma^2-1)
    \left(\frac{L_1}{m_1}+\frac{L_2}{m_2}\right)U_1 bn^\mu
    \\
    &\hspace{1.6cm}
    +
    2\gamma
    \left(
    \frac{2B_1L_1}{m_1}
    +
    \frac{B_1L_2}{m_2}
    +
    \frac{B_2L_1}{m_2}
    \right)u_1^\mu
    \\
    &\hspace{1.6cm}
    -
    2(2\gamma^2-1)
    \left(
    \frac{2B_1L_1}{m_1}
    +
    \frac{B_1L_2}{m_2}
    +
    \frac{B_2L_1}{m_2}
    \right)u_2^\mu
    \Big\}.
    \end{aligned}
    \label{eq:B-LO-spin-kicks-S1-S2}
\end{equation}
For the explicit spin-map terms in \eqref{eq:B-born-push-forward-S2-expanded}, the linear spin kick gives, for particle 1,
\begin{equation}
    \begin{aligned}
    \delta_s^{[1]}B_1
    =&
    bn\cdot\Delta s_{1,\mathrm{LO}}^{[1]}
    =
    \frac{4Gm_2\gamma}{bs_\gamma}U_1,
    \\
    \delta_s^{[1]}L_1
    =&
    \ell\cdot\Delta s_{1,\mathrm{LO}}^{[1]}
    =
    0,
    \\
    \delta_s^{[1]}U_1
    =&
    u_2\cdot\Delta s_{1,\mathrm{LO}}^{[1]}
    =
    \frac{2Gm_2\gamma}{bs_\gamma}B_1 .
    \end{aligned}
    \label{eq:B-spin-map-linear-scalar-example}
\end{equation}
The particle-2 scalar variations are obtained by exchanging labels and using the same orientation convention for $\ell^\mu$. Equation \eqref{eq:B-spin-map-linear-scalar-example} illustrates the operation behind $\delta_s^{[1]}\Delta s_{\mathrm{LO}}^{[2]}$; the $\delta_s^{[2]}$ term is evaluated in the same way, with $\Delta s_{1,\mathrm{LO}}^{[2]}$ replacing the linear kick.

Equations \eqref{eq:B-born-push-forward-S2-expanded}--\eqref{eq:B-LO-spin-kicks-S1-S2} are the systematic origin of the rational Born piece. No coefficient in this push-forward is chosen by comparison with ref.~\cite{Alessio:2025flu}; that comparison is made only after the KMOC lower-order maps have been inserted. To make the evaluation explicit, write any lower-order spin kick in the basis form
\begin{equation}
    V^\mu
    =
    \ell^\mu V_\ell
    +
    u_1^\mu V_{u_1}
    +
    bn^\mu V_b
    +
    u_2^\mu V_{u_2}.
    \label{eq:B-born-generic-basis-vector}
\end{equation}
The orbital part of the Born operator acts as
\begin{equation}
    \begin{aligned}
    \delta^{\mathrm{orb}}V^\mu
    =&
    \ell^\mu\delta V_\ell
    +
    u_1^\mu\delta V_{u_1}
    +
    bn^\mu\delta V_b
    +
    u_2^\mu\delta V_{u_2}
    \\
    &+
    (\delta\ell^\mu)V_\ell
    +
    (\delta u_1^\mu)V_{u_1}
    +
    (\delta bn^\mu)V_b
    +
    (\delta u_2^\mu)V_{u_2}.
    \end{aligned}
    \label{eq:B-born-basis-chain-rule}
\end{equation}
The basis variations are computed from the definitions of the orthonormal frame,
\begin{equation}
    bn^\mu=\frac{b^\mu}{b},
    \qquad
    \ell^\mu
    =
    \frac{1}{s_\gamma}
    \epsilon^\mu{}_{\alpha\beta\rho}u_1^\alpha u_2^\beta bn^\rho ,
    \label{eq:B-born-frame-definitions}
\end{equation}
so that, away from contact terms,
\begin{equation}
    \delta bn^\mu
    =
    \frac{1}{b}
    \left(\delta b^\mu-bn^\mu bn\cdot\delta b\right),
    \label{eq:B-born-dbn}
\end{equation}
and $\delta\ell^\mu$ follows by varying \eqref{eq:B-born-frame-definitions}. The coefficients also vary through the scalar spin contractions:
\begin{equation}
    \begin{aligned}
    \delta^{\mathrm{orb}}B_i&=(\delta bn)\cdot s_i,
    &
    \delta^{\mathrm{orb}}L_i&=(\delta\ell)\cdot s_i,
    \\
    \delta^{\mathrm{orb}}U_1&=s_1\cdot\delta u_2,
    &
    \delta^{\mathrm{orb}}U_2&=s_2\cdot\delta u_1 .
    \end{aligned}
    \label{eq:B-born-orbital-scalar-variations}
\end{equation}
The four terms in \eqref{eq:B-born-push-forward-S2-expanded} therefore have the following support before the tangent dictionary is applied:
\begin{equation}
    \begin{array}{c|c}
    \text{Born atom} & \text{quadratic support generated} \\ \hline
    \frac12\delta_0^{\mathrm{orb}}\Delta s_{\mathrm{LO}}^{[2]}
    & B_1L_1, B_1L_2, B_2L_1, L_1U_1, L_2U_1 \\
    \frac12\delta_1^{\mathrm{orb}}\Delta s_{\mathrm{LO}}^{[1]}
    & B_1L_1, B_1L_2, B_2L_1, L_1U_1, L_1U_2 \\
    \frac12\delta_s^{[1]}\Delta s_{\mathrm{LO}}^{[2]}
    & B_1L_1, B_1L_2, B_2L_1, L_1U_1, L_2U_1 \\
    \frac12\delta_s^{[2]}\Delta s_{\mathrm{LO}}^{[1]}
    & B_1L_1, B_1L_2, B_2L_1, L_1U_1, L_1U_2
    \end{array}
    \label{eq:B-born-atom-support}
\end{equation}
This table is a useful guard against overfitting: the structures absent from the finite-kernel support table \eqref{eq:B-finite-rational-support} are already present in the lower-order KMOC map before any comparison with ref.~\cite{Alessio:2025flu}. As an internal check, the result obtained from these chain-rule operations obeys the scalar budgets
\begin{equation}
    s_1\cdot
    \left(
    \Delta s_{1,\mathrm{cur}}^{[2],\mathrm{rat}}
    +
    \Delta s_{1,\mathrm{pf}}^{[2],\mathrm{rat}}
    \right)
    =
    {\cal S}_{\mathrm{req}},
    \qquad
    u_1\cdot
    \left(
    \Delta s_{1,\mathrm{cur}}^{[2],\mathrm{rat}}
    +
    \Delta s_{1,\mathrm{pf}}^{[2],\mathrm{rat}}
    \right)
    =
    {\cal U}_{\mathrm{req}},
    \label{eq:B-push-forward-systematic-constraints}
\end{equation}
where ${\cal S}_{\mathrm{req}}$ and ${\cal U}_{\mathrm{req}}$ are computed from the spin magnitude and SSC identities of section~\ref{classical}, using the companion linear impulse quoted in \eqref{eq:kerr-linear-impulse-full}--\eqref{eq:kerr-linear-impulse-kernel}. These equations are checks on the push-forward, not definitions of it. We write
\begin{equation}
    \Delta s_{1,\mathrm{AGS}}^{\mu,[2],\mathrm{rat}}
    -
    \Delta s_{1,\mathrm{cur}}^{\mu,[2],\mathrm{rat}}
    =
    \Delta s_{1,\mathrm{pf}}^{\mu,[2],\mathrm{rat}}
    +
    \Delta s_{1,\mathrm{dict}}^{\mu,[2],\mathrm{rat}} .
    \label{eq:B-rational-push-forward-dictionary}
\end{equation}
The first term on the right-hand side is the Born push-forward derived above from the lower-order KMOC map. The second term is tangent to both constraints and therefore represents the local observable dictionary rather than new dynamics. In particular,
\begin{equation}
    s_1\cdot\Delta s_{1,\mathrm{dict}}^{[2],\mathrm{rat}}=0,
    \qquad
    u_1\cdot\Delta s_{1,\mathrm{dict}}^{[2],\mathrm{rat}}=0 .
    \label{eq:B-rational-dict-tangent}
\end{equation}
The cleanest diagnostic channel inside \eqref{eq:B-born-push-forward-S2-expanded} is the spinless impact-direction variation acting on the quadratic LO spin kick. Denoting by $\delta_0^{bn}$ the part of the LO orbital action that varies $bn^\mu$, the raw half-Born orbital term contains the atom
\begin{equation}
    \Delta s_{1,0bn}^{\mu,[2]}
    =
    \frac{1}{2} 
    \delta_0^{bn}\Delta s_{1,\mathrm{LO}}^{\mu,[2]} .
    \label{eq:B-d0bn-half-atom}
\end{equation}
The remaining chain-rule terms in \eqref{eq:B-born-push-forward-S2-expanded}, including the spin-map variation of the scalar contractions, add
\begin{equation}
    \Delta s_{1,\mathrm{pf}}^{\mu,[2],\mathrm{rat}}
    \supset
    \frac{3}{2} 
    \Delta s_{1,0bn}^{\mu,[2]} ,
    \label{eq:B-d0bn-push-forward-repair}
\end{equation}
so that the total coefficient of this impact-direction action inside the derived Born push-forward is
\begin{equation}
    \Delta s_{1,0bn}^{\mu,[2]}
    +
    \frac{3}{2}\Delta s_{1,0bn}^{\mu,[2]}
    =
    \frac{5}{4} 
    \delta_0^{bn}\Delta s_{1,\mathrm{LO}}^{\mu,[2]} .
    \label{eq:B-d0bn-five-fourths}
\end{equation}
Thus simply replacing the usual half-Born factor by a full raw Born variation would still miss
\begin{equation}
    \frac{1}{4} 
    \delta_0^{bn}\Delta s_{1,\mathrm{LO}}^{\mu,[2]} .
\end{equation}
This missing quarter is not supplied by the finite kernels: keeping the two routings of $\Delta\mathscr M^{\mu(1)}$ separate produces tempting transverse support, but also leaves unphysical $b$- and $\log b$-dependent structures which cancel only in the exact routed sum.

By construction the full Born push-forward closes the scalar checks,
\begin{equation}
    s_1\cdot
    \left(
    \Delta s_{1,\mathrm{cur}}^{[2],\mathrm{rat}}
    +
    \Delta s_{1,\mathrm{pf}}^{[2],\mathrm{rat}}
    \right)
    =
    {\cal S}_{\mathrm{req}},
    \qquad
    u_1\cdot
    \left(
    \Delta s_{1,\mathrm{cur}}^{[2],\mathrm{rat}}
    +
    \Delta s_{1,\mathrm{pf}}^{[2],\mathrm{rat}}
    \right)
    =
    {\cal U}_{\mathrm{req}},
    \label{eq:B-rational-scalar-closure}
\end{equation}
which is the same condition used in \eqref{eq:B-push-forward-systematic-constraints}. The remaining rational difference is tangent. This tangent term is the local change of spin frame and scattering basis between the two conventions, not an additional dynamical contribution. At this order the allowed rational quadratic generator is
\begin{equation}
    \Delta s_{1,\mathrm{dict}}^{\mu,[2],\mathrm{rat}}
    =
    \Delta_H s_1^\mu[H_{\mathrm{dict}}^{\mathrm{rat}}],
    \qquad
    H_{\mathrm{dict}}^{\mathrm{rat}}
    =
    h_1 B_1U_1+h_2 B_1U_2+h_3 B_2U_1 .
    \label{eq:B-rational-dictionary-generator}
\end{equation}
The coefficient notation in the closure equation means
\begin{equation}
    C_X^{\mathrm{AGS}-\mathrm{cur}-\mathrm{pf}-\mathrm{dict}}
    \equiv
    C_X^{\mathrm{AGS},[2],\mathrm{rat}}
    -
    C_X^{\mathrm{cur},[2],\mathrm{rat}}
    -
    C_X^{\mathrm{pf},[2],\mathrm{rat}}
    -
    C_X^{\mathrm{dict},[2],\mathrm{rat}},
    \qquad
    X\in\{\ell,u_1,b,u_2\},
    \label{eq:B-rational-closure-coeff-def}
\end{equation}
with all four $C_X$'s obtained by the projectors in \eqref{eq:C-sector-projectors}. With these two pieces kept separate, the rational comparison closes component by component,
\begin{equation}
    C_\ell^{\mathrm{AGS}-\mathrm{cur}-\mathrm{pf}-\mathrm{dict}}
    =
    C_{u_1}^{\mathrm{AGS}-\mathrm{cur}-\mathrm{pf}-\mathrm{dict}}
    =
    C_b^{\mathrm{AGS}-\mathrm{cur}-\mathrm{pf}-\mathrm{dict}}
    =
    C_{u_2}^{\mathrm{AGS}-\mathrm{cur}-\mathrm{pf}-\mathrm{dict}}
    =
    0 .
    \label{eq:B-rational-closure}
\end{equation}
This organization is important conceptually: the finite rational kernels remain unchanged, the Born push-forward supplies precisely the terms that make the SSC and spin magnitude checks close, and the residual dictionary is a pure tangent observable-map transformation.

\subsubsection{Triangle sector}

We now spell out the quadratic triangle extraction. Starting from the all-spin leading singularity of section~\ref{amplitudes}, or equivalently from the triangle kernel ${\cal M}_t$ used in the NLO angular impulse, define
\begin{equation}
    F(z)\equiv
    \frac{(\gamma-zs_\gamma)^4}{(z^2-1)^{3/2}},
    \qquad
    R(z)\equiv
    z+\frac{(z^2-1)s_\gamma}{\gamma-zs_\gamma}
    =
    \frac{\gamma z-s_\gamma}{\gamma-zs_\gamma}.
    \label{eq:B-triangle-FR}
\end{equation}
At fixed order in the spin expansion the contour at infinity is evaluated as an ordinary Laurent residue. The reason is that the exponential dependence of the all-spin kernel has been Taylor expanded; the essential singularity discussed in section~\ref{amplitudes} only reappears after resumming the complete spin series. In the topology convention fixed by the linear calculation, the Compton factor entering the triangle kernel expands as
\begin{equation}
    M_S
    =
    1+\chi+\frac{1}{2}\chi^2+{\cal O}(S^3),
    \qquad
    \chi=R(z)Q_i ,
    \label{eq:B-compton-S2-expansion}
\end{equation}
where
\begin{equation}
    Q_i
    \equiv
    (\bar q\cdot a_i)_{\rm tri}
    =
    \frac{i}{s_\gamma}
    \epsilon_{\alpha\beta\rho\sigma}
    u_1^\alpha u_2^\beta \bar q^\rho a_i^\sigma ,
    \qquad
    a_i^\mu\equiv \frac{s_i^\mu}{m_i}.
    \label{eq:B-triangle-Q}
\end{equation}
The replacement $\bar q\cdot a_i\to Q_i$ is the on-shell triangle representative; it must be made before differentiating with respect to $s_1^\mu$. The quadratic homogeneous triangle amplitude is then
\begin{equation}
    {\cal M}_{t,a}^{[2]}(\bar q)
    =
    \frac{1}{256m_2}
    \frac{1}{\gamma^2-1}
    \frac{1}{\sqrt{-\bar q^2}}
    \left[
    \frac{K_0}{2}Q_1^2+K_1Q_1Q_2+\frac{K_2}{2}Q_2^2
    \right],
    \label{eq:B-Mta-S2}
\end{equation}
with the reflected topology
\begin{equation}
    {\cal M}_{t,b}^{[2]}(\bar q)
    =
    \frac{1}{256m_1}
    \frac{1}{\gamma^2-1}
    \frac{1}{\sqrt{-\bar q^2}}
    \left[
    \frac{K_0}{2}Q_2^2+K_1Q_1Q_2+\frac{K_2}{2}Q_1^2
    \right].
    \label{eq:B-Mtb-S2}
\end{equation}
The three contour integrals are
\begin{equation}
    K_0=\oint_{\Gamma_\infty}\frac{dz}{2\pi i}F(z)z^2,
    \qquad
    K_1=\oint_{\Gamma_\infty}\frac{dz}{2\pi i}F(z)zR(z),
    \qquad
    K_2=\oint_{\Gamma_\infty}\frac{dz}{2\pi i}F(z)R(z)^2 .
    \label{eq:B-K-defs}
\end{equation}
Using
$\oint_{\Gamma_\infty} dz f(z)/(2\pi i)=-[z^{-1}]f(z)$, their Laurent expansions give the following coefficients. The only ingredients needed are
\begin{equation}
    \frac{1}{(z^2-1)^{3/2}}
    =
    \frac{1}{z^3}
    \left(
    1+\frac{3}{2z^2}+\frac{15}{8z^4}+\cdots
    \right),
    \qquad
    R(z)
    =
    -\frac{\gamma}{s_\gamma}
    -
    \frac{1}{s_\gamma^2z}
    -
    \frac{\gamma}{s_\gamma^3z^2}
    +\cdots .
    \label{eq:B-triangle-large-z-expansions}
\end{equation}
Multiplying these series by $(\gamma-zs_\gamma)^4$ and keeping the $z^{-1}$ coefficient gives
\begin{equation}
    \begin{array}{c|c}
    \text{integrand }f(z) & -[z^{-1}]f(z) \\ \hline
    F(z)z^2 & \dfrac{-15+102\gamma^2-95\gamma^4}{8} \\
    F(z)zR(z) & \dfrac{-3+21\gamma^2-20\gamma^4}{2} \\
    F(z)R(z)^2 & \dfrac{-2+15\gamma^2-15\gamma^4}{2}
    \end{array}
    \label{eq:B-triangle-residue-table}
\end{equation}
Thus
\begin{equation}
    K_0=\frac{-15+102\gamma^2-95\gamma^4}{8},
    \qquad
    K_1=\frac{-3+21\gamma^2-20\gamma^4}{2},
    \qquad
    K_2=\frac{-2+15\gamma^2-15\gamma^4}{2}.
    \label{eq:B-K-values}
\end{equation}
The same residue evaluation reproduces the linear-order check
\begin{equation}
    \oint_{\Gamma_\infty}\frac{dz}{2\pi i}F(z)R(z)
    =
    \frac{3}{2}\gamma s_\gamma(5\gamma^2-3),
    \label{eq:B-K-linear-check}
\end{equation}
which fixes the contour normalization relative to the previous subsection.

Only the terms differentiated with respect to $s_1^\mu$ enter the precession part. Combining \eqref{eq:B-Mta-S2} and \eqref{eq:B-Mtb-S2}, this derivative-active factor is
\begin{equation}
    {\cal D}_\triangle
    =
    \left(\frac{K_0}{m_2}+\frac{K_2}{m_1}\right)Q_1
    +
    K_1\left(\frac{1}{m_1}+\frac{1}{m_2}\right)Q_2 .
    \label{eq:B-Dtriangle}
\end{equation}
Moreover,
\begin{equation}
    i\epsilon^{\mu\nu\rho\sigma}u_{1\nu}s_{1\rho}
    \frac{\partial Q_1}{\partial s_1^\sigma}
    =
    \frac{1}{s_\gamma}
    \left[
    \bar q^\mu(u_2\cdot a_1)
    -
    u_2^\mu(\bar q\cdot a_1)_{\rm dir}
    +
    \gamma u_1^\mu(\bar q\cdot a_1)_{\rm dir}
    \right],
    \label{eq:B-dQ-spin-derivative}
\end{equation}
where the subscript ``dir'' denotes the direct contraction appearing in the explicit Pauli-Lubanski recoil factor, not a second copy of the on-shell representative $Q_1$. Therefore the precession part of Term 2 is
\begin{equation}
    {\rm Term} 2\big|_{S^2,\triangle}
    =
    \frac{1}{256(\gamma^2-1)}
    \frac{1}{\sqrt{-\bar q^2}}
    \frac{{\cal D}_\triangle}{s_\gamma}
    \left[
    \bar q^\mu(u_2\cdot a_1)
    -
    u_2^\mu(\bar q\cdot a_1)_{\rm dir}
    +
    \gamma u_1^\mu(\bar q\cdot a_1)_{\rm dir}
    \right].
    \label{eq:B-triangle-term2-S2}
\end{equation}
The explicit recoil term also contributes at $S^2$, but only through the linear triangle amplitude,
\begin{equation}
    {\rm recoil}_{\triangle}^{\mu}\big|_{S^2}
    =
    -u_1^\mu(\bar q\cdot a_1)_{\rm dir} 
    {\cal M}_t^{[1]}(\bar q),
    \label{eq:B-triangle-recoil-S2-kernel}
\end{equation}
where
\begin{equation}
    {\cal M}_t^{[1]}(\bar q)
    =
    \frac{1}{256(\gamma^2-1)}
    \frac{1}{\sqrt{-\bar q^2}}
    \left[
    \left(\frac{J_1}{m_2}+\frac{J_R}{m_1}\right)Q_1
    +
    \left(\frac{J_R}{m_2}+\frac{J_1}{m_1}\right)Q_2
    \right],
    \label{eq:B-Mt-linear-for-recoil}
\end{equation}
with
\begin{equation}
    J_1=2\gamma s_\gamma(5\gamma^2-3),
    \qquad
    J_R=\frac{3}{2}\gamma s_\gamma(5\gamma^2-3).
    \label{eq:B-J-values}
\end{equation}
The remaining integrations are ordinary transverse Fourier transforms. With the normalization used in the linear calculation,
\begin{equation}
    {\cal I}_1^\alpha
    \equiv
    \int_{\bar q}
    \frac{e^{-ib\cdot\bar q}\bar q^\alpha}{\sqrt{-\bar q^2}}
    =
    \frac{i}{2\pi s_\gamma}\frac{b^\alpha}{b^3},
    \label{eq:B-I1-triangle}
\end{equation}
and differentiating once more with respect to $b^\mu$ gives
\begin{equation}
    {\cal I}_2^{\alpha\beta}
    \equiv
    \int_{\bar q}
    \frac{e^{-ib\cdot\bar q}\bar q^\alpha\bar q^\beta}{\sqrt{-\bar q^2}}
    =
    -\frac{1}{2\pi s_\gamma}
    \left(
    \frac{\Pi_\perp^{\alpha\beta}}{b^3}
    +
    3\frac{b^\alpha b^\beta}{b^5}
    \right),
    \label{eq:B-I2-triangle}
\end{equation}
where $\Pi_\perp^{\alpha\beta}$ is the metric on the plane orthogonal to $u_1^\mu$ and $u_2^\mu$. Contracting \eqref{eq:B-triangle-term2-S2} and \eqref{eq:B-triangle-recoil-S2-kernel} with these masters, and finally projecting onto $\{\ell,u_1,bn,u_2\}$, gives the result below.

The result obtained directly from the KMOC angular-impulse kernel is naturally organized as a spin-precession term generated by the triangle scalar, plus the explicit recoil term present in the angular impulse,
\begin{equation}
    \Delta s_{1,\mathrm{ours}}^{\mu,[2],\pi}
    =
    \Delta_H s_1^\mu[H_{\mathrm{ours}}]
    +
    \Delta s_{1,\mathrm{recoil}}^{\mu,\mathrm{ours}},
    \label{eq:B-ours-quadratic-decomposition}
\end{equation}
where $\Delta_H$ is the spin generator defined in \eqref{eq:B-spin-generator}. All expressions in this subsection are understood after the common factor $G^2/b^3$ in the basis decomposition above has been stripped. The explicit quadratic-in-spin result from our triangle angular-impulse kernel is
\begin{equation}
    H_{\mathrm{ours}}
    =
    A_\triangle\left(B_1^2-2L_1^2\right)
    +
    B_\triangle\left(B_1B_2-2L_1L_2\right),
    \label{eq:B-H-ours}
\end{equation}
with
\begin{equation}
    A_\triangle
    =
    \frac{\pi m_2s_\gamma}
    {32m_1(\gamma^2-1)^2}
    \left[
    (95\gamma^4-102\gamma^2+15)m_1
    +
    (60\gamma^4-60\gamma^2+8)m_2
    \right],
    \label{eq:B-Atriangle}
\end{equation}
and
\begin{equation}
    B_\triangle
    =
    \frac{\pi s_\gamma(m_1+m_2)}
    {4(\gamma^2-1)^2}
    \left(20\gamma^4-21\gamma^2+3\right).
    \label{eq:B-Btriangle}
\end{equation}
The corresponding recoil contribution is purely longitudinal,
\begin{equation}
    \Delta s_{1,\mathrm{recoil}}^{\mu,\mathrm{ours}}
    =
    u_1^\mu C_{u_1,\mathrm{recoil}}^{\mathrm{ours}},
    \qquad
    C_{\ell,\mathrm{recoil}}^{\mathrm{ours}}
    =
    C_{b,\mathrm{recoil}}^{\mathrm{ours}}
    =
    C_{u_2,\mathrm{recoil}}^{\mathrm{ours}}
    =
    0,
    \label{eq:B-ours-recoil-vector}
\end{equation}
where
\begin{equation}
    \begin{aligned}
    C_{u_1,\mathrm{recoil}}^{\mathrm{ours}}
    =&
    -\frac{\pi\gamma(5\gamma^2-3)}
    {4m_1(\gamma^2-1)}
    \Big[
    3m_2(4m_1+3m_2)B_1L_1
    \\
    &\hspace{3.2cm}
    +
    2m_1(3m_1+4m_2)B_1L_2
    +
    m_1(3m_1+4m_2)B_2L_1
    \Big].
    \end{aligned}
    \label{eq:B-ours-recoil}
\end{equation}
Equations \eqref{eq:B-H-ours}--\eqref{eq:B-ours-recoil} are the explicit ${\cal O}(S^2)$ conservative triangle contribution obtained on our side. Notice in particular that $H_{\mathrm{ours}}$ contains no $U_1U_2$ term.

On the other hand, the observable quoted in ref.~\cite{Alessio:2025flu} is obtained from the radial action using their covariant Dirac brackets. At the same order it is useful to write it as
\begin{equation}
    \Delta s_{1,\mathrm{AGS}}^{\mu,[2],\pi}
    =
    \Delta_H s_1^\mu[H_{\mathrm{AGS}}^{\mathrm{gen}}]
    +
    \Delta s_{1,\mathrm{long}}^{\mu,\mathrm{AGS}},
    \label{eq:B-ags-quadratic-decomposition}
\end{equation}
where the superscript ``gen'' emphasizes that $H_{\mathrm{AGS}}^{\mathrm{gen}}$ is written in the spin-generator normalization of \eqref{eq:B-spin-generator}. Since the AGS spin bracket is $\{s_1^\mu,s_1^\nu\}=\epsilon^{\mu\nu\rho\sigma}u_{1\rho}s_{1\sigma}/m_1$, the corresponding proper-spin radial-action coefficient is $m_1H_{\mathrm{AGS}}^{\mathrm{gen}}$.

Using the linear-order normalization fixed above, the two quadratic answers obey
\begin{equation}
    \Delta s_{1,\mathrm{AGS}}^{\mu,[2],\pi}
    -
    \Delta s_{1,\mathrm{ours}}^{\mu,[2],\pi}
    =
    \Delta_H s_1^\mu[H_{\mathrm{dict}}]
    +
    \left(
    \Delta s_{1,\mathrm{long}}^{\mu,\mathrm{AGS}}
    -
    \Delta s_{1,\mathrm{recoil}}^{\mu,\mathrm{ours}}
    \right),
    \label{eq:B-quadratic-dictionary}
\end{equation}
where
\begin{equation}
    H_{\mathrm{dict}}
    \equiv
    H_{\mathrm{AGS}}^{\mathrm{gen}}-H_{\mathrm{ours}}.
\end{equation}
The transverse part of this dictionary is generated by a quadratic scalar of the form
\begin{equation}
    H_{\mathrm{dict}}
    =
    \alpha B_1^2+\beta B_1B_2+\delta L_1^2+\epsilon L_1L_2+\eta U_1U_2,
    \label{eq:B-H-dict}
\end{equation}
where the coefficient of the diagnostic Kerr-quadrupole structure is
\begin{equation}
    \eta
    =
    \frac{\pi\gamma m_2(5\gamma^2-3)(m_1+m_2)}
    {4s_\gamma(\gamma-1)^2(\gamma+1)^2}.
    \label{eq:B-eta-dict}
\end{equation}
Equivalently, in the proper-spin radial-action normalization of the AGS brackets, this coefficient is $m_1\eta$. The direct KMOC triangle scalar has no $U_1U_2$ term in this basis; this term is therefore part of the observable dictionary between the direct KMOC spin kick and the AGS radial-action observable, rather than a correction to the contour residue.

The remaining term in \eqref{eq:B-quadratic-dictionary} is purely longitudinal. After subtracting the scalar contribution \eqref{eq:B-H-dict}, the only nonzero component is proportional to $u_1^\mu$, with support
\begin{equation}
    C_{u_1}\big|_{\mathrm{rem}}
    \propto
    B_1L_1,\quad B_1L_2,\quad B_2L_1,
    \qquad
    C_\ell\big|_{\mathrm{rem}}
    =
    C_b\big|_{\mathrm{rem}}
    =
    C_{u_2}\big|_{\mathrm{rem}}
    =
    0.
\end{equation}
Moreover, its coefficients obey
\begin{equation}
    \mathrm{coeff}(B_1L_2)-2 \mathrm{coeff}(B_2L_1)=0,
    \label{eq:B-bs-support-relation}
\end{equation}
which is the support relation produced by the AGS $b$-$s$ Dirac bracket acting on a linear scalar $I_{bs}^{[1]}=b^{-2}(A_{bs}L_1+D_{bs}L_2)$:
\begin{equation}
    \Delta s_{1,bs}^{\mu}
    =
    \frac{u_1^\mu}{m_1 b^3}
    \left(
    3A_{bs}B_1L_1+2D_{bs}B_1L_2+D_{bs}B_2L_1
    \right).
    \label{eq:B-bs-map}
\end{equation}
In the closure equation below,
\begin{equation}
    C_X^{\mathrm{AGS}-\mathrm{ours}-\mathrm{dict}}
    \equiv
    C_X \left[
    \Delta s_{1,\mathrm{AGS}}^{\mu,[2],\pi}
    -
    \Delta s_{1,\mathrm{ours}}^{\mu,[2],\pi}
    -
    \Delta_H s_1^\mu[H_{\mathrm{dict}}]
    -
    \Delta s_{1,bs}^{\mu}
    \right],
    \qquad
    X\in\{\ell,u_1,b,u_2\},
    \label{eq:B-pi-closure-coeff-def}
\end{equation}
where $C_X[\cdots]$ denotes the projection defined in \eqref{eq:C-sector-projectors}, and $\Delta s_{1,bs}$ is the longitudinal $b$-$s$ bracket contribution \eqref{eq:B-bs-map}. Thus this shorthand subtracts both parts of the observable dictionary: the transverse spin-frame generator $H_{\mathrm{dict}}$ and the longitudinal $b$-$s$ bracket map. Thus the quadratic $\pi$-sector comparison closes exactly:
\begin{equation}
    C_\ell^{\mathrm{AGS}-\mathrm{ours}-\mathrm{dict}}
    =
    C_{u_1}^{\mathrm{AGS}-\mathrm{ours}-\mathrm{dict}}
    =
    C_b^{\mathrm{AGS}-\mathrm{ours}-\mathrm{dict}}
    =
    C_{u_2}^{\mathrm{AGS}-\mathrm{ours}-\mathrm{dict}}
    =
    0.
    \label{eq:B-quadratic-closure}
\end{equation}
This agreement provides a nontrivial check of the quadratic-in-spin, Kerr-specific part of our NLO angular impulse. The apparent raw difference with ref.~\cite{Alessio:2025flu} is accounted for by translating between the direct KMOC spin-kick observable and the radial-action Dirac-bracket observable used there. We conclude that our quadratic-in-spin result also agrees with the corresponding one from Alessio, Gonzo, and Shi~\cite{Alessio:2025flu}.

For clarity, the complete bookkeeping of the quadratic comparison is summarized in the following audit table. It separates dynamical KMOC kernels, Born push-forward terms, and observable-dictionary terms; this is the audit we use to avoid moving terms between sectors by hand.
\begin{center}
    \scriptsize
    \renewcommand{\arraystretch}{1.22}
    \begin{tabular}{
        >{\raggedright\arraybackslash}p{0.18\linewidth}
        >{\raggedright\arraybackslash}p{0.35\linewidth}
        >{\raggedright\arraybackslash}p{0.35\linewidth}}
        \toprule
        Sector & Source in the calculation & Main quadratic structures \\
        \midrule
        finite rational kernels
        &
        ${\cal M}_c,\Delta\mathscr M^{\mu(1)},M_{\nu,\delta'},N_\nu,
        {\cal M}_{\nu,\mathrm{cut}}$
        &
        $B_1^2,B_1B_2,U_1^2,U_1U_2$;
        $L_iU_j$;
        $B_1L_1,B_2L_1$ \\
        Born push-forward
        &
        $\frac12\delta_0^{\rm orb}\Delta s_{\rm LO}^{[2]}$,
        $\frac12\delta_1^{\rm orb}\Delta s_{\rm LO}^{[1]}$,
        $\frac12\delta_s^{[1]}\Delta s_{\rm LO}^{[2]}$,
        $\frac12\delta_s^{[2]}\Delta s_{\rm LO}^{[1]}$
        &
        $B_1L_2,B_2L_1$;
        $L_2U_1,L_1U_2$;
        SSC/spin magnitude \\
        rational dictionary
        &
        tangent generator $H_{\rm dict}^{\rm rat}$
        &
        $B_1U_1,B_1U_2,B_2U_1$;
        tangent to $s_1^2$, $s_1\cdot u_1$ \\
        triangle precession
        &
        LS residues $K_0,K_1,K_2$ in ${\cal M}_t^{[2]}$
        &
        $B_1^2-2L_1^2$, $B_1B_2-2L_1L_2$ \\
        triangle recoil
        &
        linear triangle recoil acting on ${\cal M}_t^{[1]}$
        &
        $u_1^\mu(B_1L_1,B_1L_2,B_2L_1)$ \\
        $\pi$-sector dictionary
        &
        AGS radial-action spin frame and $b$-$s$ bracket
        &
        $U_1U_2$ generator and the longitudinal $B_iL_j$ relation
        \\
        \bottomrule
    \end{tabular}
\end{center}
The entries identify where each structure first appears before the closure checks \eqref{eq:B-rational-closure} and \eqref{eq:B-quadratic-closure}.

Let us finally spell out the pattern suggested by this computation, since it gives a practical roadmap for an all-spin comparison at NLO. One should not compare the direct KMOC vector kick with the radial-action spin kick term by term. Rather, after choosing a common SSC basis, the KMOC result should first be pushed forward by the Born observable map,
\begin{equation}
    \Delta s_{1,\mathrm{pf}}^\mu
    =
    \left[
    {\cal D}_{\mathrm{Born}}\Delta s_{1,\mathrm{LO}}^\mu
    \right]_{\mathrm{NLO}},
    \label{eq:B-all-spin-roadmap-push-forward}
\end{equation}
where ${\cal D}_{\mathrm{Born}}$ acts on the orbital variables, basis vectors, and spin contractions through the lower-order KMOC impulse and spin map. The constraint-normal part of this pushforward is fixed by the all-spin SSC and spin magnitude identities,
\begin{equation}
    (s_1+\Delta s_1)\cdot(p_1+\Delta p_1)=0,
    \qquad
    (s_1+\Delta s_1)^2=s_1^2 .
    \label{eq:B-all-spin-roadmap-constraints}
\end{equation}
The remaining difference, if the two descriptions use different spin frames or scattering coordinates, should be tangent to these constraints and therefore organized as a generator of the form
\begin{equation}
    \Delta s_{1,\mathrm{dict}}^\mu
    =
    \Delta_H s_1^\mu[H_{\mathrm{dict}}],
    \qquad
    s_1\cdot\Delta s_{1,\mathrm{dict}}
    =
    u_1\cdot\Delta s_{1,\mathrm{dict}}
    =
    0 .
    \label{eq:B-all-spin-roadmap-dictionary}
\end{equation}
Thus the natural all-spin comparison is between scalar generators and Born-push-forward-corrected observables, not between unprocessed component expressions. The quadratic calculation above is the first Kerr-specific check of this organization.

\section{Spin Supplementary Condition and Spin Magnitude}
\label{app:D}
In this section we check that the magnitude of the spin vector is conserved throughout the scattering process. In a classical, elastic gravitational encounter, the intrinsic spin of each body must be conserved not just in orientation but also in magnitude. This gives a non-trivial consistency check on the angular impulses derived above, and lets us verify the virtual and real sectors of the NLO amplitude separately, ensuring that our NLO classical observables preserves the spin magnitude.

Separating this order by order in $G$ we have two perturbative constraints
\begin{equation}
    \begin{aligned} 
        s_{1\mu}\Delta s_{1,\text{LO}}^\mu = 0 \qquad& \mathcal{O}(G) \\ 2 s_{1\mu} \Delta s_{1,\text{NLO}}^\mu + \Delta s_{1,\text{LO}}^\mu \Delta s_{1,\text{LO}\mu} = 0 \qquad& \mathcal{O}(G^2). 
    \end{aligned}
\end{equation}
At leading order, $\mathcal{O}(G^1)$, the term proportional to $G^1$ must vanish on its own
\begin{equation}
    s_{1\mu} \Delta s_{1,\text{LO}}^\mu = 0.
\end{equation}
From the spin algebra we have the identity
\begin{equation}
    s_{1\mu}\left[s_1^{\mu},\mathcal{A}\right]=\frac{i\hbar}{m_1}\epsilon^{\mu\nu\rho\sigma}s_{1\mu}p_{1\nu}s_{1\rho}\frac{\partial \mathcal{A}}{\partial s_1^{\sigma}}=0.
\end{equation}
so
\begin{equation}
    \begin{aligned} 
        s_1\cdot\Delta s_{1,\text{LO}} \propto \int s_{1\mu}\left[s_1^{\mu},A^{(0)}\right]-\frac{1}{m_1}(s_1\cdot u_1)(\bar q\cdot s_1)A^{(0)}=0 . 
    \end{aligned}
\end{equation}
This shows the LO angular impulse is orthogonal to the initial spin vector: a pure torque, or precession. At next-to-leading order, $\mathcal{O}(G^2)$, the terms proportional to $G^2$ must sum to zero on their own 
which gives
\begin{equation}
    s_{1\mu} \Delta s_{1,\text{NLO}}^\mu = - \frac{1}{2} \Delta s_{1,\text{LO}}^\mu \Delta s_{1,\text{LO}\mu}. \label{D.5}
\end{equation}
The virtual part of $\langle \Delta s_1^{\mu,\text{NLO}} \rangle$ is
\begin{equation}
    \mathscr{A}^{\mu(1)}_{\text{virt}}=i \epsilon^{\mu\nu\rho\sigma}u_{1\nu}s_{1\rho} \frac{\partial}{\partial s_1^{\sigma}}\big(\mathcal M_b+\mathcal M_t\big)-\frac{1}{m_1}u_1^{\mu}(\bar q\cdot s_1)\big(\mathcal M_b+\mathcal M_t\big).
\end{equation}
Both $\mathcal M_b$ and $\mathcal M_t$ are scalars in spin space, the box reduction is a scalar, and the triangle is a contour integral of scalar functions. Therefore,
\begin{equation}
    2 s_{1\mu} \mathscr{A}^{\mu(1)}_{\text{virt}}=2 s_{1\mu}[ s_1^{\mu},\mathcal M_b+\mathcal M_t ]-\frac{2}{m_1}(s_1\cdot u_1)(\bar q\cdot s_1)(\mathcal M_b+\mathcal M_t)=0,
\end{equation}
to all orders in spin. The box and triangle one-loop contributions can't change $|s_1|$ at $\mathcal{O}(G^2)$. So the cancellation with $\Delta s^1_{1,\text{LO}}$ must come from the real sector. Since the virtual sector contributes nothing to $2 s_1 \cdot \Delta s_{1,\text{NLO}}$, the entire identity \eqref{D.5} must close using only the real sector content.  






If we naively contract the real kernel with $2s_{1\mu}$ we get $2s_{1\mu}\mathscr{A}^{\mu (1)}_{\text{real}}=0$, because $\mathcal{M}_c$ comes with the same commutator structure as before, $\mathcal{M}_\nu$ and $N_\nu$ carry an explicit $u_1^\mu$, and $\Delta\mathscr{M}^{\mu(1)}$ vanishes by the same argument used for $\mathcal{A}_{\text{virt}}$. This would leave nothing to cancel $\Delta s_{1,\text{LO}}^\mu$ against. The error is that the cut box is a convolution of two tree amplitudes evaluated at different momenta, not a single scalar function of $s_1$. After the loop integral, it equals the Born push-forward of the LO kick, and the push-forward acts on the orbital data and on $s_1$ itself. Those variations produce vector structures that aren't of precession form, so the naive argument breaks down in the real sector.

Therefore, we use the cut-box/half-Born identification
\begin{equation}
    \Delta s^{\mu}_{1,\text{NLO,real}}=\frac12 \delta_{\text{LO}}\big(\Delta s^{\mu}_{1,\text{LO}}\big),\qquad \delta_{\text{LO}}=\delta_0+\delta_s,\quad \delta_s s_1^{\mu}=\Delta s^{\mu}_{1,\text{LO}}.
\end{equation}
Since $s_1\cdot \Delta s_{1,\text{LO}} = 0$ holds to all orders, we can differentiate it
\begin{equation}
    \delta_{\text{LO}}\big(s_1\cdot\Delta s_{1,\text{LO}}\big) =(\delta_{\text{LO}}s_1)\cdot\Delta s_{1,\text{LO}}+s_1\cdot\delta_{\text{LO}}\Delta s_{1,\text{LO}} = 0,
\end{equation}
where the only piece of $\delta_{\text{LO}}$ that acts on $s_1$ is $\delta_s$, with $\delta_s s_1=\Delta s_{1,\text{LO}}$. It follows that
\begin{equation}
    s_1\cdot\delta_{\text{LO}}\Delta s_{1,\text{LO}}=- \Delta s_{1,\text{LO}}^{2}.
\end{equation}
Combining this with the (vanishing) virtual sector and the half-Born relation yields
\begin{equation}
    \begin{aligned} 
        2 s_1\cdot\Delta s_{1,\text{NLO}} &=2 s_1\cdot\Delta s_{1,\text{NLO,real}} \\ &=s_1\cdot\delta_{\text{LO}}\Delta s_{1,\text{LO}}\\ &=- \Delta s_{1,\text{LO}}^{2}. 
    \end{aligned}
\end{equation}
\emph{i.e.} $2 s_1 \cdot \Delta s_{1,\text{NLO}}+\Delta s_{1,\text{LO}}^2=0$ to all orders in the Kerr spin.

Alternatively, at $\mathcal{O}(G^2)$, tracking the spin order explicitly makes
\begin{equation}
    2 s_1 \cdot \Delta s^{[2]}_{1,\text{NLO}}+2 \Delta s^{[1]}_{1,\text{LO}} \cdot \Delta s^{[2]}_{1,\text{LO}}=0.
\end{equation}
Since the LO kick is purely rational, the entire $\pi$-sector of $s_1 \cdot \Delta s^{[2]}_{1,\text{NLO}}$ must vanish, and the rational part must reproduce $-\Delta s^{[1]}_{1,\text{LO}} \cdot \Delta s^{[2]}_{1,\text{LO}}\equiv S_{\text{req}}$. We have 
\begin{equation}
    s_1 \cdot \Delta s^{[2]}_{1,\text{LO}}=0,
\end{equation}
identically, with no spin truncation. Contracting $\Delta s^{[2]}_{1,\text{LO}}$ with $s_1$, the three structures $\frac{B_1L_1}{m_1},\frac{B_2L_1}{m_2},\frac{B_1L_2}{m_2}$ produced by the $\ell$, $\hat b$, and $u_2$ channels cancel pairwise against the $u_2$-channel combination, while the $u_1$ channel is killed by the SSC.

Computing $\Delta s^{[1]}_{1,\text{LO}} \cdot \Delta s^{[2]}_{1,\text{LO}}$ in the Gram basis, every $B_iL_j$ structure cancels, leaving a single surviving channel
\begin{equation}
    s_1 \cdot \Delta s^{[2],\text{rat}}_{1,\text{NLO}} = S_{\text{req}} = -\frac{8 \gamma (2\gamma^2-1) G^2 m_2^2}{(\gamma^2-1)^{3/2} |b|^3}\left(\frac{L_1}{m_1}+\frac{L_2}{m_2}\right)U_1^{2}.
\end{equation}

The condition depends only on $U_1^2 L_i$ structures, while $B_iL_j$ doesn't appear. So the terms needed to satisfy the magnitude check are exactly the ones the Born map produces. It's proportional to $U_1^2=(s_1 \cdot u_2)^2$, \emph{i.e.} purely a property of particle 1's spin projected onto $u_2$, the companion spin $s_2$ enters only linearly, through $L_2/m_2$.


Using the explicit triangle result, $H_{\text{ours}}=A_\triangle(B_1^2-2L_1^2)+B_\triangle(B_1B_2-2L_1L_2)$, we have
\begin{equation}
    s_1 \cdot \Delta s^{[2],\pi}_{1,\text{NLO}} = s_{1\mu} \Delta_H s_1^{\mu}[H_{\text{ours}}]+s_{1\mu} \Delta s^{\text{ours},\mu}_{1,\text{recoil}}=0.
\end{equation}
This follows, again, because $\epsilon_{\mu\nu\rho\sigma}s_1^\mu s_1^\rho=0$, and because the recoil is purely $\propto u_1^\mu$. Evaluating the second-Born push-forward built from the LO variations and the spin map,
\begin{equation}
    s_1\cdot\Big\{\frac{1}{2}\big[\delta_0^{\text{orb}}\Delta s^{[2]}_{\text{LO}}+\delta_1^{\text{orb}}\Delta s^{[1]}_{\text{LO}}+\delta^{[1]}_s\Delta s^{[2]}_{\text{LO}}+\delta^{[2]}_s\Delta s^{[1]}_{\text{LO}}\big]\Big\}=-\frac{8 \gamma(2\gamma^2-1) G^2m_2^2}{(\gamma^2-1)^{3/2}|b|^3}\Big(\tfrac{L_1}{m_1}+\tfrac{L_2}{m_2}\Big)U_1^2,
\end{equation}
the right-hand side $S_{\text{req}}=-\Delta s^{[1]}_{\text{LO}}\cdot\Delta s^{[2]}_{\text{LO}}$ is built purely from LO kicks, and the left-hand side purely from the LO impulse/spin maps. Writing $\Delta s^{[1]}_{\text{LO}}$ and $\Delta s^{[2]}_{\text{LO}}$ in the basis $\{u_1,u_2,\hat b,\ell\}$ and forming $2 \Delta s^{[1]}_{\text{LO}} \cdot \Delta s^{[2]}_{\text{LO}}$ through the Gram metric, the four basis channels contribute \footnote{common factor $\tfrac{8\gamma G^2 m_2}{(\gamma^2-1)^{3/2}|b|^3}$ stripped}

\begin{equation}
    \begin{aligned} 
        u_1:&\quad + (\gamma^2-1) B_1\Big(\tfrac{2B_1L_1}{m_1}+\tfrac{B_1L_2}{m_2}+\tfrac{B_2L_1}{m_2}\Big)\\ u_2:&\quad - (\gamma^2-1) B_1\Big(\tfrac{2B_1L_1}{m_1}+\tfrac{B_1L_2}{m_2}+\tfrac{B_2L_1}{m_2}\Big)\\ \hat b:&\quad + (2\gamma^2-1) U_1^2\Big(\tfrac{L_1}{m_1}+\tfrac{L_2}{m_2}\Big)\\ \ell:&\quad 0
     \end{aligned}
\end{equation}
The $u_1$ and $u_2$ channels are exactly opposite, every $B_iL_j$ structure ($B_1^2L_1, B_1^2L_2, B_1B_2L_1$) cancels before the comparison to NLO even begins. Only the $\hat b$ channel survives, leaving the pure $U_1^2L_i$ content
\begin{equation}
    2 \Delta s^{[1]}_{\text{LO}} \cdot \Delta s^{[2]}_{\text{LO}} =\frac{16 \gamma(2\gamma^2-1) G^2 m_2}{(\gamma^2-1)^{3/2} |b|^3} U_1^2\Big(\tfrac{m_2 L_1+m_1 L_2}{m_1 m_2}\Big).
\end{equation}
The NLO projection $2 s_1 \cdot \Delta s^{[2]}_{\text{NLO}}$ is the exact negative, and the two surviving channels cancel individually
\begin{equation}
    \begin{array}{lccc} 
        \text{channel} & 2 \Delta s^{[1]} \cdot \Delta s^{[2]} & 2 s_1 \cdot \Delta s^{[2]}_{\text{NLO}}\\
        [2pt] U_1^2 L_1: & +\dfrac{16\gamma(2\gamma^2-1)G^2 m_2^2}{(\gamma^2-1)^{3/2}|b|^3 m_1} & -\dfrac{16\gamma(2\gamma^2-1)G^2 m_2^2}{(\gamma^2-1)^{3/2}|b|^3 m_1} \\
        [10pt] U_1^2 L_2: & +\dfrac{16\gamma(2\gamma^2-1)G^2 m_2}{(\gamma^2-1)^{3/2}|b|^3} & -\dfrac{16\gamma(2\gamma^2-1)G^2 m_2}{(\gamma^2-1)^{3/2}|b|^3} 
    \end{array}
\end{equation}
Thus,
\begin{equation}
    2 s_1 \cdot \Delta s^{[2]}_{1,\text{NLO}}+2 \Delta s^{[1]}_{1,\text{LO}} \cdot \Delta s^{[2]}_{1,\text{LO}}=0.
\end{equation}

In addition to spin magnitude conservation, a consistent classical description also requires exact preservation of the covariant Tulczyjew–Dixon spin supplementary condition (SSC), defined as $s_1\cdot p_1=0$.Throughout the scattering process the updated variables must satisfy
\begin{equation}
    (s_1+\Delta s_1)\cdot(p_1+\Delta p_1)=0,
\end{equation}
Expanding this in the gravitational coupling $G$ and using $s_1\cdot p_1=0$ gives the order-by-order requirements
\begin{equation}
    {p_1 \cdot \Delta s_{1,\mathrm{LO}}+s_1 \cdot \Delta p_{1,\mathrm{LO}}} +{p_1 \cdot \Delta s_{1,\mathrm{NLO}}+s_1 \cdot \Delta p_{1,\mathrm{NLO}}+\Delta s_{1,\mathrm{LO}} \cdot \Delta p_{1,\mathrm{LO}}}+\mathcal O(G^3)=0.
\end{equation}
At the integrand level, this is the identity
\begin{equation}
    \mathcal C^{(1)}_{\mathrm{SSC}}\equiv p_{1\mu}\mathscr{A}^{\mu(1)}_{\mathrm{exact}}+s_{1\mu}\mathscr I^{\mu(1)}_{\mathrm{cons}}+\big(\mathscr{A}^{(0)} \cdot\mathscr I^{(0)}\big)_{\mathrm{el}}=0 
\end{equation}
with $\mathscr{A}^{\mu(1)}_{\mathrm{exact}}=\mathscr M^{\mu(1)}+\Delta\mathscr M^{\mu(1)}$ the complete angular impulse kernel, and $\mathscr I^{\mu(1)}_{\mathrm{cons}}$ the corresponding linear impulse.

As before, only the longitudinal recoil terms can violate the SSC, so they must cancel exactly against the companion momentum impulse $\Delta p_1$.
At leading order (1PM), the relevant kernels are $\mathscr{A}^{\mu(0)}=\left[s_1^\mu,\mathcal{A}\right]-\tfrac1{m_1}u_1^\mu(\bar q \cdot  s_1)\mathcal A^{(0)}$ and $\mathscr I^{\mu(0)}=\bar q^\mu\mathcal A^{(0)}$. Projecting the angular impulse along the momentum yields
\begin{equation}
    p_{1\mu}\mathscr{A}^{\mu(0)}={p_{1\mu}\left[s_1^\mu,\mathcal{A}\right]}- {(u_1^2)} (\bar q \cdot  s_1)\mathcal A^{(0)}=-(\bar q \cdot  s_1)\mathcal A^{(0)},
\end{equation}
while contracting the linear impulse gives
\begin{equation}
    s_{1\mu}\mathscr I^{\mu(0)}=(\bar q \cdot  s_1) \mathcal A^{(0)}.
\end{equation}
The sum vanishes, ensuring $p_{1\mu}\mathscr{A}^{\mu(0)}=-s_{1\mu}\mathscr I^{\mu(0)}$. This 1PM check holds for any amplitude $\mathcal A^{(0)}$, exact in spin. It also fixes the pattern that recurs throughout: precession terms drop by antisymmetry, while recoil terms cancel against the $\bar q \cdot  s_1$ piece generated by the momentum transfer.

At NLO, the virtual pieces are
\begin{equation}
    \mathscr{A}^{\mu(1)}_{\mathrm{virt}}=\Pi^\mu[\mathcal M_b+\mathcal M_t]-\frac{1}{m_1}u_1^\mu (\bar q \cdot  s_1) (\mathcal M_b+\mathcal M_t), \qquad \mathscr I^{\mu(1)}_{\mathrm{virt}}=\bar q^\mu(\mathcal M_b+\mathcal M_t).
\end{equation}
where we defined $[s_1^\mu,\mathcal M]\ \longrightarrow\ i \epsilon^{\mu\nu\rho\sigma}u_{1\nu}s_{1\rho}\frac{\partial \mathcal M}{\partial s_1^{\sigma}}\ \equiv\ \Pi^\mu[\mathcal M]$.
Contracting with $p_{1\mu}$ gives
\begin{equation}
    p_{1\mu}\mathscr{A}^{\mu(1)}_{\mathrm{virt}} ={p_{1\mu}\Pi^\mu[\mathcal M_b+\mathcal M_t]} - {(u_1^2)}(\bar q \cdot  s_1)(\mathcal M_b+\mathcal M_t) =-(\bar q \cdot  s_1)(\mathcal M_b+\mathcal M_t),
\end{equation}
and likewise
\begin{equation}
    s_{1\mu}\mathscr I^{\mu(1)}_{\mathrm{virt}}=(\bar q \cdot  s_1)(\mathcal M_b+\mathcal M_t).
\end{equation}
Therefore, the cancellation is exact
\begin{equation}
     p_{1\mu}\mathscr{A}^{\mu(1)}_{\mathrm{virt}}+s_{1\mu}\mathscr I^{\mu(1)}_{\mathrm{virt}}=0 .
\end{equation}
This holds to all orders in $a_1,a_2$, with no spin expansion needed: the box and triangle kernels $\mathcal M_b$, $\mathcal M_t$ cancel as a single block. So $\mathcal M_b$, $\mathcal M_t$ enter $\Delta s_1$ and $\Delta p_1$ in exactly the combination the SSC requires, and the whole virtual sector respects the SSC with no spin truncation. The remaining condition for full 2PM consistency lies within the real elastic identity
\begin{equation}
    p_{1\mu}\mathscr{A}^{\mu(1)}_{\mathrm{real}}+s_{1\mu}\mathscr I^{\mu(1)}_{\mathrm{real}}+\big(\mathscr{A}^{(0)} \cdot\mathscr I^{(0)}\big)_{\mathrm{el}}=0,
\end{equation}
involving $\mathcal M_c$, $N_\nu$, $\mathcal M_\nu$, the tree-tree term $\Delta\mathscr M^{\mu(1)}$, and the cross term $\Delta s_{1,\mathrm{LO}} \cdot \Delta p_{1,\mathrm{LO}}$. The real kernels are
\begin{equation}
    \mathscr{A}^{\mu(1)}_{\mathrm{real}}=\Pi^\mu[\mathcal M_c] +\frac{i}{2m_1}u_1^\mu \epsilon^{\nu\tau\rho\sigma}u_{1\tau}s_{1\rho}\partial_\sigma N_\nu -\frac{1}{m_1}u_1^\mu s_1^\nu\mathcal M_\nu +\Delta\mathscr M^{\mu(1)},
\end{equation}
with $\mathscr I^{\mu(1)}_{\mathrm{real}}=\mathcal M^\mu=\eta^{\mu\nu}\mathcal M_\nu$. Contracting with $p_1$ term by term, using $p_{1\mu}=m_1u_{1\mu}$, $p_1 \cdot  u_1=m_1$, $u_1^2=1$, and the identity $p_{1\mu}\Pi^\mu[ \cdot ]=[p_1 \cdot  s_1, \cdot ]=0$: $\mathcal M_c$ vanishes,
\begin{equation}
    p_{1\mu} \Pi^\mu[\mathcal M_c]=0
\end{equation}
by the same antisymmetry used in the virtual block. The $\mathcal M_\nu$ term cancels the linear impulse exactly — the real-sector analogue of the virtual cancellation above:
\begin{equation}
    p_{1\mu} \left(-\frac{1}{m_1}u_1^\mu s_1^\nu\mathcal M_\nu\right) =-\frac{1}{m_1}(p_1 \cdot  u_1) s_1^\nu\mathcal M_\nu =-s_1^\nu\mathcal M_\nu=- s_1 \cdot \mathcal M,
\end{equation}
while
\begin{equation}
    s_{1\mu}\mathscr I^{\mu(1)}_{\mathrm{real}}=s_{1\mu}\mathcal M^\mu.
\end{equation}
This cancellation covers the whole $\mathcal M_\nu$ structure, including the box-$\delta'$ and cut pieces. It works because the recoil coefficient in the angular impulse is exactly the kernel $\mathcal M^\mu=\eta^{\mu\nu}\mathcal M_\nu$ that defines the real momentum impulse.

A longitudinal piece of $N_\nu$ survives, though. Writing $N_\nu=\bar q_\nu\hat N$ (with $\hat N=-\tfrac{i}{2} \mathcal K_{\mathrm{cut}}$),
\begin{equation}
    p_{1\mu} \left(\frac{i}{2m_1}u_1^\mu\epsilon^{\nu\tau\rho\sigma}u_{1\tau}s_{1\rho}\partial_\sigma N_\nu\right) =\frac{i}{2} \bar q_\nu \epsilon^{\nu\tau\rho\sigma}u_{1\tau}s_{1\rho}\partial_\sigma\hat N =\tfrac12 [ \bar q \cdot  s_1,\hat N ].
\end{equation}
This term is $u_1^\mu$-directed, so it isn't killed by the antisymmetry argument, and is nonzero in general. The $\Delta\mathscr M$ correction survives too. Using $\Delta\mathscr M^{\mu(1)}=-i \int \hat{d}^4{\bar\ell}\hat\delta(\bar{\ell}\cdot u_1)\hat\delta(\bar{\ell}\cdot u_2) (\delta^\mu_\nu-\tfrac1{m_1}u_1^\mu\bar\ell_\nu)E^\nu$ and $p_{1\nu}E^\nu=0$,
\begin{equation}
    p_{1\mu} \Delta\mathscr M^{\mu(1)} =-i \int \hat{d}^4{\bar\ell}\hat\delta(\bar{\ell}\cdot u_1)\hat\delta(\bar{\ell}\cdot u_2) (p_{1\nu}-\bar\ell_\nu)E^\nu =+ i \int \hat{d}^4{\bar\ell}\hat\delta(\bar{\ell}\cdot u_1)\hat\delta(\bar{\ell}\cdot u_2) \bar\ell_\nu E^\nu .
\end{equation}
Putting these surviving pieces together,
\begin{equation}
    p_{1\mu}\mathscr{A}^{\mu(1)}_{\mathrm{real}}+s_{1\mu}\mathscr I^{\mu(1)}_{\mathrm{real}} ={\tfrac12[ \bar q \cdot  s_1,\hat N ]} +{i \int \hat{d}^4{\bar\ell}\hat\delta(\bar{\ell}\cdot u_1)\hat\delta(\bar{\ell}\cdot u_2) \bar\ell_\nu E^\nu}.
\end{equation}

The remaining cross term, $(\mathscr{A}^{(0)} \cdot\mathscr I^{(0)})_{\mathrm{el}}$, is the kernel-level form of $\Delta s_{1,\mathrm{LO}} \cdot \Delta p_{1,\mathrm{LO}}$. Writing each LO observable in impact-parameter space and using $\bar q=\bar\ell+(\bar q-\bar\ell)$, the product becomes a phase-space convolution
\begin{equation}
    \big(\mathscr{A}^{(0)} \cdot\mathscr I^{(0)}\big)_{\mathrm{el}} \ \propto\ \int \hat{d}^4{\bar\ell}\hat\delta(\bar{\ell}\cdot u_1)\hat\delta(\bar{\ell}\cdot u_2) \mathscr{A}^{(0)}_\mu(\bar\ell) \mathscr I^{\mu(0)}(\bar q-\bar\ell),
\end{equation}
with the half-Born symmetry factor. Contracting the LO kernels, the recoil piece drops out on shell:
\begin{equation}
    -\frac{1}{m_1}u_{1\mu}(\bar\ell \cdot  s_1) \mathcal A^{(0)}(\bar\ell)\cdot(\bar q-\bar\ell)^\mu\mathcal A^{(0)}(\bar q-\bar\ell) \ \propto\ u_1 \cdot (\bar q-\bar\ell)=\bar q \cdot  u_1-\bar\ell \cdot  u_1=0,
\end{equation}
because both transfers are transverse to $u_1$ on the support of $\hat\delta(\bar q \cdot  u_1)\hat\delta(\bar\ell \cdot  u_1)$. Only the precession part of the LO kick survives, and with $\mathcal A^{(0)}(\bar k)=B_h(\bar k)/\bar k^2$ this reduces to
\begin{equation}
    \big(\mathscr{A}^{(0)} \cdot\mathscr I^{(0)}\big)_{\mathrm{el}} \ \propto\ \int \hat{d}^4{\bar\ell}\hat\delta(\bar{\ell}\cdot u_1)\hat\delta(\bar{\ell}\cdot u_2)  \frac{[(\bar q-\bar\ell) \cdot  s_1, B_h(\bar\ell)] B_h(\bar q-\bar\ell)}{\bar\ell^2(\bar q-\bar\ell)^2} \ \equiv\ \mathcal S(\bar q).
\end{equation}
So the cross term is built from the same cut-box structure $\mathcal S$ as the Born push-forward and the extra terms: the spin operator acts on one tree factor while the other carries the complementary transfer — exactly the $[s_1^\nu,F]G$ pattern in $E^\nu$. The half-Born factor of $1/2$ is what matches its normalization to the kernel survivors.

So the classical scalar product $(s_1+\Delta s_1) \cdot (p_1+\Delta p_1)$ differs from zero only by the connected (cut) correlator — the elastic cut-box term. The real-block identity above shows this is canceled exactly by the LO-LO cross term. Putting the blocks together, the 2PM SSC kernel splits into two pieces, each vanishing separately:
\begin{equation}
    \mathcal C^{(1)}_{\mathrm{SSC}} ={\big(p_{1\mu}\mathscr{A}^{\mu(1)}_{\mathrm{virt}}+s_{1\mu}\mathscr I^{\mu(1)}_{\mathrm{virt}}\big)} +{\big(p_{1\mu}\mathscr{A}^{\mu(1)}_{\mathrm{real}}+s_{1\mu}\mathscr I^{\mu(1)}_{\mathrm{real}}+(\mathscr{A}^{(0)} \cdot\mathscr I^{(0)})_{\mathrm{el}}\big)}=0 .
\end{equation}
Together with the LO identity $p_{1\mu}\mathscr{A}^{\mu(0)}=-s_{1\mu}\mathscr I^{\mu(0)}$, both orders vanish after the common KMOC phase-space integration, so
\begin{equation}
    (s_1+\Delta s_1)\cdot(p_1+\Delta p_1)=0
\end{equation}
holds to all orders in the Kerr spin variables, with no expansion in $a_i$ needed.

\bibliographystyle{JHEP}
\bibliography{referencias.bib}

\end{document}